\documentclass[12pt,a4paper]{article}
\usepackage{amsmath}
\usepackage{amsthm}
\usepackage{amssymb}
\usepackage{graphicx}
\usepackage{enumitem}
\usepackage{natbib}
\usepackage{url}
\usepackage[ruled]{algorithm2e}
\usepackage{algpseudocode}
\usepackage{subcaption}
\usepackage{tikz}
\usetikzlibrary{arrows,shapes.arrows,shapes.geometric,shapes.multipart,backgrounds,decorations.pathmorphing,positioning,fit,automata}
\tikzset{
	>=stealth',
	true/.style={
		rectangle,
		draw=black, very thick,
		text width=6.5em,
		minimum height=2em,
		text centered,
		fill=gray, opacity = 0.5},
	punkt/.style={
		rectangle,
		rounded corners,
		draw=black, very thick,
		text width=6.5em,
		minimum height=2em,
		text centered},
	est/.style={
		circle,
		draw=black, very thick,
		text centered},
	shade/.style={
		circle,
		draw=black, very thick, fill=gray!50,
		text centered},
	weight/.style={
		circle,
		draw=black, very thick,
		text width=6.5em,
		minimum height=2em,
		text centered},
	pil/.style={
		->,
		thick,
		shorten <=2pt,
		shorten >=2pt,},
	double/.style={
		<->,
		thick,
		shorten <=2pt,
		shorten >=2pt,},
	dash/.style={
		dashed,
		thick,
		shorten <=2pt,
		shorten >=2pt,},
	dashdouble/.style={
		<->,
		dashed,
		thick,
		shorten <=2pt,
		shorten >=2pt,}
}
\usepackage[hidelinks,hyperfootnotes=false]{hyperref}
\newcommand{\blind}{1}
\usepackage[a4paper, top=0.6in, bottom=0.6in, left=1in, right=1in]{geometry}
\usepackage{setspace}
\makeatletter
\g@addto@macro\normalsize{
  \setlength{\abovedisplayskip}{5pt}
  \setlength{\belowdisplayskip}{5pt}
  \setlength{\abovedisplayshortskip}{5pt}
  \setlength{\belowdisplayshortskip}{5pt}
}
\g@addto@macro\small{
  \setlength{\abovedisplayskip}{5pt}
  \setlength{\belowdisplayskip}{5pt}
  \setlength{\abovedisplayshortskip}{5pt}
  \setlength{\belowdisplayshortskip}{5pt}
}
\g@addto@macro\footnotesize{
  \setlength{\abovedisplayskip}{4pt}
  \setlength{\belowdisplayskip}{4pt}
  \setlength{\abovedisplayshortskip}{4pt}
  \setlength{\belowdisplayshortskip}{4pt}
}
\makeatother
\makeatletter
\renewcommand\footnotesize{
  \@setfontsize\footnotesize{9pt}{10pt}
  \abovedisplayskip=5pt
  \belowdisplayskip=5pt
}
\makeatother
\usepackage{titlesec}
\titlespacing*{\section}
  {0pt}
  {0pt}
  {0pt}
\titlespacing*{\subsection}
  {0pt}
  {10pt}
  {5pt}
\titlespacing*{\subsubsection}
  {0pt}
  {8pt}
  {4pt}
\theoremstyle{plain}
\newtheorem{theorem}{Theorem}
\newtheorem{lemma}{Lemma}
\newtheorem{corollary}{Corollary}
\theoremstyle{definition}
\newtheorem{assumption}{Assumption}
\newtheorem{remark}{Remark}
\allowdisplaybreaks
\newtheoremstyle{mystyle}
  {0pt}
  {0pt}
  {}
  {}
  {\bfseries}
  {.}
  {5pt plus 1pt minus 1pt}
  {}
\theoremstyle{mystyle}
\newtheorem{examplex}{Example}
\newenvironment{example}
  {\examplex}
  {\endexamplex}
\def\T{{ \mathrm{\scriptscriptstyle T} }}
\newcommand{\bigCI}{\!\perp\!\!\!\perp}

\newcommand{\nind}{\not\!\perp\!\!\!\perp}

\newcommand{\logit}{\text{logit}}
\newcommand{\E}{\mathbb{E}}
\begin{document}
\def\spacingset#1{\renewcommand{\baselinestretch}
{#1}\small\normalsize} \spacingset{1}
\if1\blind
{
  \title{\bf Causal Inference with a Hidden Treatment}
  \author{Ying Zhou
    \hspace{.2cm}\\
    Department of Statistics, University of Connecticut\\
    and \\
    Eric Tchetgen Tchetgen \\
    Department of Biostatistics, Epidemiology and Informatics \\
    Department of Statistics and Data Science\\
    University of Pennsylvania}
  \maketitle
} \fi
\if0\blind
{
  \bigskip
  \bigskip
  \bigskip
  \begin{center}
    {\LARGE\bf Causal Inference for a Hidden Treatment}
\end{center}
  \medskip
} \fi
\bigskip
\begin{abstract}
In many causal inference settings, the treatment of interest is not directly observed; instead, one or more error-prone proxy measurements are available, posing a fundamental identification challenge. Building on and extending existing identification methods for a hidden treatment leveraging available proxies, we develop a general semiparametric framework for causal effect estimation that accommodates several common causal estimands of interest. Notably, hidden treatment models fundamentally differ from classical missing-data settings, where existing semiparametric theory relies on a stringent positivity assumption requiring a nonzero probability of observing a complete case for each possible treatment value. In the hidden treatment setting, this assumption fails by design because the true treatment is never observed, creating substantial challenges for both semiparametric characterization and efficient estimation. To overcome these challenges, we develop a new semiparametric characterization for hidden treatment models by deriving a formal mapping between the orthogonal complement to the nuisance tangent space, which contains the set of all influence functions for a given causal functional in the oracle full-data model, and its counterpart in the observed hidden treatment model. This mapping provides an explicit closed-form expression for observed-data influence functions of the causal effect of interest, from which we derive the semiparametric efficiency bound under the hidden treatment model. This characterization leads to a class of semiparametric efficient estimators with a new form of multiple robustness or mixed bias property, facilitating inference based on nonparametric estimators of nuisance functions. A fundamental inferential challenge is the estimation of nuisance functions that depend on the hidden treatment, which prevents the direct application of standard nonparametric regression methods. To address this challenge, we introduce a novel iterative estimation algorithm and establish its large-sample theoretical properties. We investigate the finite-sample performance of the proposed estimators through simulation studies and demonstrate their practical utility in an application estimating the causal effect of Alzheimer’s disease on hippocampal volume using data from the Alzheimer’s Disease Neuroimaging Initiative.
\end{abstract}
\vfill
\newpage
\spacingset{1.9}
\section{Introduction}
\label{sec:intro}

In many practical settings where one seeks to evaluate the causal effect of a hypothetical intervention from observational data, the treatment of interest may not be directly observed without error. This can arise due to misclassification in ascertaining treatment status, or because the treatment itself is inherently latent and cannot be directly measured.
Examples of the former are common in health and social sciences, where self-reported measures are often error-prone. For instance, dietary intake in nutrition epidemiology may be misreported, and discrepancies between actual college attendance and reported education levels have been documented \citep{hu2012returns}.
An example of the latter arises in studying the causal impact of early-onset dementia on subsequent cognitive decline, as no gold standard exists for diagnosing dementia in early stages \citep{porsteinsson2021diagnosis}.

The study of measurement error models is longstanding, reflecting their broad relevance across disciplines. Much of this literature has focused on continuous treatments, where classical measurement error assumptions are commonly imposed. Numerous methods have been developed in this setting, and \cite{schennach2016recent} provides a comprehensive review. By contrast, causal inference with misclassified binary treatments has received considerably less attention \citep{bollinger1996bounding}. This disparity likely stems from the inherently nonclassical nature of binary misclassification: such error cannot, except in degenerate cases, be simultaneously additive, independent of the hidden treatment, and mean-zero (see \S{}S1 of the Supplementary Materials for details).

Several strands of work have addressed identification and estimation with misclassified binary treatments. \cite{mahajan2006identification} studies nonparametric and semiparametric conditional mean regression models with a misclassified binary regressor, using instrumental variables for identification. \cite{lewbel2007estimation} similarly employs a three-valued instrumental variable to identify and estimate the conditional average treatment effect (CATE). \cite{hu2008identification} extends this framework to general categorical treatments via a matrix diagonalization approach that solves a nonlinear system of equations. Relatedly, \cite{allman2009identifiability} establishes general identification results for unobserved discrete variables using multiple measurements, based on array decomposition techniques. Validation data, where both error-prone and error-free measurements are observed on a subsample, provide another commonly used strategy \citep{chen2005measurement, sepanski1993semiparametric}.

In the absence of auxiliary data, \cite{chen2009nonparametric} establish point identification for a nonparametric location-shift model using characteristic functions. They propose sieve maximum likelihood estimators for certain smooth functionals and develop corresponding semiparametric efficiency theory \citep{shen1997methods}. However, the resulting efficiency bounds are characterized variationally and lack closed-form expressions, complicating practical uncertainty quantification and robust estimation. A parallel literature emphasizes partial rather than point identification. For example, \cite{imai2010causal} derive bounds for the average treatment effect (ATE) under differential misclassification that may depend on the outcome, while \cite{ditraglia2019identifying} extend the model of \cite{mahajan2006identification} to allow for endogenous treatment assignment, establishing conditions for both partial and point identification.

In this paper, we develop a general framework for the identification and semiparametric estimation of a broad class of causal effects of a hidden binary treatment for which proxy variables are available, however without access to validation data. Our identification strategy builds on the matrix diagonalization technique of \cite{hu2008identification}, but shifts the focus from identifying features of the latent distribution to identifying causal effects and other causal functionals. Specifically, we leverage one \emph{strong proxy} error-prone treatment measurement, referred to as a \emph{surrogate treatment}, and a \emph{proxy variable} that provides additional exogenous variation associated with the hidden treatment. The surrogate is considered a strong proxy in the sense that it must be more accurate than random guessing, meaning that its misclassification error is known to admit a nontrivial bound, whereas no such bound is required for the proxy.

Throughout the paper, we use the ATE as a running example of our approach; other causal parameters and related theoretical extensions are discussed briefly in the main text, with additional details deferred to the Supplementary Materials. While the underlying identification strategy has a longstanding precedent, existing estimation methods face notable practical limitations. In particular, \cite{lewbel2007estimation} relies on GMM-based procedures for discrete covariates and local GMM for continuous covariates evaluated at fixed points, restricting their applicability. The plug-in estimators of \cite{mahajan2006identification} and \cite{hu2008identification} follow directly from the identification argument and are therefore sensitive to errors in estimating the observed data law, which can lead to numerical instability, including non-real estimated probabilities involving the hidden treatment.

In contrast to the estimators discussed above, our principal contribution is to develop a new semiparametric framework tailored to hidden-treatment models to construct regular and asymptotically linear (RAL) estimators for a broad class of causal parameters \citep{bickel1993efficient, newey1990semiparametric, van1991differentiable}. Rather than deriving influence functions directly in the observed-data model, we begin by characterizing all influence functions in an idealized oracle model where the treatment is observed and establish a formal mapping between the oracle and hidden-treatment settings. This characterization provides a principled way to construct influence functions under hidden treatment and leads to a broad class of new estimators. To our knowledge, this transport-based perspective has not previously been developed for hidden-treatment models. It is worth noting that in contrast with the rich existing semiparametric theory for missing data due to James Robins, Andrea Rotnitzky  and colleagues \citep{robins1994estimation}; also see \cite{tsiatis2007semiparametric} for a textbook exposition of the framework, which critically relies on a stringent positivity assumption that the probability of observing a complete-case unit for each treatment value, our framework does not require such a strong positivity assumption and as such, does not require validation data. As we show, this presents several new challenges for estimation and inference that existing methods do not contend with, for which we provide a comprehensive formal solution.

The proposed characterization of observed data influence functions yields a large class of new multiply robust semiparametric estimators for various causal functionals of interest in the hidden treatment setting.
Multiple robustness implies that the estimators remain consistent when some, but not all, nuisance functions are consistently estimated, and ensures root-$n$ consistency and asymptotic normality under mixed convergence rates \citep{robins2008higher, rotnitzky2021characterization, ghassami2022minimax}. A unique challenge we face in this setting is that nuisance functions involve the hidden treatment itself. We address this by developing an iterative, sieve-based estimation procedure inspired by minimum-distance methods \citep{chen2012estimation}, which directly targets the relevant nuisance parameters while preserving local robustness.

Importantly, our approach avoids the need to consistently estimate the full observed-data law and circumvents the inverse-problem difficulties inherent in existing plug-in estimators \citep{mahajan2006identification, hu2008identification}. We show that our estimator is semiparametric efficient for binary outcome and we characterize efficiency bounds and construct efficient or approximately efficient estimators for more general categorical and continuous outcome.

To summarize our key contributions: (1) We unify existing point identification techniques for a broad class of causal parameters in hidden treatment models where both surrogate and proxy measurements are available. (2) We develop semiparametric theory for causal functionals in hidden treatment settings by characterizing a formal mapping from the oracle tangent space, in which the treatment is observed, to the observed-data tangent space.
This mapping yields explicit relationships between oracle and observed-data influence functions, thereby providing a new characterization of the influence functions for the causal parameters of interest. (3) We characterize the semiparametric efficiency bound in the hidden treatment model by deriving closed-form expressions for the efficient influence function for a broad class of causal parameters. (4) We propose a class of multiply robust semiparametric estimators for causal functionals in the hidden treatment model, based on an iterative, sieve-based estimation procedure for nuisance functions involving the hidden treatment. We study these estimators in detail for the ATE and evaluate their performance through simulations and an application.  We emphasize that while (1) is a relatively straightforward generalization of existing identification results, our contribution is to tailor these results specifically towards the goal of causal inference; in contrast,  our main contributions (2)-(4) are entirely novel in addressing and resolving fundamental challenges in the hidden treatment problem which prior literature has not addressed at our level of generality.

The remainder of the paper is organized as follows. Section~\ref{sec:functionals} reviews common causal parameters of interest. Section~\ref{sec:ATE} introduces the hidden treatment model and presents identification results. Section~\ref{sec:eif} develops the semiparametric theory for the ATE, and Section~\ref{sec:estimation} describes the proposed estimator and its asymptotic properties. Extensions to other causal functionals are discussed in Section~\ref{sec:generalization}. We illustrate the performance of the method through simulations in Section~\ref{sec:simulation} and an application to the effect of Alzheimer’s disease on hippocampal volume under imperfect diagnosis in Section~\ref{sec:application}. Section~\ref{sec:discussion} concludes.

\section{Examples of causal parameters}\label{sec:functionals}
Let $A^*$ denote the hidden binary treatment, $Y$ be a binary, count, or continuous outcome, and $X$ be a vector of observed covariates.
The potential outcome $Y^{a^*}$ represents a unit's outcome had, possibly contrary to fact, the treatment $A^*$ been set to $a^*$. Depending on the nature of the outcome and of the scientific inquiry, a variety of causal effects might be of interest. Below we briefly review several examples of causal parameters of broad interest. Throughout, we make standard positivity, consistency and no unmeasured confounding assumptions about the hidden treatment as formally stated below.
For each parameter, we briefly review its identifying functional and corresponding class of influence functions that an oracle with access to $A^*$ might use as basis for inference.
\begin{example}\label{ex:ate}
\textbf{Average treatment effect (ATE).} The ATE measures the average additive effect of a treatment on the outcome, $\E(Y^{a^*=1}-Y^{a^*=0})$. It therefore suffices to identify $\E(Y^{a^*}), a^*=0,1$. The ATE can be identified by the oracle using the g-formula
\begin{align*}
\psi_{a^*}:=\E(Y^{a^*})=\int \E\left( Y\mid A^*=a^*,X\right) dF\left(
X\right) ,\text{ }a^*=0,1,
\end{align*}
under the following standard causal assumptions \citep{robins1986new}.
\begin{assumption}\label{asp:causal}
i. \emph{(Ignorability)} $Y^{a^*}\bigCI A^* \mid X$ for  $a^*=0,1$;
ii. \emph{(Positivity)} There exists a constant $\delta>0$ such that $\delta < \Pr(A^*=1\mid X) < 1-\delta$ almost surely;
iii. \emph{(Consistency)} $Y^{A^*}=Y$ almost surely.
\end{assumption}
Furthermore, the efficient influence function (EIF) for $\psi_{a^*}$ in a nonparametric model that places no restriction on the joint law of $(A^*,Y,X)$ is
\begin{align}\label{eq:IF_full}
   \phi_{a^*}^F = \frac{\mathbb{I}\left( A^*=a^*\right) }{\Pr\left( A^{\ast}=a^*\mid X\right) }\left\{ Y-\E\left( Y\mid A^* = a^*,X\right) \right\} + \E\left( Y\mid A^* = a^*,X\right) -\psi _{a^*},
\end{align}
where $\Pr( A^*=a^* \mid X)$ is the propensity score and $\E( Y\mid A^*,X)$ is the outcome regression function \citep{robins1994estimation}. The superscript $F$ denotes the full data model where $A^*$ is observed.
\end{example}
\begin{remark}
For categorical $Y$, the parameters of interest would be $\E\{\mathbb{I}(Y^{a^*}=y)\}$ for each category $y$. Its EIF can be obtained by replacing $Y$ with $\mathbb{I}(Y=y)$ in \eqref{eq:IF_full}.
\end{remark}
\begin{example}\textbf{The effect of treatment on the treated (ETT).} The ETT is defined as $\psi_{ett} = \E(Y^{a^*=1} - Y^{a^*=0}\mid A^*=1)$. It is often of interest when evaluating the effect of harmful exposures, in which case it may not be of interest to consider an intervention that would forcefully expose unexposed individuals, such that the causal effect of interest is that experienced by exposed individuals only.
Under a weaker version of Assumption \ref{asp:causal}, whereby (i) is assumed to hold only for $a^*=0$, and (ii) is replaced by $\Pr(A^*=1\mid X)/\Pr(A^*=0\mid X)< \infty$ almost surely, the ETT is identified by $\psi_{ett}=\E\{\E(Y\mid A^*=1,X) -  \E(Y\mid A^*=0,X) \mid A^*=1 \}$.
The EIF for $\psi_{ett}$ as derived in \cite{hahn1998role} is given by
{\small
\begin{align}\label{eq:eif_ett}
\phi^F_{ett}&=\frac{\mathbb{I}(A^*=1)}{\Pr(A^*=1)}\{Y-\E(Y\mid A^*=1,X)\}-\frac{\mathbb{I}(A^*=0)}{\Pr(A^*=1)}\frac{\Pr(A^*=1\mid X)}{1-\Pr(A^*=1\mid X)}\{Y-\E(Y\mid A^*=0,X)\} \nonumber\\
& \quad +\frac{\E(Y\mid A^*=1,X) - \E(Y\mid A^*=0,X)-\psi_{ett}}{\Pr(A^*=1)}\Pr(A^*=1\mid X).
\end{align}}
\end{example}
\begin{example}
\textbf{Quantile treatment effect (QTE).}   Quantile treatment effects examine the impact of a treatment on a given quantile of the outcome (e.g., the median) rather than the average outcome.
Suppose we are interested in the $\gamma$-quantile of a continuous potential outcome $Y^{a^*}$, i.e., the $\theta_1^{a^*}$ such that $\Pr(Y^{a^*}\le \theta_1^{a^*})=\gamma$.
Under Assumption \ref{asp:causal}, the QTE is identified. The EIF for $\theta_1$ is proportional to
\begin{align}\label{eq:qte_eif_full}
\phi_{a^*,\gamma}^F=\mathbb{I}(A^*=a^*)\frac{\mathbb{I}(Y\le\theta_1^{a^*})-\eta_1(X;\theta_1^{a^*})}{\eta_2(a^*;X)}+\eta_1(X;\theta_1^{a^*})-\gamma,
\end{align}
where $\eta_1(X;\theta_1^{a^*})=\Pr(Y\le\theta_1^{a^*}\mid X,A^*=a^*)$ and $\eta_2(a^*;X)=\Pr(A^*=a^*\mid X)$.
\end{example}
\begin{example}
\textbf{Semi-linear structural regression model.} One flexible model for estimating treatment effects while controlling for confounders is $\E(Y\mid A^*,X) =\beta_x(X) + \beta_{a^*}A^*$, where $\beta_x(\cdot)$ is an unknown function of $X$. Any influence function for $\beta_{a^*}$ in this model has the form
\begin{align}\label{eq:if_sl}
\phi^F_{sl} = h(X)\{Y-\beta_{a^*}A^* - \beta_x(X)\}\{A^*-\Pr(A^*=1\mid X)\},
\end{align}
where $h(X)$ is any user-specified function with finite variance.
\end{example}
\begin{example}
\textbf{Marginal structural generalized linear model.} Consider the model
$g\{\E(Y^{a^*}\mid V)\}=\beta_0+\beta_1a^*+\beta_2^\T V$ where $g$ is a canonical link function and $V$ is a strict subset of the covariates $X$. Any influence function for this model must be of the form
{\small
\begin{align}\label{eq:if_msm}
\phi^F_{msm}& = Wh(A^*,V)\left\{Y - g^{-1}(\beta_0+\beta_1A^*+\beta_2^\T V)  \right\}-Wh(A^*,V)\left\{\E(Y\mid A^*,X) - g^{-1}(\beta_0+\beta_1A^*+\beta_2^\T V)\right\}\nonumber\\
& \quad +\sum_{a^{*}} (-1)^{1-a^*} h(a^*,V)\left\{\E(Y\mid A^*=a^*,X) - g^{-1}(\beta_0+\beta_1a^*+\beta_2^\T V)\right\},
\end{align}}
where $W = 1/\Pr(A^*\mid X)$, and $h(A^*,V)$ is a user-specified function with the same dimension as $(\beta_0,\beta_1,\beta_2^\T)^\T$.
\end{example}
\medskip
Let $\mathbb P_n f = n^{-1}\sum_{i=1}^n f(O_i)$ denote the empirical average. Because influence functions have mean zero at the true data-generating distribution, they provide a convenient tool for constructing semiparametric estimators. For instance, in Example \ref{ex:ate}, by solving the estimating equation $\mathbb{P}_n\{\widehat{\phi}_{a^*}^F(\widehat{\psi}_{a^*})\}=0$ based on an estimate of the EIF \eqref{eq:IF_full}, one recovers the well-known augmented-inverse-probability-weighted estimator (AIPW):
\begin{align}\label{eq:ate_full_estimator}
\widehat{\psi}_{a^*} = \mathbb{P}_n \left\{ \frac{\mathbb{I}\left( A^*=a^*\right) }{\widehat{\Pr}\left( A^{\ast}=a^*\mid X\right) }\left\{ Y-\widehat{\E}\left( Y\mid a^*,X\right) \right\} +\widehat{\E}\left( Y\mid a^*,X\right) \right\},
\end{align}
where $\widehat{\Pr}( A^{\ast}=a^*\mid X)$ and $\widehat{\E}( Y\mid a^*,X)$ are estimated propensity score and outcome regression function, respectively. This estimator is known to be doubly robust \citep{scharfstein1999adjusting},
in that $\widehat{\psi}_{a^*} $ is consistent if either $\widehat{\Pr}( A^{\ast}=a^*\mid X)$ or $\widehat{\E}( Y\mid a^*,X)$ is consistent, but not necessarily both. When both are consistent, the estimator achieves the semiparametric efficiency bound for the nonparametric model.
Because the true treatment $A^*$ is not directly observed, the estimator \eqref{eq:ate_full_estimator} is infeasible and the target parameter $\psi_{a^*}$ is not identifiable without additional assumptions. In Section~\ref{sec:ATE}, we present a general identification strategy that appropriately recovers valid inferences about the causal effect of a hidden treatment using two auxiliary variables: a surrogate $A$ and a proxy $Z$ for the hidden treatment. Rather than observing $A^*$, we assume access to an error prone measurement $A$ we refer to as a \emph{surrogate} variable,   representing a genuine attempt to measure $A^*$, such as a self-reported treatment or inferred treatment from ICD-10 codes recorded in medical records, both of which are imperfect measurements of the true underlying treatment. In contrast, a \emph{proxy} variable $Z$ need only be associated with $A^*$ and need not constitute a direct measurement of the treatment.
For instance, in a mask-wearing study, the hidden treatment \(A^*\) corresponds to the individual's actual mask-wearing behavior, the surrogate \(A\) could be a self-reported mask-wearing indicator, and the proxy \(Z\) could be mask possession or recent mask purchase history. In our real data application, the hidden treatment \(A^*\) corresponds to Alzheimer's disease (AD) status, the surrogate \(A\) is the recall subscore of the MMSE, and the proxy \(Z\) is the orientation subscore.
These concepts are formalized in Section~\ref{sec:ATE}.
\section{Identification of ATE}\label{sec:ATE}
Until stated otherwise, our exposition largely focuses on the ATE.
It is instructive and straightforward to compute the bias incurred if one were to naively compute the ATE using the surrogate $A$ instead of $A^*$; mainly, the bias of the surrogate-based g-formula for $\E(Y^{a^*})$ is
given by $\int \delta(X,a^*)dF(X)$, where
$
\delta(X,a^*)=\E(Y\mid A=a^*,X)-\E(Y\mid A^*=a^*,X)
=\{\E(Y\mid A^*=1-a^*,X)-\E(Y\mid A^*=a^*,X)\}\Pr(A^*=1-a^*\mid A=a^*,X),
 a^*=0,1.
$
This bias will generally be non-null, unless the ATE of $A^*$ on $Y$ is actually null, or if misclassification error is absent. \footnote{ Although unlikely, it is possible that under certain exceptional laws, the conditional bias $\delta(X)$ given $X$ itself is not null, with opposite signs over different parts of the support of $X$, such that however it averages to zero over $X$.}
In order to identify $\psi_{a^*}$, our hidden treatment model imposes Assumptions \ref{asp:relevance} and \ref{asp:conditional_independence} below.
\begin{assumption}[Relevance]\label{asp:relevance}
$Y\nind A^*\mid X; A\nind A^*\mid X; Z\nind A^* \mid X$.
\end{assumption}
\begin{assumption}[Conditional independence]\label{asp:conditional_independence}
$Y$, $A$, and $Z$ are mutually independent conditional on $(A^*,X)$. That is, $Y\bigCI A\bigCI Z \mid A^*, X$.
\end{assumption}
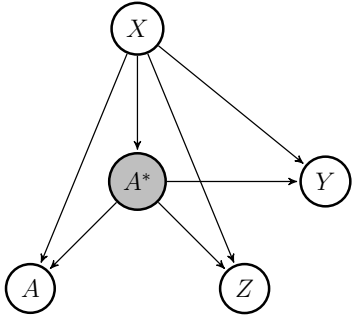
\begin{figure}[h]
\centering
\begin{subfigure}[t]{0.45\textwidth}
\centering
\scalebox{0.8}{
\begin{tikzpicture}[->,>=stealth',shorten >=1pt,auto,node distance=2.3cm,
     semithick, scale=0.50,
     pre/.style={-,>=stealth,semithick,blue,ultra thick,line width=1.5pt}]
     \tikzstyle{every state}=[fill=none,draw=black,text=black]
     \node[shade] (As)                    {$A^*$};
     \node[est] (Z) [below right =1.6 cm of As] {$Z$};
     \node[est] (A) [below left =1.6 cm of As] {$A$};
     \node[est] (Y) [right =2.2 cm of As] {$Y$};
     \node[est] (X) [above = 1.6 cm of As] {$X$};
     \path
     (X) edge node {} (As)
     (X) edge node {} (A)
     (X) edge node {} (Z)
     (X) edge node {} (Y)
     (As) edge node {} (A)
     (As) edge node {} (Z)
     (As) edge node {} (Y);
\end{tikzpicture}}
\caption{}
\label{fig:measurement_error_ATE}
\end{subfigure}
\hspace{0.05\textwidth}
\begin{subfigure}[t]{0.45\textwidth}
\centering
\scalebox{0.8}{
\begin{tikzpicture}[->,>=stealth',shorten >=1pt,auto,node distance=2.3cm,
     semithick, scale=0.50,
     pre/.style={-,>=stealth,semithick,blue,ultra thick,line width=1.5pt}]
     \tikzstyle{every state}=[fill=none,draw=black,text=black]
     \node[shade] (As)                    {$A^*$};
     \node[est] (Z) [below right =1.6 cm of As] {$Z$};
     \node[est] (A) [below left =1.6 cm of As] {$A$};
     \node[est] (Y) [right =2.2 cm of As] {$Y$};
     \node[est] (X) [above = 1.6 cm of As] {$X$};
     \path
     (X) edge node {} (As)
     (X) edge node {} (A)
     (X) edge node {} (Z)
     (X) edge node {} (Y)
     (As) edge node {} (A)
     (As) edge node {} (Z)
     (As) edge node {} (Y)
     (A) edge node {} (Y)
     (Z) edge node {} (Y)
     (A) edge node {} (Z);
\end{tikzpicture}}
\caption{}
\label{fig:oracle_model}
\end{subfigure}
\caption{(a) $A^*$ is the hidden binary exposure, $A$ is a binary surrogate, $Z$ is a binary proxy, $Y$ is the outcome, and $X$ is the confounder. (b) A DAG for the oracle nonparametric model $\mathcal{M^*}_{np}$.}
\end{figure}
Figure \ref{fig:measurement_error_ATE} provides a graphical illustration of Assumptions \ref{asp:relevance} and \ref{asp:conditional_independence}. Note that the arrow from $A^*$ to $Z$ can be reversed, without compromising Assumption \ref{asp:conditional_independence}, thus  allowing $Z$ to act as an ``instrument" in this context \citep{hu2008identification}.
The condition $Y\nind A^*\mid X$ in Assumption \ref{asp:relevance} rules out the \emph{causal null} hypothesis that $A^*$ has no causal effect on $Y$.
This is not overly restrictive, as the causal null is empirically testable under Assumption \ref{asp:conditional_independence}, which implies the stronger condition that $Y$ is independent of $(A,Z)$ given $X$.
\begin{remark}
We implicitly assume that $A\ne A^*$ and $Z\ne A^*$, a necessary implication of misclassification error. If  $A$ equals $A^*$, the matrix $P(Y,Z,A=1)$ which will be introduced shortly, would not be of full rank. Therefore, a non-zero determinant of  $P(Y,Z,A=1)$ indicates that $A\ne A^*$. A similar argument applies to $Z$.
\end{remark}
Under Assumptions \ref{asp:relevance} and \ref{asp:conditional_independence}, the joint distribution of $(A^*,Y,A,Z,X)$ is  determined up to labeling. To illustrate the idea, consider a simplified scenario where $Y$ is binary and $X=\emptyset$.
In this situation, the joint distribution of the observed data $(Y,A,Z)$ has $2^3-1=7$ degrees of freedom, and we have 7 unknown parameters: $\E(A^*), \E(Y\mid A^*=a^*), \E(A\mid A^*=a^*), \E(Z\mid A^*=a^*), a^*=0,1$.
Let $P(Y,Z)=(\Pr(Y=2-i,Z=2-j))_{i,j}$, $P(Y,Z,A=1)=(\Pr(Y=2-i,Z=2-j,A=1))_{i,j}$, $P(Y\mid A^*)=(\Pr(Y=2-i\mid A^*=2-j))_{i,j}$, $P(Z\mid A^*)=(\Pr(Z=2-i\mid A^*=2-j))_{i,j}$, $P_D(A^*)=\text{diag}(\Pr(A^*=1),\Pr(A^*=0))$, $P_D(A=1\mid A^*)=\text{diag}(\Pr(A=1\mid A^*=1),\Pr(A=1\mid A^*=0))$.
By Assumption \ref{asp:conditional_independence}, we have
$
P(Y,Z)=P(Y\mid A^*)P_D(A^*)P(Z\mid A^*)^\T,
P(Y,Z,A=1)=P(Y\mid A^*)P_D(A^*)P_D(A=1\mid A^*)P(Z\mid A^*)^\T.
$
By Assumption \ref{asp:relevance}, $P(Y,Z)$ is invertible.
Multiplying $P(Y,Z,A=1)$ by $P(Y,Z)^{-1}$ gives
$
P(Y,Z,A=1)P(Y, Z)^{-1} = P(Y\mid A^*)P_D(A=1\mid A^*)P(Y\mid A^*)^{-1},
$
which as noted in \cite{hu2008identification} is the canonical form of the eigendecomposition of the matrix on the left-hand side. Since the right-hand side is observed, we can solve for the eigenvalues and eigenvectors, which correspond to the diagonal elements of  $P_D(A=1\mid A^*)$ and the columns of $P(Y\mid A^*)$, respectively. While eigenvectors are identified up to scale, the requirement that columns of $P(Y\mid A^*)$ sum to 1 fixes it.
The remaining challenge lies in correctly matching each eigenvalue with its respective diagonal element in $P_D(A=1\mid A^*)$. To resolve this, we propose the following assumption encoding four alternative ways in which the surrogate $A$ carries sufficient information about $A^*$ to resolve the remaining identification gap.
\begin{assumption}[Surrogate]\label{asp:label}
At least one of the following holds: for any value $x$ of $X$,
\begin{enumerate}[label=Condition \arabic*, leftmargin=*, itemsep=0.5ex, topsep=.5ex, parsep=0pt, partopsep=0pt]
    \item\hspace{-0.5em}. $\Pr(A=0\mid A^*=1, X=x)+ \Pr(A=1\mid A^*=0, X=x)<1$ \label{cdn:1}
    \item\hspace{-0.5em}. $\Pr(A^*=0\mid A=1, X=x)+ \Pr(A^*=1\mid A=0, X=x)<1$ \label{cdn:2}
    \item\hspace{-0.5em}. $\Pr(A^*=1\mid A=1, X=x)>\Pr(A^*=0\mid A=1, X=x)$ \label{cdn:3}
    \item\hspace{-0.5em}. $\Pr(A^*=0\mid A=0, X=x)>\Pr(A^*=1\mid A=0, X=x)$ \label{cdn:4}
\end{enumerate}
\end{assumption}
Assumption \ref{asp:label} clarifies the different roles of $A$ and $Z$, although they appear symmetric in Figure \ref{fig:measurement_error_ATE}. While both the proxy $Z$ and the surrogate $A$ need to satisfy Assumptions \ref{asp:relevance} and \ref{asp:conditional_independence}, an extra Assumption \ref{asp:label} is imposed on the surrogate.
This assumption essentially suggests that $A$, as a surrogate, should provide more information about $A^*$ than a random guess.
The condition is expected to hold if $A$ is a meaningful but imperfect measurement of $A^*$.
\ref{cdn:1} is equivalent to $\Pr(A=1\mid A^*=0,X=x)<\Pr(A=1\mid A^*=1,X=x)$. Similarly, \ref{cdn:2} can be written as $\Pr(A^*=1\mid A=0,X=x)<\Pr(A^*=1\mid A=1,X=x)$. \ref{cdn:1} and \ref{cdn:2} are effectively of similar but symmetric nature.
\ref{cdn:3} indicates the precision $\Pr(A^*=1\mid A=1, X=x)>0.5$, while \ref{cdn:4} states that the false omission rate $\Pr(A^*=1\mid A=0, X=x)<0.5$. These conditions are reasonable in many scenarios.
For example, consider $A^*$ as the actual COVID-19 vaccination status of a patient, and $A$ as whether the patient was vaccinated at a specific hospital. If hospital records show a positive vaccination status, it must be that the patient was truly vaccinated, making $\Pr(A^*=1\mid A=1, X=x)=1$, thereby fulfilling \ref{cdn:2} (and \ref{cdn:1}).
If one can assume that the misclassification error is balanced across covariates, such that $X\bigCI A\mid A^*$, i.e., the surrogate $A$ only depends on $A^*$, then one can  safely omit $X$ from Assumption \ref{asp:label}. We note that \cite{hu2008identification} considers alternatives to Conditions 1-4.
The informal proof given above is formalized in Theorem \ref{thm:identification} below in a more general setting where $Y$ can be either binary, a count, or a continuous outcome,
and $X$ is possibly vector-valued.
For the proof of this theorem, please refer to \S{}S2 in the Supplementary Materials.
\begin{theorem}\label{thm:identification}
Under Assumptions \ref{asp:causal}, \ref{asp:relevance}, \ref{asp:conditional_independence}, and \ref{asp:label}, the potential outcome means $\psi_{a^*}=\E(Y^{a^*})$ for $a^*=0,1$ are uniquely identified, and therefore the ATE $\psi_{\mathrm{ATE}}=\psi_1-\psi_0=\E(Y^{a^*=1}-Y^{a^*=0})$ is uniquely identified whether $Y$ is binary, a count, or a continuous outcome.
\end{theorem}
A key step in the proof entails establishing point identification of the joint conditional oracle law $\Pr(A,Z,Y, A^*\mid X=x)=\Pr(A\mid A^*,X=x)\Pr(Z\mid A^*,X=x)\Pr(Y\mid A^*,X=x)\Pr(A^*\mid X=x)$, which, in principle implies point identification of any smooth functional of the oracle law.  This observation is crucial to establishing identification for more general causal functionals beyond the ATE.
The next section builds on this key identification result to construct highly robust and efficient semiparametric estimators of $\psi_{a^{*}}$ under the semiparametric model $\mathcal{M}$ for the observed data, induced by the oracle semiparametric model $\mathcal{M^*}$ defined by the conjunction of Assumptions \ref{asp:causal}--\ref{asp:label}.
\section{Semiparametric theory}\label{sec:eif}
Having established identification of $\psi_{a^*}$, we develop the semiparametric theory for the hidden-treatment model $\mathcal{M}$. We characterize its tangent space, nuisance tangent space, and their ortho-complement; this provides a characterization of the set of influence-function space for $\psi_{a^*}$ in model $\mathcal{M}$, that explicitly links oracle influence functions in the full data model to corresponding observed-data influence functions. This characterization underpins the construction of efficient semiparametric estimators for $\psi_{a^*}$ and principled inferences under $\mathcal{M}$.
Specifically, we begin with the oracle full-data model in which $A^*$ is observed and characterize its tangent space. We then transition to the observed-data model where $A^*$ is hidden through the relevant coarsening operator, which relates oracle and observed-data score functions. This relationship induces corresponding constraints linking oracle and hidden-model influence functions, thereby characterizing the influence-function space in the hidden-treatment setting. These influence functions generate a rich class of feasible moment equations with appealing robustness and efficiency properties.
Under Assumptions \ref{asp:relevance} and \ref{asp:conditional_independence}, the full data distribution is
$
f\left( Y,A,A^*,Z,X\right) =f\left( Y\mid A^*,X\right) \allowbreak f\left(
A\mid A^*,X\right)\allowbreak f\left( Z\mid A^*,X\right) f\left( A^*\mid X\right)
f\left( X\right),
$
and the tangent space is given by
\begin{align}\label{eq:full_tangant}
\Lambda ^{F}=\left\{ \Lambda _{y}^{F}\oplus \Lambda _{a}^{F}\oplus \Lambda
_{z}^{F}\oplus \Lambda _{a^*}^{F}\oplus \Lambda _{x}\right\} \cap
L_{0}^{2,F},
\end{align}
where
$\Lambda _{y}^{F} =\left\{ G_{y}=g_{y}\left( Y,A^*,X\right): \E\left(G_{y}\mid A^*,X\right)=0\right\}$,
$\Lambda _{a}^{F} =\left\{ G_{a}=g_{a}\left( A,A^*,X\right): \E\left(G_{a}\mid A^*,X\right) =0\right\} $,
$\Lambda _{z}^{F} =\left\{ G_{z}=g_{z}\left( Z,A^*,X\right): \E\left(G_{z}\mid A^*,X\right) =0\right\} $,
$\Lambda _{a^*}^{F} =\left\{ G_{a^*}=g_{a^*}\left(A^*,X\right): \E\left( G_{a^*}\mid X\right) =0\right\} $,
$\Lambda _{x} =\left\{ G_{x}=g_{x}\left( X\right): \E\left( G_{x}\right)=0\right\}$. We use $L^2_0(V)$ to denote the Hilbert space of measurable functions of $V$ with mean zero and finite variance equipped with the covariance inner product, and $L_{0}^{2,F}$ is a shorthand for $L^2_0(Y,A,A^*,Z,X)$.
The ortho-complement of the tangent space \eqref{eq:full_tangant} is characterized by
\begin{align}\label{eq:tangentspace_full}
\begin{split}
\Lambda ^{F,\bot }=\bigl\{ K=&k\left( Y,A,A^*,Z,X\right): \E\left(
K\mid Y,A^*,X\right)\\
&=\E\left( K\mid A,A^*,X\right) =\E\left( K\mid Z,A^{\ast
},X\right) =0\bigr\} \cap L_{0}^{2,F},
\end{split}
\end{align}
which can equivalently be expressed as
\begin{align}\label{eq:tangent_space}
\begin{split}
\Lambda ^{F,\bot }=\bigl\{ W=M-&\E\left( M\mid Y,A^*,X\right) -\E\left(
M\mid A,A^*,X\right)-\E\left( M\mid Z,A^*,X\right) \\
& +2 \E\left( M\mid A^{\ast
},X\right): M=m\left( Y,A,A^*,Z,X\right) \in L_{0}^{2,F}\bigr\}.
\end{split}
\end{align}
The derivation of \eqref{eq:tangentspace_full} and \eqref{eq:tangent_space} is deferred to \S{}S3 in the Supplementary Materials.  We refer to the oracle nonparametric model $\mathcal{M}_{np}^*$ as the collection of oracle laws compatible with the DAG in Figure \ref{fig:oracle_model}, which places no restriction on the joint distribution of $(A^*,Y,Z,A,X)$. The following result then holds.
\begin{lemma}\label{lem:eif}
The EIF of $\psi_{a^*}$ in the oracle hidden treatment model $\mathcal{M^*}$ is equal to the  EIF in the oracle nonparametric model $\mathcal{M}_{np}^*$.
\end{lemma}
The proof of Lemma \ref{lem:eif} is relegated to the Supplementary Materials, \S{}S4.
The EIF for $\psi_{a^*}$ in  $\mathcal{M^*}_{np}$, denoted by $\phi_{a^*}^{F, \text{ eff}}$ is well known and given by \eqref{eq:IF_full}.
Therefore, the set of influence functions for $\psi_{a^*}$ in the full data model $\mathcal{M^*}$ is the linear variety
$
 \phi_{a^*}^{F, \text{ eff}} \oplus \Lambda ^{F,\bot }
$
\citep{tsiatis2007semiparametric}.
So far, we have characterized the tangent space, the set of influence functions and the EIF for $\psi_{a^*}$ in $\mathcal{M^*}$. We now turn to the observed data model $\mathcal{M}$.
The following result summarizes our characterization of the tangent space of the observed data hidden treatment model, and its ortho-complement, obtained by applying a coarsening operator which maps the tangent space for the full data model, $\Lambda^F$, onto $L^2_0(Y,A,Z,X)$. The proof is deferred to the Supplementary Materials \S{}S5.
\begin{theorem} \label{thm:ortho_tangent_space}
The tangent space  $\Lambda$ for the observed data hidden treatment model is $\Lambda = \{\Lambda_y + \Lambda_a + \Lambda_z + \Lambda_{a^*} + \Lambda_x\}\cap L^{2}_0$, where $\Lambda_y = \E(\Lambda_y^F\mid Y,A,Z,X)$, $\Lambda_a = \E(\Lambda_a^F\mid Y,A,Z,X)$, $\Lambda_z = \E(\Lambda_z^F\mid Y,A,Z,X)$, $\Lambda_{a^*} = \E(\Lambda_{a^*}^F\mid Y,A,Z,X)$, $\Lambda_x = \E(\Lambda_x^F\mid Y,A,Z,X)$. The ortho-complement of the observed data tangent space, $\Lambda^\bot$, is characterized by
$$
\Lambda^{\bot} = \{L=l(Y,A,Z,X):\E(L\mid Y,A^*,X) = \E(L\mid A,A^*,X) = \E(L\mid Z,A^*,X)=0\}\cap L^{2}_0,
$$
which is equivalently  expressed as
{\small
\begin{flalign}\label{eq:ortho}
&\Lambda^{\bot} = \bigl\{\{A-\E(A\mid A^*=0,X)\}\{Z-\E(Z\mid A^*=1,X)\}h_1(Y,X) +\{A-\E(A\mid A^*=1,X)\} && \nonumber \\
&\{Z-\E(Z\mid A^*=0,X)\}h_2(Y,X): \E\{h_1(Y,X)\mid A^*,X\}=\E\{h_2(Y,X)\mid A^*,X\}=0\bigr\}\cap L^{2}_0. && \raisetag{0.9\baselineskip}
\end{flalign}
}
\end{theorem}
We illustrate Theorem \ref{thm:ortho_tangent_space} in key special cases. Suppose $Y$ is categorical, taking values in $\{1,\ldots, k\}$. In this case, the functions $h_1(Y,X)$ and $h_2(Y,X)$ must satisfy
\begin{align}\label{eq:null_space}
[\, h_j(Y=1,X)\ \cdots\ h_j(Y=k,X) \,]
\ M_{Y\mid A^*,X} =
[\, 0\ \ 0 \,],
\quad j=1,2,
\end{align}
where $M_{Y\mid A^*,X}=(v_1,v_2)$, with $v_1 = \left(\Pr(Y=1\mid A^*=0,X),\cdots,\Pr(Y=k\mid A^*=0,X)\right)^\T$ and $v_2 = \left(\Pr(Y=1\mid A^*=1,X),\cdots,\Pr(Y=k\mid A^*=1,X)\right)^\T$.
Thus, $h_1$ and $h_2$ lie in the null space of the $k\times 2$ matrix $M_{Y\mid A^*,X}$. Under Assumption \ref{asp:relevance}, the null space has dimension $k-2$ when $k\ge 2$.
When $Y$ is binary, i.e., $k=2$, $M_{Y\mid A^*,X}$ is full rank,
leading to the unique solution of \eqref{eq:null_space}: $h_j(Y,X) = 0, j=1,2$.
This is further detailed in Corollary \ref{cor:tangent_space}.
\begin{corollary}\label{cor:tangent_space}
Suppose that  $A^*,A,Z,Y$ are all binary variables, then the observed data hidden treatment model is locally nonparametric in the sense that the tangent space of $\mathcal{M}$ is $L^2_0$, i.e., $\Lambda = \{L=l(Y,A,Z,X)\in L^2_0\}$.
\end{corollary}
For more general $Y$, i.e., categorical or continuous, the following theorem provides a closed-form characterization of $\Lambda$ which facilitates influence function-based inference. In this vein, let $P_1 = \{ h(Y,X): \E\{ h(Y,X)\mid A^*,X \}=0\}\cap L^2_0$ and the space of functions
\begin{align*}
& P_2 = \Biggl\{v(Y,X) = \frac{v_1(Y,X)-\E\{v_1(Y,X)\mid A^*=0,X\}}{\E\{v_1(Y,X)\mid A^*=1,X\}-\E\{v_1(Y,X)\mid A^*=0,X\}} \\
& \quad +  \frac{v_0(Y,X)-\E\{v_0(Y,X)\mid A^*=1,X\}}{\E\{v_0(Y,X)\mid A^*=0,X\}-\E\{v_0(Y,X)\mid A^*=1,X\}} - 1 : v_1(Y,X),v_0(Y,X)\in L^2 \Biggr\}.
\end{align*}
\begin{theorem}\label{thm:representation} Suppose that $Y$ is bounded, i.e., $|Y|<L$; then we have that $P_1=P_2$; and
\begin{align*}
&\Lambda^{\bot} = \bigl\{\{A-\E(A\mid A^*=0,X)\}\{Z-\E(Z\mid A^*=1,X)\}w_1(Y,X)  \nonumber\\
&\qquad +\{A-\E(A\mid A^*=1,X)\}\{Z-\E(Z\mid A^*=0,X)\}w_2(Y,X): w_1,w_2 \in P_2 \bigr\}.
\end{align*}
\end{theorem}
Such equivalence provides a closed-form characterization of $P_1$ and hence a closed-form characterization of $\Lambda^\bot$.
See \S{}S6 in the Supplementary Materials for the proof.
Having characterized the observed data tangent space, we note that the set of all observed data influence functions for $\psi _{a^*}$ in the hidden treatment model can be written as $\phi+\Lambda^\bot$ for any influence function $\phi$. Therefore, it suffices to identify a single influence function.
To this end, we establish a relationship between influence functions in the observed-data and oracle models.
Since we have already characterized the influence functions for the oracle model, we may leverage this relationship to derive the corresponding influence functions in the observed hidden treatment model. The following result formalizes this characterization for  $\psi_{a^*}$.
\begin{lemma}\label{lem:Q}
A function $Q=q(Y,A,Z,X)$ is an influence function for $\psi_{a^*}$ in $\mathcal{M}$ if and only if there exists an oracle influence function $\text{IF}^{F}$ for $\psi_{a^*}$ in $\mathcal{M^*}$ such that
\begin{align}\label{eq:if_condition}
\begin{split}
\E\left( Q\mid Y,A^*,X\right)  &=\E\left( \text{IF}^{F}\mid Y,A^*,X\right), \\
\E\left( Q\mid A,A^*,X\right)  &=\E\left( \text{IF}^{F}\mid A,A^*,X\right), \\
\E\left( Q\mid Z,A^*,X\right)  &=\E\left( \text{IF}^{F}\mid Z,A^*,X\right).
\end{split}
\end{align}
\end{lemma}
Once such a $Q$ is identified, the set of influence functions for $\psi_{a^*}$ in the observed data model is given by $Q + \Lambda^\bot$.  Theorem \ref{thm:eif} constructs one such influence function.
\begin{theorem}\label{thm:eif} The following expression provides a closed-form influence function for $\psi_{a^*}$ in the hidden treatment model $\mathcal{M}$, applicable when $Y$ is binary, count or continuous:
\begin{align}\label{eq:IF_observed}
\phi_{a^*}(Y,A,Z,X) = &\frac{A-\E(A\mid 1-a^*,X)}{\E(A\mid a^*,X)-\E(A\mid 1-a^*,X)}\frac{Z-\E(Z\mid 1-a^*,X)}{\E(Z\mid a^*,X)-\E(Z\mid 1-a^*,X)}\frac{1}{\Pr(A^*=a^*\mid X)}  \nonumber\\
&\{Y-\E(Y\mid a^*,X)\}+ \E(Y\mid a^*,X) - \psi_{a^*},
\end{align}
with $Y$ replaced by $\mathbb{I}(Y=y)$ for a categorical $Y$ where $\psi_{a^*}=\E\{ \mathbb{I}(Y^{a^*}=y)\}$.
 Equation \eqref{eq:IF_observed} admits a triple robustness property: $\E(\phi_{a^*})=0$ if any one of the following holds: \\
1. $\E(A\mid A^*, X)$ and $\E(Y\mid A^*, X)$ are correctly specified.\\
2. $\E(Z\mid A^*, X)$ and $\E(Y\mid A^*, X)$ are correctly specified.\\
3. $\E(A\mid A^*, X)$, $\E(Z\mid A^*, X)$ and $\E(A^*\mid X)$ are correctly specified.
The EIF $\phi_{a^*}^\text{eff}$ is given by the orthogonal projection of  $\phi_{a^*}$ onto the observed data tangent space of the hidden treatment model, i.e.,  $\phi_{a^*}^\text{eff}=\Pi(\phi_{a^*}\mid \Lambda)$
    where $\Pi$ is the orthogonal projection operator. Therefore,
   \begin{align*}
    \phi_{a^*}^\text{eff}&=\phi_{a^*}(Y,A,Z,X) - \{A-\E(A\mid A^*=0,X)\}\{Z-\E(Z\mid A^*=1,X)\}w^{opt}_1(Y,X) \\
   & \qquad  -\{A-\E(A\mid A^*=1,X)\}
\{Z-\E(Z\mid A^*=0,X)\}w^{opt}_2(Y,X).
  \end{align*}
When $Y$ is binary, $w^{opt}_1=w^{opt}_2=0$, and thus \eqref{eq:IF_observed} is the EIF for $\psi_{a^*}$ in the hidden treatment model $\mathcal{M}$. Closed-form expressions for $\{w^{opt}_j:j=1,2\}$ in the categorical outcome case are given in the Supplementary Materials \S{}S8.2.
For continuous $Y$, closed-form expressions for $\{w^{opt}_j:j=1,2\}$ and thus for the EIF are not generally available; in this case, we give in the Supplementary Materials a simple approach for obtaining an approximate EIF based on a basis function approximation of  $\{w^{opt}_j:j=1,2\}$.
\end{theorem}
Theorem \ref{thm:eif} reveals an insightful connection between an oracle influence function and its observed-data counterpart. Specifically, comparing \eqref{eq:IF_observed} with the full data influence function \eqref{eq:IF_full}, we observe that the former can be obtained by substituting the indicator function $\mathbb{I}_{a^*}(A^*)$ in the latter with:
\begin{align}\label{eq:substitute}
M(A,Z,X)=\frac{A-\E(A\mid 1-a^*,X)}{\E(A\mid a^*,X)-\E(A\mid 1-a^*,X)}\frac{Z-\E(Z\mid 1-a^*,X)}{\E(Z\mid a^*,X)-\E(Z\mid 1-a^*,X)}.
\end{align}
Notably, \eqref{eq:substitute} is conditionally unbiased for $\mathbb{I}_{a^*}(A^*)$, satisfying $\E\{M(A,Z,X)\mid A^*,X\} = \mathbb{I}_{a^*}(A^*)$.
See \S{}S9 for the derivation.
This provides an elegant link between \eqref{eq:IF_full} and \eqref{eq:IF_observed}. In \eqref{eq:IF_full}, the unobserved $A^*$ appears only through $\mathbb{I}_{a^*}(A^*)$, whereas in \eqref{eq:IF_observed}, $\mathbb{I}_{a^*}(A^*)$ is replaced by the observable $M(A,Z,X)$, whose conditional expectation recovers $\mathbb{I}_{a^*}(A^*)$.
This observation suggests a general approach for constructing influence functions in hidden treatment models, which can be extended to other causal functionals of $(A^*,X,Y)$ discussed in \S\ref{sec:functionals}.
See \S\ref{sec:generalization} for details.
\section{Semiparametric estimation and inference}\label{sec:estimation}
In this section, we develop semiparametric estimators for $\psi_{a^*}$, based on the influence function   \eqref{eq:IF_observed}. A standard approach involves the estimation of nuisance parameters, which are then substituted in the influence function to obtain a substitution estimator:
\begin{align}\label{eq:plugin_estimator}
\begin{split}
\widehat{\psi}_{a^*} = \mathbb{P}_n\Biggl\{\frac{A-\widehat{\E}(A\mid 1-a^*,X)}{\widehat{\E}(A\mid a^*,X)-\widehat{\E}(A\mid 1-a^*,X)}\frac{Z-\widehat{\E}(Z\mid 1-a^*,X)}{\widehat{\E}(Z\mid a^*,X)-\widehat{\E}(Z\mid 1-a^*,X)}  \\
\frac{1}{\widehat{\Pr}(a^*\mid X)}\{Y-\widehat{\E}(Y\mid a^*,X)\}+ \widehat{\E}(Y\mid a^*,X) \Biggr\}.
\end{split}
\end{align}
The estimator depends on seven nuisance functions
$p_{z,1}(X), p_{z,0}(X), p_{a,1}(X), p_{a,0}(X), \allowbreak p_{y,1}(X), p_{y,0}(X),$ and $p_{a^*}(X),$
where
$p_{z,a^*}(X)=\E(Z\mid A^*=a^*,X)$, $p_{a,a^*}(X)=\E(A\mid A^*=a^*,X)$,
$p_{y,a^*}(X)=\E(Y\mid A^*=a^*,X)$ for $a^*=0,1$, and $p_{a^*}(X)=\Pr(A^*=1\mid X)$. For notational convenience, we collect these nuisance functions into the vector
$\theta(X) =\bigl(p_{z,1}(X), p_{z,0}(X), p_{a,1}(X), p_{a,0}(X), p_{y,1}(X),\allowbreak p_{y,0}(X), p_{a^*}(X)\bigr)^\top$.
Typically, nuisance functions are estimated in an initial step using the observed data.  However, because the nuisance functions appearing in \eqref{eq:plugin_estimator} involve the unobserved $A^*$, their estimation is challenging, especially when  using nonparametric estimators. One possible strategy is to first estimate the observed data law $\Pr(Y,A,Z\mid X)$ and then recover $\theta(X)$ via a standard plug-in principle; see \S{}S10 of the Supplementary Materials.
A major limitation of this approach is its heavy dependence on accurate estimation of the observed data distribution.
The nuisance functions arise as solutions to systems of polynomial equations that are smooth in the observed data distribution, so even moderate estimation error in $\Pr(Y,A,Z\mid X)$ can propagate nonlinearly and lead to unstable or infeasible nuisance estimates that violate model constraints.
To address this challenge, we propose a novel semiparametric iterative algorithm that directly targets the nuisance functions and avoids the need to first estimate the observed data law.
Define, for $W\in\{ Z,A,Y \}$,
\begin{align}\label{eq:AZY}
W_{a^*} &=\frac{ W-\E\left( W\mid A^*=1-a^*,X\right)  }{ \E\left(
W\mid A^*=a^*,X\right) -\E\left( W\mid A^*=1-a^*,X\right)  }, \quad a^*\in \{ 0,1 \}.
\end{align}
Then $\E\left( W_{a^*}\mid A^*,X\right) =\mathbb{I}( A^*=a^*) $.
We now construct moment equations for the nuisance functions. As an illustration, consider $p_{z,1}(X) = \E(Z\mid A^* = 1, X)$.
In the full data model where $A^*$ is observed, we have the conditional moment restriction
$\E\left [\mathbb{I}(A^*=1)\{p_{a^*}(X)\}^{-1}\{ Z - p_{z,1}(X) \}  \ \Big| \ X \right] = 0.$
When $A^*$ is not observed but $A,Y$ are observed, we replace the indicator $\mathbb{I}(A^*=1)$ with the observable $A_1Y_1$, yielding the observed-data conditional moment equation
\begin{align}\label{eq:condition1}
\E\left [\{ Z - p_{z,1}(X) \} A_1Y_1 \{ p_{a^*}(X) \}^{-1}  \ \Big| \ X \right] = 0.
\end{align}
Applying the same construction yields the following conditional moment equations for the remaining nuisance functions:
\begin{align}
\E\left\{ \left( Z-p_{z,0}(X) \right) A_{0}Y_{0} \{1-p_{a^*}(X)\}^{-1}\mid X\right\}
&=0, \\
\E\left\{ \left( A-p_{a,1}(X) \right) Z_{1}Y_{1} p_{a^*}(X)^{-1}\mid X\right\}
&=0, \\
\E\left\{ \left( A- p_{a,0}(X) \right) Z_{0}Y_{0} \{1-p_{a^*}(X)\}^{-1}\mid X\right\}
&=0, \\
\E\left\{ \left( Y-p_{y,1}(X) \right) Z_{1}A_{1} p_{a^*}(X)^{-1}\mid X\right\}
&=0, \\
\E\left\{ \left( Y- p_{y,0}(X)\right) Z_{0}A_{0} \{1-p_{a^*}(X)\}^{-1}\mid X\right\}
&=0.
\end{align}
For $p_{a^*}(X)$, a natural moment condition is
$
\E\left\{ Z_{1}A_{1}Y_{1}-p_{a^*}(X) \mid X\right\} =0.
$
We however adopt the following adjusted moment equation
\begin{align}\label{eq:condition7}
\E\left\{ \left.
\begin{array}{c}
Z_{1}A_{1}Y_{1} - p_{a^*}(X)
-\frac{Z_{1}A_{1}}{p_{y,1}(X) -p_{y,0}(X)
 }\left\{ Y-p_{y,1}(X) \right\}  \\
-\frac{A_{1}Y_{1}}{p_{z,1}(X) -p_{z,0}(X)}\left\{ Z-p_{z,1}(X) \right\}
-\frac{Z_{1}Y_{1}}{p_{a,1}(X) -p_{a,0}(X)}\left\{ A-p_{a,1}(X) \right\}
\end{array}
\right\vert X\right\} =0,
\end{align}
which enjoys additional robustness: the moment condition remains unbiased if any one of the three models $p_{z,a^*}(X), p_{a,a^*}(X), p_{y,a^*}(X)$ is misspecified.
To summarize, our goal is to recover $\theta\in\Theta$, the vector of seven nuisance functions, from the seven conditional moment restrictions \eqref{eq:condition1}--\eqref{eq:condition7}, where $\Theta$ denotes the parameter space.  For each equation $j=1,\ldots,7$, let $\rho_j(O,\theta)$ denote the function inside the conditional expectation, where $O=(Y,A,Z,X)$, and define $m_j(X,\theta)=\mathbb{E}\{\rho_j(O,\theta)\mid X\}$. The moment restrictions are therefore equivalent to $m_j(X,\theta)=0$, almost surely.
We impose the following positivity and separation condition on the nuisance functions.
\begin{assumption}[Positivity and separation]\label{asp:separation}
There exists $\epsilon_1>0$ such that, almost surely,
(a) $p_{a^*}(X), p_{z,1}(X), p_{z,0}(X), p_{a,1}(X), p_{a,0}(X), p_{y,1}(X), p_{y,0}(X)\in[\epsilon_1,1-\epsilon_1]$;
(b) $\min \big\{ |p_{z,1}(X)-p_{z,0}(X)|,
      |p_{a,1}(X)-p_{a,0}(X)|,
      |p_{y,1}(X)-p_{y,0}(X)| \big\} \ge \epsilon_1$.
\end{assumption}
Part (a) is a standard positivity condition ensuring overlap of the latent classes. Part (b) guarantees sufficient separation across latent classes, preventing near-singularity of the conditional moment equations and ensuring stable local identification.
Let $\mathcal H$ be the space of square–integrable functions of $X$, so that
\(\Theta\subset\mathcal H^7\).
Let $\mathcal H_{k(n)}\subset\mathcal H$ be a $k(n)$-dimensional sieve space spanned by basis functions $\{p_1,\ldots,p_{k(n)}\}$, and define
$\bar{p}_{k(n)}(x) = (p_1(x),\ldots,p_{k(n)}(x))^\top \in \mathbb{R}^{k(n)}.
$
We assume that $\{\mathcal H_{k(n)}\}$ is an increasing sequence of spaces whose union is dense in $\mathcal H$.
Rather than solving $m(X,\theta)=0$ exactly, we seek $\theta\in\Theta_n:=\Theta\cap\mathcal H_{k(n)}^{7}$ such that the population projection of $m(X,\theta)$ onto the sieve space vanishes:
\begin{align}\label{eq:sieve_population}
\mathcal P_{k(n)}[m(X,\theta)] = 0,
\end{align}
where $\mathcal P_{k(n)}$ denotes the projection operator onto the sieve space $\mathcal H_{k(n)}$.
As $k(n)\to\infty$, solutions in the sieve space $\Theta_n$ approximate the solution to the full conditional moment problem in $\Theta$.
A natural estimator for solving \eqref{eq:sieve_population} is the sieve minimum distance (MD) estimator,
\begin{align}\label{eq:sieve_md_estimator}
\widehat\theta^{\mathrm{MD}}
    = \arg\min_{\theta \in \Theta_n} \widehat Q_n^{\mathrm{MD}}(\theta),
    \qquad
    \widehat Q_n^{\mathrm{MD}}(\theta) = \frac{1}{n}\sum_{i=1}^{n}\|\widehat m(X_i,\theta)\|_2^{2},
\end{align}
where $\widehat m(X,\theta)$ denotes the series least-squares (LS) estimator of $m(X,\theta)$ based on the sieve basis $\{p_1,\ldots,p_{k(n)}\}$, and \(\|\cdot\|_2\) denotes the Euclidean norm in \(\mathbb R^7\).
Since $m(X,\theta)=\E\{\rho(O,\theta)\mid X\}$, the estimator $\widehat{m}(X,\theta)$ can be obtained by regressing $\rho(O_i,\theta)$ on $\bar{p}_{k(n)}(X_i)$.  To express this compactly, we define the orthogonal projection operator $\mathcal P_{k(n),n}$ by
$
\mathcal P_{k(n),n}[f](X)
:= \bar{p}_{k(n)}(X)^\top (P^\top P)^{-} P^\top f_n,
$
where $P=(\bar{p}_{k(n)}(X_1),\ldots,\bar{p}_{k(n)}(X_n))^\top \in \mathbb{R}^{n\times k(n)}$, $f_n=(f(O_1),\ldots,f(O_n))^\top$, and $(P^\top P)^{-}$ denotes a generalized inverse of $P^\top P$.
For vector-valued \(f\), \(\mathcal P_{k(n),n}\) is applied componentwise. With this notation,
$
\widehat m(X,\theta)= \mathcal P_{k(n),n}[\rho(O,\theta)](X).
$
By construction, $\widehat Q_n^{\mathrm{MD}}(\theta)\ge 0$, with equality if and only if
\begin{equation}\label{eq:sample}
\mathcal P_{k(n),n}[\rho(O,\theta)](X_i) = 0, \quad  i=1,\ldots,n.
\end{equation}
That is, the empirical projection of $\rho(O,\theta)$ onto $\mathcal H_{k(n)}$ vanishes on the sample support of $X$.  Condition \eqref{eq:sample} is therefore the sample analog of the population-level projected moment equation \eqref{eq:sieve_population}.
Sieve MD estimators of the form \eqref{eq:sieve_md_estimator} can, in principle, be used to solve conditional moment restrictions of the type considered here.  However, our problem requires jointly solving seven coupled conditional moment equations, and directly minimizing the sieve MD criterion can be numerically unstable and may result in noisy estimates.
Moreover, implementing sieve MD estimation for multivariate conditional moments of this complexity requires substantially customized numerical optimization.
To address these challenges, we propose an iterative algorithm that is straightforward to implement and directly targets the empirical projected moment equation \eqref{eq:sample}.  As shown below, any limit point of the algorithm satisfies \eqref{eq:sample}. Consequently, it attains the minimum value zero of the sieve MD criterion and is therefore a sieve MD solution. This provides a practical computational approach while allowing us to leverage existing asymptotic theory for sieve MD estimators.
We now describe the iterative procedure for estimating the nuisance functions $\theta$. Starting from an initial value $\theta^{(0)}\in\Theta_n$, the population-level update is given by
\begin{equation*}
\theta^{(t+1)} = \mathcal{P}_{k(n)}\!\left[ \theta^{(t)} + \eta\,\rho\big(O,\theta^{(t)}\big) \right],
\end{equation*}
where $\eta>0$ is a step size.
If the sequence $\{\theta^{(t)}\}$ converges to a limit $\theta^{*}\in\Theta_n$, then $\theta^{*}$ must satisfy
$\theta^{*} = \mathcal{P}_{k(n)}\!\left[\theta^{*} + \eta\,\rho(O,\theta^{*})\right].$
Since $\mathcal{P}_{k(n)}[\theta^{*}]=\theta^{*}$ for any $\theta^{*}\in\mathcal{H}_{k(n)}^7$, subtracting $\theta^{*}$ from both sides gives
$
\mathcal{P}_{k(n)}\!\left[\rho(O,\theta^{*})\right] = 0,
$
implying the population projected moment condition \eqref{eq:sieve_population}.
At the sample level, the iterative algorithm becomes
\begin{align}\label{eq:iterative_sample}
\theta^{(t+1)} = \mathcal P_{k(n),n}\!\big[\theta^{(t)} + \eta\,\rho(O,\theta^{(t)})\big].
\end{align}
To implement \eqref{eq:iterative_sample}, each component $\theta_j$ of $\theta$ is updated by regressing $\theta^{(t)}_j(X_i) + \eta\,\rho_j(O_i,\theta^{(t)})$ on $X_i$ using a series LS estimator with basis functions $\{p_1,\ldots,p_{k(n)}\}$.
If the iteration converges to a limit $\theta^* \in \mathcal H^7_{k(n)}$, then the fixed-point relation
$
\theta^* = \mathcal P_{k(n),n}\!\left[\theta^* + \eta\,\rho(O,\theta^*)\right]
$
must hold on the sample points of $X$.  Since $\mathcal P_{k(n),n}[\theta^*] = \theta^*$ for $\theta^* \in \mathcal H^7_{k(n)}$, subtracting $\theta^*$ yields
$
\mathcal P_{k(n),n}\!\big[\rho(O,\theta^*)\big] = 0
$
over the sample support of $X$.
Therefore, upon convergence, the iterative estimator solves the same empirical projected moment equation \eqref{eq:sample} as the sieve MD estimator. In this sense, the iterative procedure provides a constructive algorithm for solving $\mathcal P_{k(n),n}[\rho(O,\theta)] = 0$, whereas the sieve MD estimator targets the same condition by minimizing $\widehat Q_n^{MD}(\theta)$.
Because $A$, $Z$, and sometimes $Y$ are binary, it is convenient to perform the updates on the log-odds scale so that all probability estimates remain in $[0,1]$.
Let $\ell=\logit(\theta)$ denote the component-wise logit transform.  Then the update rule becomes
\begin{align}\label{eq:logit_update}
\ell^{(t+1)} = \mathcal P_{k(n),n}\!\big[\ell^{(t)}+\eta\,L(O,\theta^{(t)})\big],
\end{align}
where $L(O,\theta^{(t)})$ is the logit-scale counterpart of $\rho(O,\theta^{(t)})$. For example, writing  $\ell^{(t)}_{y,1}(x)=\logit(p_{y,1}^{(t)}(x))$, the update for $\ell_{y,1}$ is
$
\ell^{(t+1)}_{y,1} = \mathcal P_{k(n),n}\!\big[\ell^{(t)}_{y,1}+\eta\,L^{(t)}_{y,1}\big],
$
where
\[
L^{(t)}_{y,1} = \frac{(Y-p_{y,1}^{(t)}(X))\,Z_1^{(t)}A_1^{(t)}}
       {p_{a^*}^{(t)}(X)\,p_{y,1}^{(t)}(X)\{1-p_{y,1}^{(t)}(X)\}}.
\]
Analogous expressions for the remaining components are provided in Supplementary Materials \S{}S11.
We next formalize the connection between the proposed iterative algorithm and the sieve MD estimator. Specifically, if the iterative sequence converges and the empirical projected moment equation has a unique solution in the sieve space, then the iterative limit coincides with the sieve MD estimator. Consequently, the iterative estimator inherits the asymptotic properties of the sieve MD estimator under the same regularity conditions.
For notational simplicity, Theorem~\ref{thm:finite_dim_equi} and Corollary~\ref{cor:CP_rate} are stated using the original-scale update \eqref{eq:iterative_sample}; all results apply equally to the logit-scale iterative algorithm.
\begin{theorem}\label{thm:finite_dim_equi}
Consider the iterative sequence $\theta^{(t)}$ defined by \eqref{eq:iterative_sample} and suppose that $\theta^{(t)} \to \widehat\theta^{\mathrm{it}} \in \Theta_n$.
If the empirical projected moment equation
$
\mathcal P_{k(n),n}\!\big[\rho(O,\theta)\big]=0
$
has a unique solution $\theta^\star \in \Theta_n$, then
$
\widehat\theta^{\mathrm{it}}= \widehat\theta^{\mathrm{MD}}= \theta^\star,
$
where $\widehat\theta^{\mathrm{MD}}$ is the sieve MD estimator defined in \eqref{eq:sieve_md_estimator}.
\end{theorem}
The proof is deferred to the Supplementary Materials \S{}S12. Under the conditions of Theorem \ref{thm:finite_dim_equi}, the iterative estimator $\widehat\theta^{\mathrm{it}}$ coincides with the sieve MD estimator $\widehat\theta^{\mathrm{MD}}$ as a finite-sample function of the data. Consequently,
it suffices to study the large-sample behavior of the sieve MD estimator.
Under standard regularity conditions, the sieve MD estimator $\widehat\theta^{\mathrm{MD}}$ is consistent for the target solution $\theta_0$; see, for example, Theorem~3.1 of \citet{chen2012estimation}. A formal proof is provided in the Supplementary Materials \S{}S13.
We now establish the convergence rate of the sieve MD estimator in \eqref{eq:sieve_md_estimator} by appealing to a general result of \cite{chen2012estimation}. To keep the presentation self-contained, we restate a simplified version of their Corollary 5.1, specialized to the unpenalized and well-posed setting relevant for our analysis.
Let \(\delta_{m,n}\to0\) be such that, uniformly over \(\theta\in\Theta_n\),
\begin{align}\label{eq:sample_criterion}
\frac{1}{n}\sum_{i=1}^n\|\widehat m(X_i,\theta)\|_2^2
\ge c \|m(\cdot,\theta)\|^2-O_p(\delta_{m,n}^2)
\end{align}
for some constant \(c>0\), where $\|\cdot\|$ denotes the $L^2(P_X)$ norm for functions of $X$.
\begin{corollary}\label{cor:CP_rate}
Let $\theta_0\in\Theta$ denote the target solution to
$
m(X,\theta)=\E\{\rho(O,\theta)\mid X\}=0 \text{ a.s.}
$.
Suppose the following conditions hold:
(a) There exist $\alpha>0$ and a sequence $v_k\uparrow\infty$ such that
$
\inf_{\theta\in\Theta_n}\|\theta-\theta_0\|  = O(v_{k(n)}^{-\alpha}).
$
(b)
The series LS estimator \(\widehat m\) satisfies the sample criterion condition \eqref{eq:sample_criterion} with \(\delta_{m,n}^2 \asymp k(n)/n.\)
(c) The Fr\'echet derivative $T_{\theta_0}$ of $m(\cdot,\theta)$ at $\theta_0$ satisfies
$
\|T_{\theta_0}h\|  \gtrsim \|h\|  \text{ for all } h \in (\Theta_n-\theta_0) \cap \mathcal{N}
$
for some neighborhood $\mathcal{N}$ of 0.
If $k(n)\to\infty$ and $k(n)/n\to0$, then
\[
\|\widehat\theta^{\mathrm{MD}}-\theta_0\|
=O_p \left(v_{k(n)}^{-\alpha} + \sqrt{\frac{k(n)}{n}}\right).
\]
\end{corollary}
In our setting, condition (c) follows from Lemma S2 in the Supplementary Materials, \S{}S14. Indeed, \(T_{\theta_0}h=-h\), and hence
\(\|T_{\theta_0}h\|=\|h\|\)
under the \(L^2(P_X)\) norm.  Therefore, the problem is locally well-posed and no ill-posedness factor appears in the rate.
Next, we verify condition (b) of Corollary \ref{cor:CP_rate}. The estimator \(\widehat m\) is obtained by projecting \(\rho(O,\theta)\) onto the finite-dimensional sieve basis. By the series LS result of \cite{chen2012estimation}, the sample criterion lower-bound condition holds with \(\delta_{m,n}^2\asymp k(n)/n\), provided the squared approximation bias of the series projection of \(m(\cdot,\theta)\) is no larger than the variance term \(k(n)/n\). Thus condition (b) holds for the series LS estimator used here.
We are now ready to obtain the convergence rate of the sieve MD estimator.
\begin{theorem}[Rate of the Sieve MD estimator]\label{thm:rate}
Suppose that Assumption~\ref{asp:separation} holds and that the sieve parameter space $\Theta_n$ satisfies condition (a) in Corollary~\ref{cor:CP_rate}. Suppose also that \(\widehat m\) is the series LS estimator described above, so that condition (b) of Corollary~\ref{cor:CP_rate} holds.
If
$k(n)\to\infty$ and $k(n)/n\to0$, then
$
\|\widehat\theta^{\mathrm{MD}}-\theta_0\|  = O_p\!\left(v_{k(n)}^{-\alpha} + \sqrt{k(n)/n}
\right).
$
In particular, if $v_k\asymp k^{1/d}$ and $k(n)\asymp n^{d/(2\alpha+d)}$, then
$
\|\widehat\theta^{\mathrm{MD}}-\theta_0\|  = O_p\!\left(n^{-\alpha/(2\alpha+d)}\right).
$
\end{theorem}
It remains to verify that the iterative estimator is itself well-defined, in the sense that the sample update \eqref{eq:iterative_sample} converges to a fixed point locally. This is a computational stability property of the algorithm and is distinct from the large-sample properties of the resulting estimator.
Let
$
U_n(\theta)=\mathcal P_{k(n),n}\!\big[\theta+\eta\,\rho(O,\theta)\big]
$
denote the sample update map. Under the regularity conditions given in the Supplementary Materials, $U_n$ is locally contractive in a neighborhood of the empirical solution, with probability approaching one. Consequently, for any initialization in this neighborhood, the iteration converges to a unique fixed point \(\widehat\theta^{\rm it}\in\Theta_n\).
By Theorem~\ref{thm:finite_dim_equi}, this fixed point coincides with the sieve MD estimator $\widehat\theta^{\mathrm{MD}}$.
A formal statement of local convergence and its full proof are provided in the Supplementary Materials \S{}S15.
In practice, we initialize $\theta^{(0)}$ using the marginal estimators for $\E(A \mid A^*=a^*)$, $\E(Z \mid A^*=a^*)$, $\E(Y \mid A^*=a^*)$ for $a^*=0,1$, and $\E(A^*)$, so that every sample point is assigned the same set of starting values. The details of this initialization are provided in the Supplementary Materials \S{}S16. See Algorithm \ref{alg:po} for implementation details of estimating $\theta$ for the binary $Y$ case. The case when $Y$ is continuous is similar and is omitted here.
\RestyleAlgo{boxruled}
\DecMargin{0.5em}
\begin{algorithm}[t]
\caption{An iterative algorithm to estimate nuisance functions in \eqref{eq:plugin_estimator} for binary $Y$}
\label{alg:po}
{\small
\begin{itemize}[label={},leftmargin=0pt,itemsep=0ex,topsep=0ex]
\item \textbf{Input:} data $\{(Y_i,A_i,Z_i,X_i)\}_{i=1}^n$, step size $\eta>0$,
maximum iterations $T$, tolerances $\tau_\Delta, \tau_{\mathrm{mean}}, \tau_{\mathrm{cond}}$, and maximum restarts $R$.
\item \textbf{Initialize:} set restart counter $r\gets 0$ and \texttt{converged}$\gets$False.
\item \textbf{While} $r<R$ \textbf{and} \texttt{converged}$=$False \textbf{do:}
\begin{itemize}[leftmargin=1.5em, itemsep=0.5ex]
\item Get a warm start $\ell^{(0)}=\big(\ell^{(0)}_{z,1},\ell^{(0)}_{z,0},
\ell^{(0)}_{a,1},\ell^{(0)}_{a,0},\ell^{(0)}_{y,1},\ell^{(0)}_{y,0},
\ell^{(0)}_{a^*}\big)$.
\item \textbf{For} $t=1,\ldots,T$ \textbf{do:}
\begin{itemize}[leftmargin=1.5em, itemsep=0.5ex]
        \item Compute $(Z_1^{(t)},Z_0^{(t)},A_1^{(t)},A_0^{(t)},Y_1^{(t)},Y_0^{(t)})$
        defined in \eqref{eq:AZY} using $\ell^{(t-1)}$.
        \item Compute $L^{(t)} = (L_{z,1}^{(t)},L_{z,0}^{(t)},L_{a,1}^{(t)},L_{a,0}^{(t)},L_{y,1}^{(t)},L_{y,0}^{(t)}, L_{a^*}^{(t)})$.
        \item Update logits on the log-odds scale: $\widetilde{\ell}^{(t)}\;=\;\ell^{(t-1)} + \epsilon\, L^{(t)}.$
        \item Smooth $\widetilde{\ell}^{(t)}$ via a series estimator fitted on the full sample,
        producing $\ell^{(t)}(X_i)$ for all $i=1,\ldots,n$.
\item \textbf{Stopping check:} if, for several consecutive iterations after burn-in,
\[
\max_{1\le i\le n}\left\|\ell^{(t)}(X_i)-\ell^{(t-1)}(X_i)\right\|_{\infty}<\tau_{\Delta},
\qquad
\mathrm{eq}_{\mathrm{mean}}^{(t)}<\tau_{\mathrm{mean}},
\qquad
\mathrm{eq}_{\mathrm{cond}}^{(t)}<\tau_{\mathrm{cond}},
\]
where
\[
\mathrm{eq}_{\mathrm{mean}}^{(t)}=\max_{1\le j\le 7}\left|\mathbb P_n \rho_j(O,\theta^{(t)})\right|,
\quad
\mathrm{eq}_{\mathrm{cond}}^{(t)}=\max_{1\le j\le 7}\max_{1\le i\le n}
\left|\left[\mathcal P_{k(n),n}\{\rho_j(O,\theta^{(t)})\}\right](X_i)\right|,
\]
then set \texttt{converged = TRUE} and \textbf{break}.
\end{itemize}
\item If \texttt{converged}$=$False, set $r\gets r+1$ and restart with a new warm start.
\end{itemize}
\item \textbf{Output:} $\theta^{(t)}=\text{expit}(\ell^{(t)})$ at termination.
If \texttt{converged}$=$False after $R$ restarts, return the \\ last iterate and report non-convergence.
\end{itemize}
}
\end{algorithm}
\IncMargin{0.5em}
The above algorithm is subject to the same label-switching ambiguity associated with latent classes as in identification. Although we write $\theta =\bigl(p_{z,1}, p_{z,0}, p_{a,1}, p_{a,0}, p_{y,1}, p_{y,0}, p_{a^*}\bigr)^\top$, the output may converge either to $\theta$ or to its label-switched counterpart $\bigl(p_{z,0}, p_{z,1}, p_{a,0}, p_{a,1}, \allowbreak p_{y,0}, p_{y,1}, 1-p_{a^*}\bigr)^\top$. Consequently, substituting the estimated nuisance functions into the plug-in estimator \eqref{eq:plugin_estimator} yields two candidate values, $\widehat{\psi}_{a^*}$ and $\widehat{\psi}_{1-a^*}$. An additional step is therefore required to determine which quantity corresponds to the target parameter of interest. We resolve this ambiguity by invoking Assumption~\ref{asp:label}. Among the four alternative conditions in Assumption~\ref{asp:label}, we focus on conditions~\ref{cdn:1} and~\ref{cdn:2}, both translating to $\E(A\mid A^*=0,X)<\E(A\mid A^*=1,X)$. Accordingly, we conclude that $\widehat{a}^*=1$ if $\sum^n_{i=1}\widehat{\E}(A\mid a^*,X_i)> \sum^n_{i=1}\widehat{\E}(A\mid 1-a^*,X_i)$, and $\widehat{a}^*=0$ otherwise.
We now propose an estimator for $\psi_{1} = \E(Y^{a^*=1})$ under \ref{cdn:1} or \ref{cdn:2} of Assumption \ref{asp:label}.
Let $d = \arg\max_{j=1,2}\E\{\E(A\mid A^*=a^*_j,X)\}$,
$S_j = \{A-\E(A\mid 1-a^*_j,X)\}\{\E(A\mid a^*_j,X)-\E(A\mid 1-a^*_j,X)\}^{-1}\{Z-\E(Z\mid 1-a^*_j,X)\}
\{\E(Z\mid a^*_j,X)-\E(Z\mid 1-a^*_j,X)\}^{-1}\{\Pr(a^*_j\mid X)\}^{-1}\{Y-\E(Y\mid a^*_j,X)\}   + \E(Y\mid a^*_j,X)$,
$\widehat{S}_j = \{A-\widehat{\E}(A\mid 1-a^*_j,X)\}\{\widehat{\E}(A\mid a^*_j,X)-\widehat{\E}(A\mid 1-a^*_j,X)\}^{-1}\{Z-\widehat{\E}(Z\mid 1-a^*_j,X)\}
\{\widehat{\E}(Z\mid a^*_j,X)-\widehat{\E}(Z\mid 1-a^*_j,X)\}^{-1}
\{\widehat{\Pr}(a^*_j\mid X)\}^{-1}\{Y-\widehat{\E}(Y\mid a^*_j,X)\} + \widehat{\E}(Y\mid a^*_j,X)$,
we have that
\begin{align*}
\begin{split}
\psi_{1} &= \sum^2_{j=1}\E\Bigl[\mathbb{I}\{d=j\}S_j\Bigr]
= \mathbb{P} \left[\sum^2_{j=1}  \mathbb{I}\{d=j\}S_j\right].
\end{split}
\end{align*}
Plugging in the estimated nuisance functions from Algorithm \ref{alg:po} and $\widehat{d}= \arg\max_{j=1,2}\mathbb{P}_n\{\widehat{\E}(A\mid A^*=a^*_j,X)\}$ gives
\begin{align}\label{eq:estimator}
\begin{split}
\widehat{\psi}_{1} &= \sum^2_{j=1}\mathbb{P}_n\Bigl[\mathbb{I}\{\widehat{d}=j\}\widehat{S}_j\Bigr] = \mathbb{P}_n \left[\sum^2_{j=1}  \mathbb{I}\{\widehat{d}=j\}\widehat{S}_j\right].
\end{split}
\end{align}
\begin{theorem}\label{thm:consistency}
Suppose that the class of functions obtained by evaluating the influence function \eqref{eq:IF_observed} at admissible nuisance estimators is $P$-Donsker.
Moreover, assume Assumptions \ref{asp:relevance}-\ref{asp:separation} hold, $Y$ is bounded, and
\begin{enumerate}[itemsep=0.5ex, topsep=.5ex, parsep=0pt, partopsep=0pt]
    \item $\|\widehat{\E}(Y\mid A^*=a^*,X) - \E(Y\mid A^*=a^*,X)\|=o_p(1)$, $\|\widehat{\E}(A\mid A^*=a^*,X) - \E(A\mid A^*=a^*,X)\|=o_p(1)$, $\|\widehat{\E}(Z\mid A^*=a^*,X) - \E(Z\mid A^*=a^*,X)\|=o_p(1)$, $\|\widehat{\Pr}(A^*=a^*\mid X) - \Pr(A^*=a^*\mid X)\|=o_p(1)$ for $a^*=0,1$.
    \item There exists $\epsilon_2>0$
    such that $\min \{\widehat{\E}(A\mid A^*=1,X)-\widehat{\E}(A\mid A^*=0,X), \bigl|\widehat{\E}(Z\mid A^*=1,X)-\widehat{\E}(Z\mid A^*=0,X)\bigr|, \widehat{\Pr}(A^*=1\mid X) \}> \epsilon_2$ almost surely.
\end{enumerate}
Then we have
\begin{align*}
\widehat{\psi}_{1} - \psi_{1} &= (\mathbb{P}_n-\mathbb{P})\left[\sum^2_{j=1} \mathbb{I}\{d=j\}\phi_{a_j^*}\right] \\
&+ \max_{a^*\in\{0,1\}} \Biggl\{O_p(\|\widehat{\Pr}(a^*\mid X) - \Pr(a^*\mid X)\| \| \widehat{\E}(Y\mid a^*,X)- \E(Y\mid a^*,X)\|)\\
&+ O_p(\|\widehat{\E}(A\mid a^*,X) - \E(A\mid a^*,X)\|\| \widehat{\E}(Y\mid a^*,X)- \E(Y\mid a^*,X)\|) \\
&+ O_p(\|\widehat{\E}(Z\mid a^*,X) - \E(Z\mid a^*,X)\|\| \widehat{\E}(Y\mid a^*,X)- \E(Y\mid a^*,X)\|) \\
&+O_p(\|\widehat{\E}(A\mid 1-a^*,X) - \E(A\mid 1-a^*,X)\|\|\widehat{\E}(Z\mid 1-a^*,X) - \E(Z\mid 1-a^*,X)\|) \\
& + O_p(\|\widehat{\E}(A\mid a^*_j,X) - \E(A\mid a^*_j,X) \|^r_\infty) \Biggr\} +o_p(n^{-1/2}),
\end{align*}
where $r>0$.
If it further holds that $\|\widehat{\Pr}(a^*\mid X) - \Pr(a^*\mid X)\| \| \widehat{\E}(Y\mid a^*,X)- \E(Y\mid a^*,X)\|=o_p(n^{-1/2})$, $\|\widehat{\E}(A\mid a^*,X) - \E(A\mid a^*,X)\|\| \widehat{\E}(Y\mid a^*,X)- \E(Y\mid a^*,X)\|=o_p(n^{-1/2})$, $\|\widehat{\E}(Z\mid a^*,X) - \E(Z\mid a^*,X)\|\| \widehat{\E}(Y\mid a^*,X)- \E(Y\mid a^*,X)\|=o_p(n^{-1/2})$, $\|\widehat{\E}(A\mid 1-a^*,X) - \E(A\mid 1-a^*,X)\|\|\widehat{\E}(Z\mid 1-a^*,X) - \E(Z\mid 1-a^*,X)\|=o_p(n^{-1/2})$, $\|\widehat{\E}(A\mid a^*_j,X) - \E(A\mid a^*_j,X) \|^r_\infty=o_p(n^{-1/2})$, then
$
\sqrt{n}\left(\widehat{\psi}_{1} - \psi_{1}\right)\overset{d}{\to} \mathcal{N}(0,\E(\phi_{1}^2)).
$
\end{theorem}
Remarkably, Theorem \ref{thm:consistency} does not require nuisance functions to converge at root-$n$ rates as it suffices for  their pairwise product of convergence rates to be faster than root-$n$, and for the converge rate of $\widehat{\E}(A\mid a^*_j,X)$  raised to the $r$ power to likewise be faster than root-$n$, where $r$ is a user-specified positive number.
In our setting, convergence-rate results for the relevant nuisance estimators are available under standard smoothness and sieve regularity conditions; see Theorem \ref{thm:rate}. These results ensure that the product conditions in Theorem~\ref{thm:consistency} are satisfied for appropriate choices of sieve dimension $k(n)$.
The choice of $r$ appears to offer a cost-free advantage; however, there is indeed a trade-off between $r$ and the margin constant $\epsilon_1$ in Assumption \ref{asp:separation}. If $\epsilon_1$ tends to zero, $r$ must also tend to zero, otherwise the constant term $\epsilon_1^{-r}$ scaling $\|\widehat{\E}(A\mid a^*_j,X) - \E(A\mid a^*_j,X) \|^r_\infty$ will become unbounded (see the proof of Theorem \ref{thm:consistency} in the Supplementary Materials \S{}S17).
\section{Extension to other functionals}\label{sec:generalization}
In this section, we revisit the examples of \S\ref{sec:functionals} and derive corresponding influence functions in the hidden treatment model. Specifically, we continue to assume that the analyst has access to a surrogate $A$ and a proxy $Z$ for the hidden treatment $A^*$,  that  satisfy Assumptions \ref{asp:relevance}, \ref{asp:conditional_independence}, and \ref{asp:label}.
We establish that one can readily leverage the same techniques as in \S\ref{sec:ATE}-\ref{sec:eif} for identification, and  we generically denote as $\psi^*$ the corresponding observed data functional for which we derive influence functions.
The estimation of nuisance functions then follows in an analogous manner as above, details of which we omit for brevity.
To present our results in a unified fashion, we note that all examples given in \S\ref{sec:functionals} have full data influence functions of the bi-linear form
\begin{align}\label{eq:dr_if}
\begin{split}
\phi_{\mathrm{g}}^F(Y,A^*,X;&\psi^*) =\sum_{a^*=0,1} IF_{a^*;\psi^*}=\sum_{a^*=0,1}\Bigl\{q_{0,a^*}(X) h_{0,a^*}(X) g_{1,a^*}(Y,X) \mathbb{I}( A^*=a^*)  \\
& +q_{0,a^*}(X) g_{2,a^*}( Y,X) \mathbb{I}(A^*=a^*) +h_{0,a^*}(X) g_{3,a^*}( Y,X) +g_{4,a^*}( Y,X) \Bigr\},
\end{split}
\end{align}
where $q_{0,a^*}(X)$ and $h_{0,a^*}(X) $ are unknown nuisance functions, $g_{j,a^*}$ are known functions of $(Y,X)$, and $IF_{a^*;\psi^*}, a^*=0,1$, belong to the doubly robust class of influence functions studied in \cite{robins2008higher}. To simplify the exposition, we suppress the influence function dependence on $\psi^*$ throughout as $\E[IF_{a^*;\psi^*}]=0,  a^*=0,1$. The next theorem characterizes for each influence function in the class above, a corresponding observed data influence function in the hidden treatment model. While we state the result for the class given above, the proof is more general as it considers an even larger class of doubly robust influence functions introduced in \cite{ghassami2022minimax}, an important generalization which includes influence functions for causal functionals identified using the proximal causal inference framework \citep{miao2016identifying,miao2018identifying,tchetgen2024introduction}
that do not strictly fall within the above doubly robust class \citep{cui2024semiparametric}.
The proof is deferred to the Supplementary Materials \S{}S18.
\begin{theorem}\label{thm:extension}
Suppose that the influence functions for the parameter of interest $\psi^*$ in the full data model $\mathcal{M}^*$ are of the form in \eqref{eq:dr_if}, indexed by functions $\mathrm{g}=(g_{1,a^*},g_{2,a^*},g_{3,a^*},g_{4,a^*}, \allowbreak a^*=0,1) \in \mathcal{G}$ for a certain space of functions $\mathcal{G}$ in $L_0^{2} (X,Y)$ implied by the semiparametric model; then
\begin{align*}
\begin{split}
\phi_{\mathrm{g}}&(Y,A,Z,X;\psi^*) =\sum_{a^*=0,1}\Bigl\{q_{0,a^*}(X) h_{0,a^*}(X) g_{1,a^*}\left(Y,X\right) M_{a^*,A} M_{a^*,Z} \\
& +q_{0,a^*}(X) g_{2,a^*}\left( Y,X\right) M_{a^*,A} M_{a^*,Z} +h_{0,a^*}(X) g_{3,a^*}( Y,X) +g_{4,a^*}( Y,X) \Bigr\}
\end{split}
\end{align*}
is an influence function for $\psi^*$ in the hidden treatment model $\mathcal{M}$, where
{\small
\begin{align}\label{eq:m_indicator}
\begin{split}
M_{a^*,A} = \{A-\E(A\mid A^*=1-a^*,X)\}\{\E(A\mid A^*=a^*,X)-\E(A\mid A^*=1-a^*,X)\}^{-1},\\
M_{a^*,Z} = \{Z-\E(Z\mid A^*=1-a^*,X)\}\{\E(Z\mid A^*=a^*,X)-\E(Z\mid A^*=1-a^*,X)\}^{-1}.
\end{split}
\end{align}}
The set of influence functions in $\mathcal{M}$ is $\{\phi_{\mathrm{g}}:\mathrm{g}\in \mathcal{G}\} + \Lambda^\bot$, where $\Lambda^\bot$ is given in \eqref{eq:ortho}.
\end{theorem}
Upon deriving the set of influence functions, an efficient estimator for $\psi^*$ in the hidden treatment model can be constructed as follows:
\begin{enumerate}[itemsep=0.5ex, topsep=.5ex, parsep=0pt, partopsep=0pt]
    \item Obtain a preliminary (potentially inefficient) estimator $\widehat{\psi}$ that solves $0 = \mathbb{P}_n\widehat{\phi}_{\mathrm{g}}(Y,A,Z,X)$ by using an estimated influence function $\widehat{\phi}_{\mathrm{g}}(Y,A,Z,X)$ as an estimating function for a fixed choice $\mathrm{g}$.
    \item Perform a one-step update in the direction of estimated EIF
    $
    \widehat{\psi}^\text{eff} = \widehat{\psi} +\mathbb{P}_n \widehat{\phi}_\text{eff}(Y,A,Z,X),
    $
    where $\widehat{\phi}_\text{eff}(Y,A,Z,X)$ is an estimator of the EIF of ${\phi}_\text{eff}(Y,A,Z,X)=\Pi(\phi_{\mathrm{g}}\mid \Lambda)$ evaluated at $\widehat{\psi}$ given in (S23) in the Supplementary Materials \S{}S8.2.
\end{enumerate}
\paragraph{Example 2. ETT (continued).}
An observed data influence function is derived by replacing the unobserved indicator $\mathbb{I}(A^*=a^*), a^*=0,1$ in \eqref{eq:eif_ett} with $M_{a^*,A}(A,X)M_{a^*,Z}(Z,X)$ in \eqref{eq:m_indicator},   which we can write as
\begin{align*}
\phi_{ett}&=\frac{1}{\Pr(A^*=1)}M_{1,A}(A,X)M_{1,Z}(Z,X)\{Y-\E(Y\mid A^*=1,X)\}\\
&\quad -\frac{1}{\Pr(A^*=1)}\frac{\Pr(A^*=1\mid X)}{1-\Pr(A^*=1\mid X)}M_{0,A}(A,X)M_{0,Z}(Z,X)\{Y-\E(Y\mid A^*=0,X)\}\\
&\quad +\frac{\E(Y\mid A^*=1,X) - \E(Y\mid A^*=0,X)-\psi_{ett}}{\Pr(A^*=1)}\Pr(A^*=1\mid X).
\end{align*}
\paragraph{Example 3. QTE (continued).}
The nuisance parameters $\eta_1(Y,X;\theta_1)=\Pr(Y\le\theta_1\mid X,A^*=a^*)$ and $\eta_2(X)=\Pr(A^*=a^*\mid X)$ can both be identified by the approach in \S\ref{sec:ATE} by treating $\mathbb{I}(Y\le\theta_1^{a^*})$ as a binary outcome.
The influence function in the observed data model is obtained by substituting $\mathbb{I}(A^*=a^*)$ in \eqref{eq:qte_eif_full} with $M_{a^*,A}(A,X)M_{a^*,Z}(Z,X)$ in \eqref{eq:m_indicator},
which gives
\begin{align*}
\phi_{a^*,\gamma}=M_{a^*,A}(A,X)M_{a^*,Z}(Z,X)\frac{\mathbb{I}(Y\le\theta_1^{a^*})-\eta_1(X;\theta_1^{a^*})}{\eta_2(X)}+\eta_1(X;\theta_1^{a^*})-\gamma.
\end{align*}
\paragraph{Example 4. Semi-linear structural regression model (continued).} The full data influence function in \eqref{eq:if_sl} can be reformulated as
$
\phi^F_{sl} = \mathbb{I}(A^*=1)h(X)\{Y-\beta_{a^*}-\beta_x(X) \}\{ 1-\Pr(A^*=1\mid X) \} + \mathbb{I}(A^*=0)h(X)\{Y-\beta_x(X) \}\{ -\Pr(A^*=1\mid X) \}.
$
An influence function in the observed data model can be obtained by replacing $\mathbb{I}(A^*=a^*), a^*=0,1$ with
$M_{a^*,A}(A,X)M_{a^*,Z}(Z,X)$ in \eqref{eq:m_indicator}, which gives
\begin{align*}
\phi_{sl} &= M_{1,A}(A,X)M_{1,Z}(Z,X)h(X)\{Y-\beta_{a^*}-\beta_x(X) \}\{ 1-\Pr(A^*=1\mid X) \} \\
&\quad + M_{0,A}(A,X)M_{0,Z}(Z,X)h(X)\{Y-\beta_x(X) \}\{ -\Pr(A^*=1\mid X) \}.
\end{align*}
\paragraph{Example 5. Marginal structural generalized linear model (continued).}
The full data influence function in \eqref{eq:if_msm} can be rewritten as
\begin{align*}
\phi^F_{msm}
&=\mathbb{I}(A^*=1)\frac{1}{\Pr(A^*=1\mid X)}h(1,V)\left\{Y - \E(Y\mid A^*=1,X)  \right\}\\
& \quad + \mathbb{I}(A^*=0)\frac{1}{\Pr(A^*=0\mid X)}h(0,V)\left\{Y - \E(Y\mid A^*=0,X)  \right\}\\
& \quad + \sum_{a^*} (-1)^{1-a^*} h(a^*,V)\left\{\E(Y\mid A^*=a^*,X) - g^{-1}(\beta_0+\beta_1a^*+\beta_2^\T V)\right\}.
\end{align*}
An influence function in the observed data model can be derived by replacing $\mathbb{I}(A^*=a^*), a^*=0,1$ with $M_{a^*,A}(A,X)M_{a^*,Z}(Z,X)$ in \eqref{eq:m_indicator}, which gives
\begin{align*}
\phi_{msm}
&=\frac{M_{1,A}(A,X)M_{1,Z}(Z,X)}{\Pr(A^*=1\mid X)}h(1,V)\left\{Y - \E(Y\mid A^*=1,X)  \right\}\\
& \quad + \frac{M_{0,A}(A,X)M_{0,Z}(Z,X)}{\Pr(A^*=0\mid X)}h(0,V)\left\{Y - \E(Y\mid A^*=0,X)  \right\}\\
& \quad + \sum_{a^*} (-1)^{1-a^*} h(a^*,V)\left\{\E(Y\mid A^*=a^*,X) - g^{-1}(\beta_0+\beta_1a^*+\beta_2^\T V)\right\}.
\end{align*}
\section{Simulation study}\label{sec:simulation}
In this section, we evaluate the performance of our proposed estimator for both  binary and continuous outcomes. The covariates $X=(X_1,X_2)$ are generated from a 2-dimensional uniform $[0,1]^2$ distribution. We then generate $A^*$ conditional on $X$ from a Bernoulli distribution with $\Pr(A^*=1\mid X)=\text{logit}^{-1}(X_1 + X_2 - 1)$. Next, we generate $A$ and $Z$ from
$\Pr(A=1\mid A^*=1,X) = \text{logit}^{-1}(1 + 2X_1)$,
$\Pr(A=1\mid A^*=0,X) = \text{logit}^{-1}(-1 + 0.5X_1 -X_2)$,
$\Pr(Z=1\mid A^*=1,X) = \text{logit}^{-1}(-1 - X_1 + 0.5X_2)$,
$\Pr(Z=1\mid A^*=0,X) = \text{logit}^{-1}(2 + X_1)$,
where $\text{logit}(p) = \ln p(1-p)^{-1}$.
We consider two outcome regression models:
\begin{align*}
\text{i. (binary outcome) }\quad  &\Pr(Y=1\mid A^*=1,X) = \text{logit}^{-1}(-2 + 0.5X_1 + X_2), \\
&\Pr(Y=1\mid A^*=0,X) = \text{logit}^{-1}(1 - 2X_1 + 0.5 X_2); \\
\text{ii. (continuous outcome) }\quad &Y = \sin(\pi X_1) + (A^*-0.5)(X_1+X_2) + \varepsilon, \quad \varepsilon\sim\mathcal{N}(0,1).
\end{align*}
For comparison, we also present results for
an infeasible estimator which substitutes the true values of nuisance parameters into \eqref{eq:estimator}, and the naive estimator, which treats $A$ as $A^*$ and fits a logistic regression of $Y$ on $(A, X)$ for binary $Y$, and a kernel regression for continuous $Y$. In our setting, $A$ matches $A^*$ around 82\% of the time. Detailed tuning-parameter settings are provided in the Supplementary Materials \S{}S19.
The results are displayed in Table \ref{tab:simu}. From the results, one can see that our proposed estimator shows robust performance, achieving coverage close to the nominal rate of 95\%, while the naive estimator is severely biased.
\begin{table}[h]
    \centering
    \begin{tabular}{ccrrrc}
         && \multicolumn{3}{c}{bias(s.e.)($\times 100$)} &  coverage   \\
         $Y$& $n$ & \multicolumn{1}{c}{proposed} & \multicolumn{1}{c}{infeasible} & \multicolumn{1}{c}{naive} & \multicolumn{1}{c}{proposed} \\
         \hline
    &$500$ & $-0.72 (0.16)$ & $0.14 (0.16)$ &  $12.67(0.13)$ & 95.1\% \\
    Binary &$1000$ &  $-0.46(0.13)$ & $0.14(0.12)$ & $12.55(0.10)$ & 93.7\% \\
    &$2000$ &  $-0.33 (0.09)$ & $0.14(0.09)$ &  $12.49(0.07)$ & 93.4\% \\
    \hline
    &$500$ &  $3.54 (0.39)$ & $0.21 (0.38)$ &  $-31.35(0.33)$ & 93.7\% \\
    Continuous & $1000$& $2.49 (0.26)$ & $0.22(0.26)$ & $-31.37(0.22)$ & 95.3\% \\
    &$2000$ &  $2.65 (0.20)$ & $0.20(0.20)$ &  $-31.52(0.17)$ & 93.4\% \\
    \end{tabular}
    \caption{Simulation results when $Y$ is binary or continuous. The ``proposed'' column is for our proposed ATE estimator in \eqref{eq:estimator}. The ``infeasible'' column is for the estimator using true values of nuisance parameters. The ``naive'' column is for the estimator that treats $A$ as $A^*$. The coverage for the proposed estimator is given in the last column.}
    \label{tab:simu}
\end{table}
\section{Real data example}\label{sec:application}
Alzheimer's Disease (AD) is a progressive neurodegenerative disorder that primarily affects the elderly, characterized by significant cognitive decline and memory impairment. One of the hallmark changes observed in AD is the atrophy of the hippocampus, a critical brain region involved in memory formation and spatial navigation. As the disease advances, neuronal damage occurs, resulting in brain atrophy that is especially pronounced in the hippocampus during the early stages of AD \citep{fox1996presymptomatic,yu2022mapping}.
We aim to analyze the impact of AD on hippocampal volume using data from the Alzheimer's Disease Neuroimaging Initiative (ADNI). The data usage acknowledgment is included in \S{}S20 of the Supplementary Materials.
Diagnosing AD involves a comprehensive approach, combining cognitive assessments such as the Mini-Mental State Examination (MMSE) and neuroimaging techniques like MRI and PET scans to detect brain changes. However, AD diagnosis may not reflect the actual latent disease status \citep{hunter2015medical}. The diagnosis essentially serves as an approximation for the unobserved actual state of the disease.
Our approach, as outlined in Section \ref{sec:estimation}, requires selecting a surrogate and a proxy that are conditionally independent given a participant's true but hidden AD status. Because AD diagnosis is essentially a composite of several measured factors including test scores and biomarker measurements, using it as a surrogate would invalidate any such measurements as potential proxies given that they would fail to be independent.
Instead, we use the recall and orientation subscores of the MMSE as indicators: one as a surrogate for underlying disease status and the other as a proxy. Specifically, the recall component evaluates the ability to remember three previously mentioned objects, while the orientation component assesses the awareness of the current time and place.
We posit that the recall and orientation scores, which assess distinct cognitive functions, are independent given disease status and other covariates.
This claim finds support in \cite{yew2013lost}, which revealed distinct neural correlates of AD: performance on orientation scores was associated with the posterior hippocampus, whereas memory performance was more affected by the anterior hippocampal regions. The scores are converted into binary variables, where 0 indicates a perfect score in a subcategory, and 1 indicates any deviation. In our analysis, we also account for important observed confounders such as age, sex, and number of years of education.
We consider the underlying AD status at Month 6 as the latent exposure, and the outcome of interest is hippocampal volume at Month 12. The MMSE recall subscore at Month 6 serves as the surrogate, while the MMSE orientation subscore at the same time point acts as the proxy.
We include 1,172 subjects with complete MMSE scores at Month 6 and measurements of hippocampal volume at Month 12. Among these subjects, the hippocampal volume at Month 12 ranges from 2,839 to 11,041 mm$^3$, with an average of 6,627 mm$^3$ (SD = 1,258).  Regarding the MMSE subscores, 50.4\% of participants achieved a full score for orientation, and 45.3\% achieved a full score for recall.
The average baseline age is 73.4 years old (SD = 7.0); 44.4\% of the subjects are female.
We conduct separate analyses for the female and male groups, adjusting for age and years of education as potential confounders within each group. We consider the estimator proposed in equation \eqref{eq:estimator} in \S\ref{sec:estimation}.
Our results indicate that for females, AD leads to a decrease of 1,652 mm$^3$ in hippocampal volume, with 95\% confidence interval of $[1,487, 1,816]$. For males, AD leads to a decrease of 1,345 mm$^3$ in hippocampal volume, with 95\% confidence interval of $[1,173, 1,518]$ \citep{fox1996presymptomatic}. These findings appear to be aligned with prior research indicating that females are affected by AD more severely than males, and AD-control brain weight differences were significantly greater for females \citep{filon2016gender}. For comparison, we also performed linear regression for both groups, with hippocampal volume as the outcome, error-prone diagnosis as the main exposure, age, and education years as regressors.
These analyses indicate an AD-related decrease of 1,162 mm$^3$ for females (95\% CI: $[976, 1,347]$) and 971 mm$^3$ for males (95\% CI: $[793, 1,151]$), which are of smaller magnitude than our estimates, suggesting the potential for effect attenuation, a common consequence of exposure measurement error. In addition to linear regression, we also implemented the more complex doubly robust Targeted Maximum Likelihood Estimator \citep{benkeser2017doubly}, which we expect to be biased in this context and yields results similar to those of linear regression.
\section{Discussion}\label{sec:discussion}
In this paper, we considered the fundamental problem of causal inference with a hidden binary treatment, an all too common problem in epidemiology and the social sciences. We proposed a unified solution which delivers nonparametric identification of a rich class of  causal parameters, by drawing on a strategy which leverages a surrogate and a proxy variable of the hidden treatment to restore identification, while bypassing the need for validation data or restrictive functional form assumptions prominent in prior measurement error literature.
An important contribution of the paper is to develop formal semiparametric theory which gives rise to influence function-based nonparametric estimators with important robustness and efficiency properties. The approach provides a generic roadmap for constructing novel semiparametric estimators in a large class of hidden treatment models.
The paper also addresses the challenge of estimating high-dimensional nuisance functions defined in terms of the hidden treatment, therefore ruling out direct application of standard machine learning and nonparametric estimators. To resolve this difficulty, we proposed a novel approach for estimating nuisance functions.
We establish the corresponding  semiparametric efficiency bound for discrete outcomes, and provide an approximation for continuous outcomes.
The proposed methods may be improved or extended in several directions. First, our results can be extended to scenarios where the hidden treatment is a categorical variable with $k$ levels, which we anticipate will present several new challenges both for identification and semiparametric estimation.
Additionally, extending these methods to continuous treatments presents a promising avenue, particularly given the practical relevance of continuous treatments in a wide range of scientific inquiries, such as for instance evaluating the impact of the magnitude of air pollution (e.g. levels of ambient PM2.5) on health outcomes.


\begin{thebibliography}{}
\bibitem[Allman et~al., 2009]{allman2009identifiability}
Allman, E.~S., Matias, C., and Rhodes, J.~A. (2009).
\newblock Identifiability of parameters in latent structure models with many observed variables.
\newblock {\em The Annals of Statistics}, 37(6A):3099--3132.
\bibitem[Benkeser et~al., 2017]{benkeser2017doubly}
Benkeser, D., Carone, M., Laan, M. V.~D., and Gilbert, P.~B. (2017).
\newblock Doubly robust nonparametric inference on the average treatment effect.
\newblock {\em Biometrika}, 104(4):863--880.
\bibitem[Bickel et~al., 1993]{bickel1993efficient}
Bickel, P., Klaassen, C.~A., Ritov, Y., and Wellner, J.~A. (1993).
\newblock {\em Efficient and adaptive estimation for semiparametric models}, volume~4.
\newblock Springer.
\bibitem[Bollinger, 1996]{bollinger1996bounding}
Bollinger, C.~R. (1996).
\newblock Bounding mean regressions when a binary regressor is mismeasured.
\newblock {\em Journal of Econometrics}, 73(2):387--399.
\bibitem[Chen et~al., 2005]{chen2005measurement}
Chen, X., Hong, H., and Tamer, E. (2005).
\newblock Measurement error models with auxiliary data.
\newblock {\em The Review of Economic Studies}, 72(2):343--366.
\bibitem[Chen et~al., 2009]{chen2009nonparametric}
Chen, X., Hu, Y., and Lewbel, A. (2009).
\newblock Nonparametric identification and estimation of nonclassical errors-in-variables models without additional information.
\newblock {\em Statistica Sinica}, 19(3):949--968.
\bibitem[Chen and Pouzo, 2012]{chen2012estimation}
Chen, X. and Pouzo, D. (2012).
\newblock Estimation of nonparametric conditional moment models with possibly nonsmooth generalized residuals.
\newblock {\em Econometrica}, 80(1):277--321.
\bibitem[Cui et~al., 2024]{cui2024semiparametric}
Cui, Y., Pu, H., Shi, X., Miao, W., and Tchetgen~Tchetgen, E. (2024).
\newblock Semiparametric proximal causal inference.
\newblock {\em Journal of the American Statistical Association}, 119(546):1348--1359.
\bibitem[DiTraglia and Garcia-Jimeno, 2019]{ditraglia2019identifying}
DiTraglia, F.~J. and Garcia-Jimeno, C. (2019).
\newblock Identifying the effect of a mis-classified, binary, endogenous regressor.
\newblock {\em Journal of Econometrics}, 209(2):376--390.
\bibitem[Filon et~al., 2016]{filon2016gender}
Filon, J.~R., Intorcia, A.~J., Sue, L.~I., Vazquez~Arreola, E., et~al. (2016).
\newblock Gender differences in {Alzheimer} disease: Brain atrophy, histopathology burden, and cognition.
\newblock {\em Journal of Neuropathology \& Experimental Neurology}, 75(8):748--754.
\bibitem[Fox et~al., 1996]{fox1996presymptomatic}
Fox, N., Warrington, E., Freeborough, P., Hartikainen, P., Kennedy, A., Stevens, J., and Rossor, M.~N. (1996).
\newblock Presymptomatic hippocampal atrophy in {Alzheimer's} disease: A longitudinal {MRI} study.
\newblock {\em Brain}, 119(6):2001--2007.
\bibitem[Ghassami et~al., 2022]{ghassami2022minimax}
Ghassami, A., Ying, A., Shpitser, I., and Tchetgen~Tchetgen, E. (2022).
\newblock Minimax kernel machine learning for a class of doubly robust functionals with application to proximal causal inference.
\newblock In {\em International conference on artificial intelligence and statistics}, pages 7210--7239. PMLR.
\bibitem[Hahn, 1998]{hahn1998role}
Hahn, J. (1998).
\newblock On the role of the propensity score in efficient semiparametric estimation of average treatment effects.
\newblock {\em Econometrica}, 66:315--331.
\bibitem[Hu, 2008]{hu2008identification}
Hu, Y. (2008).
\newblock Identification and estimation of nonlinear models with misclassification error using instrumental variables: A general solution.
\newblock {\em Journal of Econometrics}, 144(1):27--61.
\bibitem[Hu and Lewbel, 2012]{hu2012returns}
Hu, Y. and Lewbel, A. (2012).
\newblock Returns to lying? {Identifying} the effects of misreporting when the truth is unobserved.
\newblock {\em Frontiers of Economics in China}, 7(2):163--192.
\bibitem[Hunter et~al., 2015]{hunter2015medical}
Hunter, C.~A., Kirson, N., Desai, U., Cummings, A. K.~G., Faries, D.~E., and Birnbaum, H.~G. (2015).
\newblock Medical costs of {Alzheimer's} disease misdiagnosis among us medicare beneficiaries.
\newblock {\em Alzheimer's \& Dementia}, 11:887--895.
\bibitem[Imai and Yamamoto, 2010]{imai2010causal}
Imai, K. and Yamamoto, T. (2010).
\newblock Causal inference with differential measurement error: Nonparametric identification and sensitivity analysis.
\newblock {\em American Journal of Political Science}, 54(2):543--560.
\bibitem[Lewbel, 2007]{lewbel2007estimation}
Lewbel, A. (2007).
\newblock Estimation of average treatment effects with misclassification.
\newblock {\em Econometrica}, 75(2):537--551.
\bibitem[Mahajan, 2006]{mahajan2006identification}
Mahajan, A. (2006).
\newblock Identification and estimation of regression models with misclassification.
\newblock {\em Econometrica}, 74(3):631--665.
\bibitem[Miao et~al., 2016]{miao2016identifying}
Miao, W., Geng, Z., and Tchetgen~Tchetgen, E. (2016).
\newblock Identifying causal effects with proxy variables of an unmeasured confounder.
\newblock {\em arXiv preprint arXiv:1609.08816}.
\bibitem[Miao et~al., 2018]{miao2018identifying}
Miao, W., Geng, Z., and Tchetgen~Tchetgen, E. (2018).
\newblock Identifying causal effects with proxy variables of an unmeasured confounder.
\newblock {\em Biometrika}, 105(4):987--993.
\bibitem[Newey, 1990]{newey1990semiparametric}
Newey, W.~K. (1990).
\newblock Semiparametric efficiency bounds.
\newblock {\em Journal of Applied Econometrics}, 5(2):99--135.
\bibitem[Porsteinsson et~al., 2021]{porsteinsson2021diagnosis}
Porsteinsson, A.~P., Isaacson, R., Knox, S., Sabbagh, M.~N., and Rubino, I. (2021).
\newblock Diagnosis of early {Alzheimer’s} disease: Clinical practice in 2021.
\newblock {\em The Journal of Prevention of Alzheimer's Disease}, 8(3):371--386.
\bibitem[Robins, 1986]{robins1986new}
Robins, J. (1986).
\newblock A new approach to causal inference in mortality studies with a sustained exposure period—application to control of the healthy worker survivor effect.
\newblock {\em Mathematical Modelling}, 7(9-12):1393--1512.
\bibitem[Robins et~al., 2008]{robins2008higher}
Robins, J., Li, L., Tchetgen~Tchetgen, E., van~der Vaart, A., et~al. (2008).
\newblock Higher order influence functions and minimax estimation of nonlinear functionals.
\newblock In {\em Probability and statistics: essays in honor of David A. Freedman}, volume~2, pages 335--422. Institute of Mathematical Statistics.
\bibitem[Robins et~al., 1994]{robins1994estimation}
Robins, J.~M., Rotnitzky, A., and Zhao, L.~P. (1994).
\newblock Estimation of regression coefficients when some regressors are not always observed.
\newblock {\em Journal of the American Statistical Association}, 89(427):846--866.
\bibitem[Rotnitzky et~al., 2021]{rotnitzky2021characterization}
Rotnitzky, A., Smucler, E., and Robins, J.~M. (2021).
\newblock Characterization of parameters with a mixed bias property.
\newblock {\em Biometrika}, 108(1):231--238.
\bibitem[Scharfstein et~al., 1999]{scharfstein1999adjusting}
Scharfstein, D.~O., Rotnitzky, A., and Robins, J.~M. (1999).
\newblock Adjusting for nonignorable drop-out using semiparametric nonresponse models.
\newblock {\em Journal of the American Statistical Association}, 94(448):1096--1120 (with Rejoinder, 1135--1146).
\bibitem[Schennach, 2016]{schennach2016recent}
Schennach, S.~M. (2016).
\newblock Recent advances in the measurement error literature.
\newblock {\em Annual Review of Economics}, 8:341--377.
\bibitem[Sepanski and Carroll, 1993]{sepanski1993semiparametric}
Sepanski, J.~H. and Carroll, R.~J. (1993).
\newblock Semiparametric quasilikelihood and variance function estimation in measurement error models.
\newblock {\em Journal of Econometrics}, 58(1-2):223--256.
\bibitem[Shen, 1997]{shen1997methods}
Shen, X. (1997).
\newblock On methods of sieves and penalization.
\newblock {\em The Annals of Statistics}, 25(6):2555--2591.
\bibitem[Tchetgen~Tchetgen et~al., 2024]{tchetgen2024introduction}
Tchetgen~Tchetgen, E., Ying, A., Cui, Y., Shi, X., and Miao, W. (2024).
\newblock An introduction to proximal causal learning.
\newblock {\em Statistical Science}, 39(3):375 -- 390.
\bibitem[Tsiatis, 2007]{tsiatis2007semiparametric}
Tsiatis, A. (2007).
\newblock {\em Semiparametric Theory and Missing Data}.
\newblock Springer Series in Statistics. Springer New York.
\bibitem[van~der Vaart, 1991]{van1991differentiable}
van~der Vaart, A. (1991).
\newblock On differentiable functionals.
\newblock {\em The Annals of Statistics}, 19(1):178--204.
\bibitem[Yew et~al., 2013]{yew2013lost}
Yew, B., Alladi, S., Shailaja, M., Hodges, J.~R., and Hornberger, M. (2013).
\newblock Lost and forgotten? {Orientation} versus memory in {Alzheimer's} disease and frontotemporal dementia.
\newblock {\em Journal of Alzheimer's Disease}, 33(2):473--481.
\bibitem[Yu et~al., 2022]{yu2022mapping}
Yu, D., Wang, L., Kong, D., and Zhu, H. (2022).
\newblock Mapping the genetic-imaging-clinical pathway with applications to {Alzheimer’s} disease.
\newblock {\em Journal of the American Statistical Association}, 117(540):1656--1668.
\end{thebibliography}
\end{document}


\def\spacingset#1{\renewcommand{\baselinestretch}
{#1}\small\normalsize} \spacingset{1}
 \title{Supplementary Materials for ``Causal Inference with a Hidden Treatment''}
 \date{}
 \maketitle
\spacingset{1.9}
\section{Classical and nonclassical measurement errors}\label{sec:classical_error}
Let $X^*$ be the unobserved variable with an observed but mismeasured counterpart $X$ satisfying $X = X^* + \epsilon$.
Classical measurement errors usually mean $\epsilon\bigCI X^*$ and $\E(\epsilon)=0$. When $X^*$ is binary, $\epsilon$ is usually called misclassification error. In this case, $\epsilon$ takes value from $\{0,1,-1\}$, and it is considered nonclassical because $\E(\epsilon\mid X^*=0) = \Pr(\epsilon=1\mid X^*=0)\ge 0$ and $\E(\epsilon\mid X^*=1)= -\Pr(\epsilon = -1\mid X^*=1)\le 0$, which generally violates the assumption that
$
\E(\epsilon\mid X^*)  = \E(\epsilon) = 0
$
unless in the degenerate case where $\Pr(\epsilon=1\mid X^*=0)= \Pr(\epsilon = -1\mid X^*=1) =0$, which implies $\epsilon = 0$ almost surely.
\section{Proof of Theorem \ref{main-thm:identification}}\label{sec:thm_identification}
When $Y$ is binary, let $P(Y,Z\mid x)=(\Pr(Y=2-i,Z=2-j\mid X=x))_{i,j}$, $P(Y,Z,A=1\mid x)=(\Pr(Y=2-i,Z=2-j,A=1\mid x))_{i,j}$, $P(Y\mid A^*,x)=(\Pr(Y=2-i\mid A^*=2-j,X=x))_{i,j}$, $P(Z\mid A^*,x)=(\Pr(Z=2-i\mid A^*=2-j,X=x))_{i,j}$, $P_D(A^*\mid x)=\text{diag}(\Pr(A^*=1\mid X=x),\Pr(A^*=0\mid X=x))$, $P_D(A=1\mid A^*,x)=\text{diag}(\Pr(A=1\mid A^*=1,X=x),\Pr(A=1\mid A^*=0,X=x))$.
By Assumption \ref{main-asp:conditional_independence}, we have
\begin{align*}
P(Y,Z\mid x)&=P(Y\mid A^*,x)P_D(A^*\mid x)P(Z\mid A^*,x)^\T,\\
P(Y,Z,A=1\mid x)&=P(Y\mid A^*,x)P_D(A^*\mid x)P_D(A=1\mid A^*,x)P(Z\mid A^*,x)^\T.
\end{align*}
By Assumption \ref{main-asp:relevance}, $P(Y,Z\mid x)$ is invertible.
Multiplying $P(Y,Z,A=1\mid x)$ by $P(Y,Z\mid x)^{-1}$ gives
\begin{align*}
P(Y,Z,A=1\mid x)P(Y, Z\mid x)^{-1} = P(Y\mid A^*,x)P_D(A=1\mid A^*,x)P(Y\mid A^*,x)^{-1},
\end{align*}
which is in the canonical form of matrix eigendecomposition. Since the right-hand side is observed, we are able to solve for the eigenvalues and eigenvectors, which correspond to the diagonal elements of $P_D(A=1\mid A^*,x)$ and the columns of $P(Y\mid A^*,x)$, respectively. While eigenvectors are identified up to scale, the requirement that columns of $P(Y\mid A^*,x)$ sum to 1 fixes it. To decide which one of the two eigenvalues is $\Pr(A=1\mid A^*=1,x)$, we consider the 4 conditions in Assumption \ref{main-asp:label}:
1. $\Pr(A=0 \mid A^*=1, X=x) + \Pr(A=1\mid A^*=0,X=x)<1$. This inequality is equivalent to $\Pr(A=1\mid A^*=0,X=x)<\Pr(A=1\mid A^*=1,X=x)$. Therefore, the larger one of the two eigenvalues is $\Pr(A=1\mid A^*=1,X=x)$.
2. $\Pr(A^*=0 \mid A=1, X=x) + \Pr(A^*=1\mid A=0,X=x)<1$. This inequality is equivalent to $\Pr(A=1\mid A^*=0,X=x)<\Pr(A=1\mid A^*=1,X=x)$. Therefore, the larger one of the two eigenvalues is $\Pr(A=1\mid A^*=1,X=x)$.
3. $\Pr(A^*=1\mid A=1,X=x)>\Pr(A^*=0\mid A=1,X=x)$. This inequality is equivalent to $\Pr(A=1\mid A^*=1,X=x)\Pr(A^*=1\mid x) > \Pr(A=1\mid A^*=0,X=x)\Pr(A^*=0\mid x)$. Denote the two eigenvalues as $e_1$ and $e_2$. Let $l_1 = (\Pr(A=1\mid x)-e_2)/(e_1-e_2)$ and $l_2 = 1 - l_1$, then
\begin{align*}
\Pr(A=1\mid A^*=1,x) = e_j, \text{ where } j = \arg\max_j e_jl_j.
\end{align*}
4. $\Pr(A^*=0\mid A=0,X=x)>\Pr(A^*=1\mid A=0,X=x)$. This inequality is equivalent to $\Pr(A=0\mid A^*=0,X=x)\Pr(A^*=0\mid x) > \Pr(A=0\mid A^*=1,X=x)\Pr(A^*=1\mid x)$. Denote the two eigenvalues as $e_1$ and $e_2$. Let $l_1 = (\Pr(A=1\mid x)-e_2)/(e_1-e_2)$ and $l_2 = 1 - l_1$, then
\begin{align*}
\Pr(A=1\mid A^*=1,x) = e_j, \text{ where } j = \arg\min_j (1-e_j)l_j.
\end{align*}
Once the labeling of $\Pr(A=1\mid A^*,x)$ is determined, the labeling of $\Pr(Y\mid A^*,x)$ is determined automatically by the link between eigenvalues and eigenfunctions. Therefore, $\E(Y^{a^*}) = \E\{\E(Y\mid A^*=a^*,X)\}$ is identified, as we have identified $\E(Y\mid A^*=a^*,X=x)$ for all values $x$ of $X$.
When $Y$ is a count variable, the proof is similar to the binary case by treating $\mathbb{I}(Y=k)$ as a binary outcome, and repeat it for every level $k$.
When $Y$ is continuous, we let $P(A,Z\mid x)=(\Pr(A=2-i,Z=2-j\mid X=x))_{i,j}$, $\E(Y\mid A,Z,x) = (\E(Y\mid A=2-i,Z=2-j, x))_{i,j}$, $P(A\mid A^*,x)=(\Pr(A=2-i\mid A^*=2-j,X=x))_{i,j}$, $P(Z\mid A^*,x)=(\Pr(Z=2-i\mid A^*=2-j,X=x))_{i,j}$, $P_D(A^*\mid x)=\text{diag}(\Pr(A^*=1\mid X=x),\Pr(A^*=0\mid X=x))$, $\E_D(Y\mid A^*,x) = \text{diag}(\E(Y\mid A^*=1,X=x),\E(Y\mid A^*=0,X=x))$. By Assumption \ref{main-asp:conditional_independence}, we have
\begin{align*}
P(A,Z\mid x)&=P(A\mid A^*,x)P_D(A^*\mid x)P(Z\mid A^*,x)^\T,\\
\E(Y\mid A,Z,x)\cdot P(A,Z\mid x) &=P(A\mid A^*,x)P_D(A^*\mid x)\E_D(Y\mid A^*,x)P(Z\mid A^*,x)^\T,
\end{align*}
where the left-hand side of the second equation is an elementwise product. Therefore,
\begin{align*}
\{\E(Y\mid A,Z,x)\cdot P(A,Z\mid x)\}P(A, Z\mid x)^{-1} = P(A\mid A^*,x)\E_D(Y\mid A^*,x)P(A\mid A^*,x)^{-1}.
\end{align*}
Following a similar argument for binary $Y$, we can identify $\E(Y\mid A^*,x)$, and thus $\E(Y^{a^*})$.
\section{Derivation of \texorpdfstring{\eqref{main-eq:tangentspace_full}}{equation main-eq:tangentspace-full} and \texorpdfstring{\eqref{main-eq:tangent_space}}{equation main-eq:tangent-space}}
\label{sec:proof_tangentspace_full}
\subsection{Proof of \texorpdfstring{\eqref{main-eq:tangentspace_full}}{equation main-eq:tangentspace-full}}
Denote the right-hand side of \eqref{main-eq:tangentspace_full} as $\Lambda_0$, we first show $\Lambda^{F,\bot}\subset \Lambda_0$.
For any $K\in\Lambda^{F,\bot}$, $G_y\in\Lambda^F_y$, $G_a\in\Lambda^F_a$, $G_z\in\Lambda^F_z$, $G_{a^*}\in\Lambda^F_{a^*}$, and $G_x\in\Lambda^F_x$, we have
\begin{align*}
0&=\E(KG_y)=\E\{\E(KG_y\mid Y,A^*,X)\}=\E\{G_y\E(K\mid Y,A^*,X)\}, \\
0&=\E(KG_a)=\E\{\E(KG_a\mid A,A^*,X)\}=\E\{G_a\E(K\mid A,A^*,X)\}, \\
0&=\E(KG_z)=\E\{\E(KG_z\mid Z,A^*,X)\}=\E\{G_z\E(K\mid Z,A^*,X)\}, \\
0&=\E(KG_{a^*})=\E\{\E(KG_{a^*}\mid A^*,X)\}=\E\{G_{a^*}\E(K\mid A^*,X)\}, \\
0&=\E(KG_x)=\E\{\E(KG_x\mid X)\}=\E\{G_x\E(K\mid X)\}.
\end{align*}
Let $G_x = \E(K\mid X)$, we can check that $\E(G_x) = \E\{\E(K\mid X)\}=\E(K)=0$. Then $\E\{G_x\E(K\mid X)\} =\E\{\E(K\mid X)^2\}=0$ implies $\E(K\mid X)=0$.
Let $G_{a^*} = \E(K\mid A^*,X)$, we can check that $\E(G_{a^*}\mid X)=\E(K\mid X) = 0$. Then $\E\{G_{a^*}\E(K\mid A^*,X)\} =\E\{\E(K\mid A^*,X)^2\}=0$ implies $\E(K\mid A^*, X)=0$.
Let $G_z = \E(K\mid Z,A^*,X)$, we can check that $\E(G_z\mid A^*,X)=\E(K\mid A^*,X) = 0$. Then $\E\{G_z\E(K\mid Z,A^*,X)\} =\E\{\E(K\mid Z,A^*,X)^2\}=0$ implies $\E(K\mid Z,A^*, X)=0$.
Let $G_a = \E(K\mid A,A^*,X)$, we can check that $\E(G_a\mid A^*,X)=\E(K\mid A^*,X) = 0$. Then $\E\{G_a\E(K\mid A,A^*,X)\} =\E\{\E(K\mid A,A^*,X)^2\}=0$ implies $\E(K\mid A,A^*, X)=0$.
Let $G_y = \E(K\mid Y,A^*,X)$, we can check that $\E(G_y\mid A^*,X)=\E(K\mid A^*,X) = 0$. Then $\E\{G_y\E(K\mid Y,A^*,X)\} =\E\{\E(K\mid Y,A^*,X)^2\}=0$ implies $\E(K\mid Y,A^*, X)=0$.
Therefore, $\Lambda^{F,\bot}\subset \Lambda_0$.
To show $\Lambda_0\subset\Lambda^{F,\bot}$, note that for any $L\in\Lambda_0$,
\begin{align*}
\E(LG_y) &= \E\{\E(LG_y\mid Y,A^*,X)\} = \E\{G_y\E(L\mid Y,A^*,X)\} = 0,\\
\E(LG_a) &= \E\{\E(LG_a\mid A,A^*,X)\} = \E\{G_a\E(L\mid A,A^*,X)\} = 0,\\
\E(LG_z) &= \E\{\E(LG_z\mid Z,A^*,X)\} = \E\{G_z\E(L\mid Z,A^*,X)\} = 0,\\
\E(LG_{a^*}) &= \E\{\E(LG_{a^*}\mid A^*,X)\} = \E\{G_{a^*}\E(L\mid A^*,X)\} = 0,\\
\E(LG_x) &= \E\{\E(LG_x\mid X)\} = \E\{G_x\E(L\mid X)\} = 0,
\end{align*}
which implies $L\in \Lambda^{F,\bot}$. Therefore, $\Lambda_0\subset\Lambda^{F,\bot}$.
\subsection{Equivalence of \texorpdfstring{\eqref{main-eq:tangentspace_full}}{equation main-eq:tangentspace-full} and \texorpdfstring{\eqref{main-eq:tangent_space}}{equation main-eq:tangent-space}}
The space $\Lambda_1 = \{K=k(Y,A,A^*,Z,X): \E(K\mid Y,A^*,X)=0\}$ can be written as
$$
\Lambda_1 = \{W = M-\E(M\mid Y,A^*,X):M=m(Y,A,A^*,Z,X)\}.
$$
Similarly, the space $\Lambda_2 = \{K=k(Y,A,A^*,Z,X): \E(K\mid A,A^*,X)=0\}$ can be written as
$$
\Lambda_2 = \{W = M-\E(M\mid A,A^*,X):M=m(Y,A,A^*,Z,X)\},
$$
and the space $\Lambda_3 = \{K=k(Y,A,A^*,Z,X): \E(K\mid Z,A^*,X)=0\}$ can be written as
$$
\Lambda_3 = \{W = M-\E(M\mid Z,A^*,X):M=m(Y,A,A^*,Z,X)\}.
$$
The joint space $\Lambda_1\cap\Lambda_2$ can be written as
\begin{align*}
\Lambda_1\cap\Lambda_2
&= \bigl\{W = M-\E(M\mid Y,A^*,X) \\
&\qquad - \E\{M-\E(M\mid Y,A^*,X)\mid A,A^*,X\}:\\
&\qquad M=m(Y,A,A^*,Z,X)\bigr\} \\
&= \bigl\{W = M - \E(M\mid Y,A^*,X) - \E(M\mid A,A^*,X)\\
&\qquad + \E(M\mid A^*,X): M=m(Y,A,A^*,Z,X)\bigr\}.
\end{align*}
The joint space $\Lambda_1\cap\Lambda_2\cap\Lambda_3$ can be written as
\begin{align*}
&\Lambda_1\cap\Lambda_2\cap\Lambda_3 \\
=& \bigl\{W = M - \E(M\mid Y,A^*,X) - \E(M\mid A,A^*,X)\\
&\quad + \E(M\mid A^*,X)
 - \E\{M - \E(M\mid Y,A^*,X) - \E(M\mid A,A^*,X)\\
&\quad + \E(M\mid A^*,X)\mid Z,A^*,X\}: M=m(Y,A,A^*,Z,X)\bigr\} \\
=& \bigl\{W = M - \E(M\mid Y,A^*,X) - \E(M\mid A,A^*,X)\\
&\quad - \E(M\mid Z,A^*,X) + 2\E(M\mid A^*,X):
M=m(Y,A,A^*,Z,X)\bigr\}.
\end{align*}
\section{Proof of Lemma \ref{main-lem:eif}}\label{sec:eif_lemma}
Recall the EIF in the oracle nonparametric model of $(Y,A^*,X, A, Z)$ given in \eqref{main-eq:IF_full},
\begin{align}\label{eq:eif_lemma}
   \phi_{a^*}^F = \frac{\mathbb{I}\left( A^*=a^*\right) }{\Pr\left( A^{\ast}=a^*\mid X\right) }\left\{ Y-\E\left( Y\mid A^* = a^*,X\right) \right\} +\E\left( Y\mid A^* = a^*,X\right) -\psi _{a^*}.
\end{align}
To show that \eqref{eq:eif_lemma} is also the EIF in the full data hidden treatment model $\mathcal{M}^*$, it suffices to show \eqref{eq:eif_lemma} is orthogonal to $\Lambda^{F,\perp}$, therefore in the tangent space of $\mathcal{M}^*$.
By \eqref{main-eq:tangent_space}, we just need to show for any $M\in L_0^{2,F}$, the following expectation is zero.
\begin{align*}
&\E\left[\phi_{a^*}^F \left\{  M-\E( M\mid Y,A^*,X) -\E(M\mid A,A^*,X) -\E( M\mid Z,A^*,X) +2\E( M\mid A^*,X)  \right\}  \right] \\
=& \E(\phi_{a^*}^F M) - \E\{\phi_{a^*}^F \E(M\mid Y,A^*,X) \} - \E\{\phi_{a^*}^F \E(M\mid A,A^*,X) \} - \E\{\phi_{a^*}^F \E(M\mid Z,A^*,X) \} \\
&\qquad + 2 \E\{\phi_{a^*}^F \E( M\mid A^*,X) \} \\
=& \E(\phi_{a^*}^F M) - \E\{ \E(\phi_{a^*}^F M\mid Y,A^*,X) \} - \E\{\phi_{a^*}^F \E(M\mid A,A^*,X) \} - \E\{\phi_{a^*}^F \E(M\mid Z,A^*,X) \} \\
&\qquad + 2 \E\{\phi_{a^*}^F \E( M\mid A^*,X) \} \\
=& - \E\{\phi_{a^*}^F \E(M\mid A,A^*,X) \} - \E\{\phi_{a^*}^F \E(M\mid Z,A^*,X) \} + 2 \E\{\phi_{a^*}^F \E( M\mid A^*,X) \}.
\end{align*}
Note that
\begin{align*}
\E\{\phi_{a^*}^F \E(M\mid A,A^*,X) \} &= \E\left[\E\{\phi_{a^*}^F \E(M\mid A,A^*,X) \mid A^*,X\}\right] \\
&= \E\left[\E(\phi_{a^*}^F\mid A^*,X)\E\{ \E(M\mid A,A^*,X) \mid A^*,X\}\right] \\
&= \E\left[\E(\phi_{a^*}^F\mid A^*,X) \E(M\mid A^*,X)\right] \\
&= \E\left[\E\{\phi_{a^*}^F\E(M\mid A^*,X) \mid A^*,X\} \right] \\
&= \E\{\phi_{a^*}^F \E(M\mid A^*,X) \},
\end{align*}
where the second equality makes use of $Y\bigCI A\mid A^*,X$.
Similarly, $\E\{\phi_{a^*}^F \E(M\mid Z,A^*,X) \} = \E\{\phi_{a^*}^F \E(M\mid A^*,X)\}$.
Therefore,
$$
\E\left[\phi_{a^*}^F \left\{  M-\E( M\mid Y,A^*,X) -\E(M\mid A,A^*,X) -\E( M\mid Z,A^*,X) +2\E( M\mid A^*,X)  \right\}  \right] = 0,
$$
which concludes the proof.
\section{Proof of Theorem \ref{main-thm:ortho_tangent_space}}\label{sec:ortho_tangent_space}
We begin by proving the first part of the theorem.
The observed data tangent space $\Lambda$ is the projection of the full data tangent space onto the observed data. Let
\begin{align*}
\Lambda_{x}&= \E(\Lambda^F_{x}\mid Y,A,Z,X) = \Lambda^F_{x}, \\
\Lambda_{a^*}&= \E(\Lambda^F_{a^*}\mid Y,A,Z,X) = \{\E[g_{a^*}(A^*,X)\mid Y,A,Z,X]:g_{a^*}(A^*,X)\in \Lambda^F_{a^*}\},\\
\Lambda_{a}&= \E(\Lambda^F_{a}\mid Y,A,Z,X) = \{\E[g_a(A,A^*,X)\mid Y,A,Z,X]:g_a(A,A^*,X)\in \Lambda^F_a\},\\
\Lambda_{z}&= \E(\Lambda^F_{z}\mid Y,A,Z,X) = \{\E[g_z(Z,A^*,X)\mid Y,A,Z,X]:g_z(Z,A^*,X)\in \Lambda^F_z\},\\
\Lambda_{y}&= \E(\Lambda^F_{y}\mid Y,A,Z,X) = \{\E[g_y(Y,A^*,X)\mid Y,A,Z,X]:g_y(Y,A^*,X)\in \Lambda^F_y\},
\end{align*}
then $\Lambda = \Lambda_{x} + \Lambda_{a^*} + \Lambda_{y} +\Lambda_{a} + \Lambda_{z}$ .
By the definition of $\Lambda$, $l(Y,A,Z,X)$ is orthogonal to $\Lambda$ if and only if
\begin{align*}
\E\bigg[l(Y,A,Z,X)
\Big(g_x(X) + \E\{g_{a^*}(A^*,X)\mid Y,A,Z,X\}+\E\{g_y(Y, A^*,X)\mid Y,A,Z,X\} \\
+\E\{g_a(A, A^*,X)\mid Y,A,Z,X\}+\E\{g_z(Z, A^*,X)\mid Y,A,Z,X\}\Big)\bigg]=0
\end{align*}
for all  $g_x(X)\in \Lambda^{F}_{x}, g_{a^*}(A^*,X)\in\Lambda^{F}_{a^*}, g_{y}(Y, A^*,X)\in\Lambda^{F}_{y}, g_{a}(A, A^*,X)\in\Lambda^{F}_{a}, g_{z}(Z, A^*,X)\in\Lambda^{F}_{z}.$ That is, for all $g_x(X)$ such that $\E\{g_x(X) \}=0$, $g_{a^*}(A^*,X)$ such that $\E\{g_{a^*}(A^*,X)\}=0$,
all $g_y(Y,A^*,X)$ such that $\E\{g_y(Y,A^*,X)\mid A^*,X\}=0$, all $g_a(A,A^*,X)$ such that $\E\{g_a(A,A^*,X)\mid A^*,X\}=0$, and all
$g_z(Z,A^*,X)$ such that $\E\{g_z(Z,A^*,X)\mid A^*,X\}=0$.
For any $g_x(X)\in \Lambda^{F}_{x}$, we have
\begin{align*}
0 &= \E[l(Y,A,Z,X)g_x(X)] \\
&= \E[g_x(X)\E\{l(Y,A,Z,X)\mid X\}].
\end{align*}
Let $g_x(X) = \E\{l(Y,A,Z,X)\mid X \}$, which satisfies $\E\{g_x(X)\} = \E\{l(Y,A,Z,X)\} = 0$, then
$$
\E\left[\E\left\{l(Y,A,Z,X)\mid X\right\}\E\left\{l(Y,A,Z,X)\mid X\right\}\right]=0
$$
implies
\begin{align}\label{eq:x}
\E\left\{l(Y,A,Z,X)\mid X\right\}=0.
\end{align}
For any $g_{a^*}(A^*,X)\in\Lambda^{F}_{a^*}$, we have
\begin{align*}
0&=\E[l(Y,A,Z,X)\E\{g_{a^*}(A^*,X)\mid Y,A,Z,X\}]\\
&=\E[l(Y,A,Z,X)g_{a^*}(A^*,X)]\\
&= \E[\E\left\{l(Y,A,Z,X)\mid A^*,X\right\}g_{a^*}(A^*,X)].
\end{align*}
Let $g_{a^*}(A^*,X)=\E\{l(Y,A,Z,X)\mid A^*,X\}$, which satisfies
$$
\E\{g_{a^*}(A^*,X)\}=\E[\E\{l(Y,A,Z,X)\mid A^*,X\}]=\E\{l(Y,A,Z,X)\}=0,
$$
then
$$
\E\left[\E\left\{l(Y,A,Z,X)\mid A^*,X\right\}\E\left\{l(Y,A,Z,X)\mid A^*,X\right\}\right]=0
$$
implies
\begin{align}\label{eq:a_star}
\E\left\{l(Y,A,Z,X)\mid A^*,X\right\}=0.
\end{align}
For any $g_y(Y,A^*,X)\in\Lambda^{F}_y$ such that $\E\{g_y(Y,A^*,X)\mid A^*,X\}=0$,
\begin{align*}
0&=\E[l(Y,A,Z,X)\E\{g_y(Y, A^*,X)\mid Y,A,Z,X\}]\\
&=\E[l(Y,A,Z,X)g_y(Y, A^*,X)]\\
&=\E[\E\{l(Y,A,Z,X)\mid Y,A^*,X\}g_y(Y,A^*,X)].
\end{align*}
Let $g_y = \E\{l(Y,A,Z,X)\mid Y,A^*,X\}$, one can check that
\begin{align*}
\E[g_y\mid A^*,X] &= \E[\E\{l(Y,A,Z,X)\mid Y,A^*,X\}\mid A^*,X]\\
&=\E\{l(Y,A,Z,X)\mid A^*,X\}=0
\end{align*}
by \eqref{eq:a_star}.
Therefore, $\E[\E\{l(Y,A,Z,X)\mid Y,A^*,X\}g_y(Y,A^*,X)]=0$ implies
\begin{align}\label{eq:w_a_star}
\E\{l(Y,A,Z,X)\mid Y,A^*,X\}=0.
\end{align}
Similarly, we can derive
\begin{align}
\E\{l(Y,A,Z,X\mid A,A^*,X\}=0 \label{eq:a_a_star},\\
\E\{l(Y,A,Z,X\mid Z,A^*,X\}=0 \label{eq:z_a_star}.
\end{align}
Note that \eqref{eq:x} and \eqref{eq:a_star} are implied by \eqref{eq:w_a_star},  \eqref{eq:a_a_star}, or \eqref{eq:z_a_star}.
Therefore,
\begin{align*}
\Lambda^{\bot} = \{L=l(Y,A,Z,X):\E(L\mid Y,A^*,X) = \E(L\mid A,A^*,X) = \E(L\mid Z,A^*,X)=0\}\cap L^{2}_0.
\end{align*}
We then prove the second part of the theorem.
Since $A$ and $Z$ are binary, any $L\in\Lambda^\bot$ can be written in the form of
\begin{align}\label{eq:L}
L = AZh_{11}(Y,X) + A(1-Z)h_{10}(Y,X) + (1-A)Zh_{01}(Y,X)+ (1-A)(1-Z)h_{00}(Y,X).
\end{align}
By $\E(L\mid A,A^*,X)=0$, we have
{\small
\begin{align*}
& \qquad A\E(Z\mid A^*,X)\E\{h_{11}(Y,X)\mid A^*,X\} + A\{1-\E(Z\mid A^*,X)\}\E\{h_{10}(Y,X)\mid A^*,X\} \\
&  + (1-A)\E(Z\mid A^*,X)\E\{h_{01}(Y,X)\mid A^*,X\} + (1-A)\{1-\E(Z\mid A^*,X)\}\E\{h_{00}(Y,X)\mid A^*,X\} = 0.
\end{align*}}
Let $A=1,0$ in the above equation, we get
\begin{align}
\E(Z\mid A^*,X)\E\{h_{11}(Y,X)\mid A^*,X\} + \{1-\E(Z\mid A^*,X)\}\E\{h_{10}(Y,X)\mid A^*,X\} &= 0, \label{eq:obs_ortho1}\\
\E(Z\mid A^*,X)\E\{h_{01}(Y,X)\mid A^*,X\} + \{1-\E(Z\mid A^*,X)\}\E\{h_{00}(Y,X)\mid A^*,X\} &= 0. \label{eq:obs_ortho2}
\end{align}
Similarly, we set $Z=1,0$ in $\E(L\mid Z,A^*,X)=0$ to get
\begin{align}
\E(A\mid A^*,X)\E\{h_{11}(Y,X)\mid A^*,X\} + \{1-\E(A\mid A^*,X)\}\E\{h_{10}(Y,X)\mid A^*,X\} &= 0, \label{eq:obs_ortho3}\\
\E(A\mid A^*,X)\E\{h_{01}(Y,X)\mid A^*,X\} + \{1-\E(A\mid A^*,X)\}\E\{h_{00}(Y,X)\mid A^*,X\} &= 0. \label{eq:obs_ortho4}
\end{align}
From $\E(L\mid Y,A^*,X)=0$, we have
{\footnotesize
\begin{align}
\begin{split} \label{eq:obs_ortho5}
& \qquad \E(A\mid A^*,X)\E(Z\mid A^*,X)h_{11}(Y,X) + \E(A\mid A^*,X)\{1-\E(Z\mid A^*,X)\}h_{10}(Y,X) \\
& + \{1-\E(A\mid A^*,X)\}\E(Z\mid A^*,X)h_{01}(Y,X) + \{1-\E(A\mid A^*,X)\}\{1-\E(Z\mid A^*,X)\}h_{00}(Y,X) = 0.
\end{split}
\end{align}}
Thus, \eqref{eq:obs_ortho1}-\eqref{eq:obs_ortho5} are all the constraints on $L$.
Without loss of generality, we assume $\E(Z\mid A^*=a^*,X)\ne \E(A\mid A^*=a^*,X)$ where $a^*=0,1$. This is a reasonable assumption because $\E(Z\mid A^*=a^*,X)= \E(A\mid A^*=a^*,X)$ is not a priori known to hold, so it should be not used as a restriction when we characterize the tangent space.
We will discuss the case when $\E(Z\mid A^*=a^*,X)= \E(A\mid A^*=a^*,X)$ later for completeness of the proof.
Under this event, \eqref{eq:obs_ortho1} and \eqref{eq:obs_ortho3} imply
$$
\E\{h_{11}(Y,X)\mid A^*,X\}=\E\{h_{10}(Y,X)\mid A^*,X\} = 0.
$$
Similarly, \eqref{eq:obs_ortho2} and \eqref{eq:obs_ortho4} imply
$$
\E\{h_{01}(Y,X)\mid A^*,X\}=\E\{h_{00}(Y,X)\mid A^*,X\} = 0.
$$
Set $A^*=1$ and $A^*=0$ in \eqref{eq:obs_ortho5} respectively, and let $a_j(X) = \E(A\mid A^*=j,X), z_j(X) = \E(Z\mid A^*=j,X), j=0,1$, we have
\begin{align*}
a_1z_1h_{11}(Y,X)+a_1(1-z_1)h_{10}(Y,X)+(1-a_1)z_1h_{01}(Y,X)+(1-a_1)(1-z_1)h_{00}(Y,X) = 0, \\
a_0z_0h_{11}(Y,X)+a_0(1-z_0)h_{10}(Y,X)+(1-a_0)z_0h_{01}(Y,X)+(1-a_0)(1-z_0)h_{00}(Y,X) = 0,
\end{align*}
from which we solve
\begin{align*}
h_{11}(Y,X) = -\frac{1-z_1}{z_1}h_{10}(Y,X) - \frac{1-a_1}{a_1}h_{01}(Y,X) - \frac{(1-a_1)(1-z_1)}{a_1z_1}h_{00}(Y,X), \\
h_{11}(Y,X) = -\frac{1-z_0}{z_0}h_{10}(Y,X) - \frac{1-a_0}{a_0}h_{01}(Y,X) - \frac{(1-a_0)(1-z_0)}{a_0z_0}h_{00}(Y,X).
\end{align*}
By eliminating $h_{11}(Y,X)$, we get
\begin{align*}
\left\{\frac{(1-a_1)(1-z_1)}{a_1z_1} - \frac{(1-a_0)(1-z_0)}{a_0z_0}\right\}h_{00}(Y,X) = \frac{z_1-z_0}{z_1z_0}h_{10}(Y,X)+\frac{a_1-a_0}{a_1a_0}h_{01}(Y,X).
\end{align*}
Therefore, $h_{00}(Y,X)$ and $h_{11}(Y,X)$ are functions of $h_{10}(Y,X)$ and $h_{01}(Y,X)$:
\begin{align}
\label{eq:h00}
\begin{split}
h_{00}(Y,X) &= \frac{a_1a_0(z_1-z_0)}{a_0a_1(z_1-z_0)+z_0z_1(a_1-a_0)+a_0z_0-a_1z_1}h_{10}(Y,X) \\
&\qquad + \frac{z_1z_0(a_1-a_0)}{a_0a_1(z_1-z_0)+z_0z_1(a_1-a_0)+a_0z_0-a_1z_1}h_{01}(Y,X),
\end{split}\\
\label{eq:h11}
\begin{split}
h_{11}(Y,X)&=\frac{(1-z_1)(1-z_0)(a_1-a_0)}{a_0a_1(z_1-z_0)+z_0z_1(a_1-a_0)+a_0z_0-a_1z_1}h_{10}(Y,X) \\
&\qquad + \frac{(1-a_1)(1-a_0)(z_1-z_0)}{a_0a_1(z_1-z_0)+z_0z_1(a_1-a_0)+a_0z_0-a_1z_1}h_{01}(Y,X).
\end{split}
\end{align}
Plugging \eqref{eq:h00} and \eqref{eq:h11} into \eqref{eq:L}, we get
\begin{align*}
L(Y,A,Z,X) &= H_{10}+H_{01},
\end{align*}
where
\begin{align*}
H_{10} &= h_{10}(Y,X)\frac{a_1(1-z_0)(A-a_0)(Z-z_1) - a_0(1-z_1)(A-a_1)(Z-z_0)}{a_0a_1(z_1-z_0)+z_0z_1(a_1-a_0)+a_0z_0-a_1z_1}, \\
H_{01} &= h_{01}(Y,X)\frac{z_1(1-a_0)(Z-z_0)(A-a_1) - z_0(1-a_1)(Z-z_1)(A-a_0)}{a_0a_1(z_1-z_0)+z_0z_1(a_1-a_0)+a_0z_0-a_1z_1}, \\
& \qquad \qquad \E\{h_{10}(Y,X)\mid A^*,X\} = \E\{h_{01}(Y,X)\mid A^*,X\} = 0.
\end{align*}
Note that the space $\Lambda^{\bot} = \{H_{10}+H_{01}\}$ can also be written as
\begin{align}\label{eq:ortho_tangent_space}
\begin{split}
    \Lambda^{\bot} &= \{(A-a_0)(Z-z_1)h_1(Y,X)+(A-a_1)(Z-z_0)h_2(Y,X):\\
    &\qquad\qquad \E\{h_1(Y,X)\mid A^*,X\}=\E\{h_2(Y,X)\mid A^*,X\}=0\}.
\end{split}
\end{align}
In the events (1) $D_0:=\{\E(Z\mid A^*=0,X)= \E(A\mid A^*=0,X)\}$, (2)$D_1:=\{\E(Z\mid A^*=1,X)= \E(A\mid A^*=1,X)\}$, (3)$D_0\cap D_1$, there are more constraints on the observed data tangent space $\Lambda$. As a result, the orthocomplement $\Lambda^\bot$ will be larger. For completeness of the proof, we outline the constraints in these events.
Under the event $D_1\cap D_0^C$, that is,
when $\E(Z\mid A^*=1,X)= \E(A\mid A^*=1,X)$ and $\E(Z\mid A^*=0,X)\ne \E(A\mid A^*=0,X)$, we have
\begin{align}\label{eq:tangent_space_relax1}
\begin{split}
&\E\{h_{11}(Y,X)\mid A^*=0,X\} = \E\{h_{10}(Y,X)\mid A^*=0,X\} \\
&\qquad = \E\{h_{01}(Y,X)\mid A^*=0,X\}
= \E\{h_{00}(Y,X)\mid A^*=0,X\} = 0
\end{split}
\end{align}
and
\begin{align}\label{eq:tangent_space_relax2}
\begin{split}
a_1\E\{h_{11}(Y,X)\mid A^*=1,X\} + (1-a_1)\E\{h_{10}(Y,X)\mid A^*=1,X\} &= 0, \\
a_1\E\{h_{01}(Y,X)\mid A^*=1,X\} + (1-a_1)\E\{h_{00}(Y,X)\mid A^*=1,X\} &= 0.
\end{split}
\end{align}
One can check the above equations are compatible with \eqref{eq:h00} and \eqref{eq:h11} when $a_1=z_1$. Besides, the requirement $\E\{h_1(Y,X)\mid A^*,X\}=\E\{h_2(Y,X)\mid A^*,X\}=0$ in \eqref{eq:ortho_tangent_space} is relaxed to \eqref{eq:tangent_space_relax1} and \eqref{eq:tangent_space_relax2}.
This means that the right-hand side of \eqref{eq:ortho_tangent_space} is a subspace of $\Lambda^\bot$ under the event $D_1\cap D_0^C$.
Similarly, under the event $D_0\cap D_1^C$, i.e., $\E(Z\mid A^*=1,X)\ne \E(A\mid A^*=1,X)$ and $\E(Z\mid A^*=0,X)= \E(A\mid A^*=0,X)$, we have
\begin{align*}
\begin{split}
&\E\{h_{11}(Y,X)\mid A^*=1,X\} = \E\{h_{10}(Y,X)\mid A^*=1,X\} \\
&\qquad = \E\{h_{01}(Y,X)\mid A^*=1,X\} = \E\{h_{00}(Y,X)\mid A^*=1,X\} = 0
\end{split}
\end{align*}
and
\begin{align*}
a_0\E\{h_{11}(Y,X)\mid A^*=0,X\} + (1-a_0)\E\{h_{10}(Y,X)\mid A^*=0,X\} &= 0, \\
a_0\E\{h_{01}(Y,X)\mid A^*=0,X\} + (1-a_0)\E\{h_{00}(Y,X)\mid A^*=0,X\} &= 0.
\end{align*}
One can check the above equations are compatible with \eqref{eq:h00} and \eqref{eq:h11} when $a_0=z_0$, and the right-hand side of \eqref{eq:ortho_tangent_space} is a subspace of $\Lambda^\bot$ under the event $D_0\cap D_1^C$.
When $\E(Z\mid A^*,X)= \E(A\mid A^*,X), A^*=0,1$, i.e., under the event $D_0\cap D_1$, \eqref{eq:h00} and \eqref{eq:h11} reduces to
\begin{align*}
    h_{00}(Y,X) &= -\frac{cd}{c+d} \{h_{10}(Y,X) + h_{01}(Y,X)\},  \\
    h_{11}(Y,X) &= -\frac{1}{c+d} \{h_{10}(Y,X) + h_{01}(Y,X)\},
\end{align*}
where $c = a_1(1-a_1)^{-1}$, $d = a_0(1-a_0)^{-1}$.
Therefore,
\begin{align*}
    \E\{h_{00}(Y,X)\mid A^*=1,X\} &= -\frac{cd}{c+d} [\E\{h_{10}(Y,X)\mid A^*=1,X\} + \E\{h_{01}(Y,X)\mid A^*=1,X\}],  \\
    \E\{h_{00}(Y,X)\mid A^*=0,X\} &= -\frac{cd}{c+d} [\E\{h_{10}(Y,X)\mid A^*=0,X\} + \E\{h_{01}(Y,X)\mid A^*=0,X\}],  \\
    \E\{h_{11}(Y,X)\mid A^*=1,X\} &= -\frac{1}{c+d} [\E\{h_{10}(Y,X)\mid A^*=1,X\} + \E\{h_{01}(Y,X)\mid A^*=1,X\}], \\
    \E\{h_{11}(Y,X)\mid A^*=0,X\} &= -\frac{1}{c+d} [\E\{h_{10}(Y,X)\mid A^*=0,X\} + \E\{h_{01}(Y,X)\mid A^*=0,X\}].
\end{align*}
From \eqref{eq:obs_ortho1}-\eqref{eq:obs_ortho4},
\begin{align*}
a_1\E\{h_{11}(Y,X)\mid A^*=1,X\} + (1-a_1)\E\{h_{10}(Y,X)\mid A^*=1,X\} = 0, \\
a_1\E\{h_{01}(Y,X)\mid A^*=1,X\} + (1-a_1)\E\{h_{00}(Y,X)\mid A^*=1,X\} = 0, \\
a_0\E\{h_{11}(Y,X)\mid A^*=0,X\} + (1-a_0)\E\{h_{10}(Y,X)\mid A^*=0,X\} = 0, \\
a_0\E\{h_{01}(Y,X)\mid A^*=0,X\} + (1-a_0)\E\{h_{00}(Y,X)\mid A^*=0,X\} = 0,
\end{align*}
that is,
\begin{align*}
c\E\{h_{11}(Y,X)\mid A^*=1,X\} + \E\{h_{10}(Y,X)\mid A^*=1,X\} = 0, \\
c\E\{h_{01}(Y,X)\mid A^*=1,X\} + \E\{h_{00}(Y,X)\mid A^*=1,X\} = 0, \\
d\E\{h_{11}(Y,X)\mid A^*=0,X\} + \E\{h_{10}(Y,X)\mid A^*=0,X\} = 0, \\
d\E\{h_{01}(Y,X)\mid A^*=0,X\} + \E\{h_{00}(Y,X)\mid A^*=0,X\} = 0.
\end{align*}
One can check the above equations are compatible with \eqref{eq:h00} and \eqref{eq:h11} with the constraint
\begin{align*}
d\E\{h_{10}(Y,X)\mid A^*=1,X\} - c\E\{h_{01}(Y,X)\mid A^*=1,X\} = 0, \\
c\E\{h_{10}(Y,X)\mid A^*=0,X\} - d\E\{h_{01}(Y,X)\mid A^*=0,X\} = 0.
\end{align*}
So the right-hand side of \eqref{eq:ortho_tangent_space} is a subspace of $\Lambda^\bot$ under the event $D_1\cap D_0$.
\section{Proof of Theorem \ref{main-thm:representation}}\label{sec:orthocomplement}
We aim to establish an equivalence between the space of functions $P_1 = \{ h(Y,X): \E\{ h(Y,X)\mid A^*,X \}=0\}\cap L^2_0$ and the space of functions
\begin{align*}
& P_2 = \Biggl\{v(Y,X) = \frac{v_1(Y,X)-\E\{v_1(Y,X)\mid A^*=0,X\}}{\E\{v_1(Y,X)\mid A^*=1,X\}-\E\{v_1(Y,X)\mid A^*=0,X\}} \\
& \qquad \qquad +  \frac{v_0(Y,X)-\E\{v_0(Y,X)\mid A^*=1,X\}}{\E\{v_0(Y,X)\mid A^*=0,X\}-\E\{v_0(Y,X)\mid A^*=1,X\}} - 1 : v_1(Y,X),v_0(Y,X)\in L^2 \Biggr\},
\end{align*}
Such equivalence provides a closed-form characterization of $P_1$. Based on this, we obtain a closed-form characterization  $\Lambda^\bot$.
We separately consider the categorical and continuous cases.
\subsection{Categorical \texorpdfstring{$Y$}{Y} Case}\label{sec:orthocomplement_categorical}
We first show that $P_2\subset P_1$. It is straightforward to verify that
\begin{align*}
\E\left\{ \frac{v_1(Y,X)-\E\{v_1(Y,X)\mid A^*=0,X\}}{\E\{v_1(Y,X)\mid A^*=1,X\}-\E\{v_1(Y,X)\mid A^*=0,X\}} \Biggm| A^*,X\right\} = \mathbb{I}(A^*=1), \\
\E\left\{ \frac{v_0(Y,X)-\E\{v_0(Y,X)\mid A^*=1,X\}}{\E\{v_0(Y,X)\mid A^*=0,X\}-\E\{v_0(Y,X)\mid A^*=1,X\}} \Biggm| A^*,X\right\} = \mathbb{I}(A^*=0).
\end{align*}
Therefore,
\begin{align*}
\E\{ v(Y,X)\mid A^*,X \} = \mathbb{I}(A^*=1) + \mathbb{I}(A^*=0) - 1 = 0.
\end{align*}
So $P_2\subset P_1$.
We then show $P_1\subset P_2$. Note that any $h(Y,X)\in P_1$ can be rewritten as
\begin{align*}
h(Y,X) = \sum^k_{j=1} \mathbb{I}(Y=j)h_j(X),
\end{align*}
and by $\E\{ h(Y,X)\mid A^*,X \}=0$, we have
\begin{align*}
\sum^k_{j=1}h_j(X)\Pr(Y=j\mid A^*,X) = 0.
\end{align*}
For $j = 1,\ldots, k$, we let
\begin{align*}
v_1(Y=j,X)
&= \frac{1}{2}\{ \Pr(Y=k\mid A^*=1,X) - \Pr(Y=k\mid A^*=0,X) \}h_j(X) \\
&\qquad + \mathbb{I}(j=k), \\
v_0(Y=j,X)
&= \frac{1}{2}\{ \Pr(Y=k\mid A^*=1,X) - \Pr(Y=k\mid A^*=0,X) \}h_j(X) \\
&\qquad - \mathbb{I}(j=k).
\end{align*}
Write $\delta_k(X)=\Pr(Y=k\mid A^*=1,X)-\Pr(Y=k\mid A^*=0,X)$.
Then we have
\begin{align*}
\E\{v_1(Y,X)\mid A^*,X\} &= \sum^k_{j=1}v_1(Y=j,X)\Pr(Y=j\mid A^*,X) = \Pr(Y=k\mid A^*,X), \\
\E\{v_0(Y,X)\mid A^*,X\} &= \sum^k_{j=1}v_0(Y=j,X)\Pr(Y=j\mid A^*,X) = -\Pr(Y=k\mid A^*,X),
\end{align*}
which leads to
\begin{align*}
\E\{v_1(Y,X)\mid A^*=1,X\}-\E\{v_1(Y,X)\mid A^*=0,X\}
&=  \delta_k(X), \\
\E\{v_0(Y,X)\mid A^*=0,X\}-\E\{v_0(Y,X)\mid A^*=1,X\}
&=  \delta_k(X).
\end{align*}
Therefore,
\begingroup
\allowdisplaybreaks
\begin{align*}
v(Y,X) &= \frac{v_1(Y,X)-\E\{v_1(Y,X)\mid A^*=0,X\}}{\E\{v_1(Y,X)\mid A^*=1,X\}-\E\{v_1(Y,X)\mid A^*=0,X\}} \\
& \qquad \qquad +  \frac{v_0(Y,X)-\E\{v_0(Y,X)\mid A^*=1,X\}}{\E\{v_0(Y,X)\mid A^*=0,X\}-\E\{v_0(Y,X)\mid A^*=1,X\}} - 1 \\
&= \frac{v_1(Y,X)-\Pr(Y=k\mid A^*=0,X)}{\delta_k(X)} \\
& \qquad \qquad + \frac{v_0(Y,X)+\Pr(Y=k\mid A^*=1,X)}{\delta_k(X)} - 1 \\
&= \frac{v_1(Y,X)+ v_0(Y,X)}{\delta_k(X)} \\
&= \sum^k_{j=1}\mathbb{I}(Y=j)h_j(X) \\
&= h(Y,X).
\end{align*}
\endgroup
This shows that for any $h(Y,X)\in P_1$, we can find $v_1(Y,X)$ and $v_0(Y,X)$ such that
\begin{align*}
h(Y,X) &= \frac{v_1(Y,X)-\E\{v_1(Y,X)\mid A^*=0,X\}}{\E\{v_1(Y,X)\mid A^*=1,X\}-\E\{v_1(Y,X)\mid A^*=0,X\}} \\
& \qquad \qquad +  \frac{v_0(Y,X)-\E\{v_0(Y,X)\mid A^*=1,X\}}{\E\{v_0(Y,X)\mid A^*=0,X\}-\E\{v_0(Y,X)\mid A^*=1,X\}} - 1,
\end{align*}
so $P_1\subset P_2$.
Therefore, $P_1=P_2$.
\subsection{Continuous \texorpdfstring{$Y$}{Y} Case}
The direction $P_2\subset P_1$ is easy to check. To show $P_1\subset P_2$, we
let $C\subset\real$ be a subset of the real line such that $\Pr(Y\in C\mid A^*=1,X=x) - \Pr(Y\in C\mid A^*=0,X=x)\ne 0$ for any $x$, and let
\begin{align*}
v_1(Y,X)
&= \frac{1}{2}\left\{\Pr(Y\in C\mid A^*=1,X) - \Pr(Y\in C\mid A^*=0,X) \right\}h(Y,X) \\
&\qquad + \mathbb{I}(Y\in C), \\
v_0(Y,X)
&= \frac{1}{2}\left\{\Pr(Y\in C\mid A^*=1,X) - \Pr(Y\in C\mid A^*=0,X) \right\}h(Y,X) \\
&\qquad - \mathbb{I}(Y\in C).
\end{align*}
Write $\delta_C(X)=\Pr(Y\in C\mid A^*=1,X)-\Pr(Y\in C\mid A^*=0,X)$.
Then we have
\begin{align*}
\E\{v_1(Y,X)\mid A^*,X \} &= \int v_1(y,X)f(y\mid A^*,X)\dif y = \Pr(Y\in C\mid A^*,X), \\
\E\{v_0(Y,X)\mid A^*,X \} &= \int v_0(y,X)f(y\mid A^*,X)\dif y = -\Pr(Y\in C\mid A^*,X),
\end{align*}
which leads to
\begin{align*}
\E\{v_1(Y,X)\mid A^*=1,X \} - \E\{v_1(Y,X)\mid A^*=0,X \}
&= \delta_C(X), \\
\E\{v_0(Y,X)\mid A^*=0,X \} - \E\{v_0(Y,X)\mid A^*=1,X \}
&= \delta_C(X).
\end{align*}
Therefore
\begin{align*}
v(Y,X) &= \frac{v_1(Y,X)-\E\{v_1(Y,X)\mid A^*=0,X\}}{\E\{v_1(Y,X)\mid A^*=1,X\}-\E\{v_1(Y,X)\mid A^*=0,X\}} \\
& \qquad \qquad +  \frac{v_0(Y,X)-\E\{v_0(Y,X)\mid A^*=1,X\}}{\E\{v_0(Y,X)\mid A^*=0,X\}-\E\{v_0(Y,X)\mid A^*=1,X\}} - 1 \\
&= \frac{v_1(Y,X)-\Pr(Y\in C\mid A^*=0,X)}{\delta_C(X)} \\
& \qquad \qquad + \frac{v_0(Y,X)+\Pr(Y\in C\mid A^*=1,X)}{\delta_C(X)} - 1 \\
&= \frac{v_1(Y,X)+ v_0(Y,X)}{\delta_C(X)} \\
&= h(Y,X).
\end{align*}
This completes the proof of $P_1\subset P_2$.
\section{Proof of Lemma \ref{main-lem:Q}} \label{sec:lem_Q_proof}
Let $S^F \in \Lambda^F$ be any score function of the oracle model tangent space,
we note that any influence function in the observed data model, denoted by  $Q = q\left( Y,A,Z,X\right)$,  must satisfy the following condition:
{\footnotesize
\begin{align}
\nabla _{t}\psi _{a^*,t} &=\E\left\{ Q\E\left( S^{F}\mid Y,A,Z,X\right)
\right\} \nonumber \\
&=\E\left\{ QS^{F}\right\} \nonumber \\
&=\E\left\{ Q\left( g_{y}\left( Y,A^*,X\right) +g_{a}\left( A,A^{\ast
},X\right) +g_{z}\left( Z,A^*,X\right) +g_{a^*}\left( A^{\ast
},X\right) +g_{x}\left( X\right) \right) \right\}  \nonumber \\
&=\E\left\{ \left(
\begin{array}{c}
\E\left( Q\mid Y,A^*,X\right) g_{y}\left( Y,A^*,X\right) +\E\left(
Q\mid A,A^*,X\right) g_{a}\left( A,A^*,X\right)  \\
+\E\left( Q\mid Z,A^*,X\right) g_{z}\left( Z,A^*,X\right) +\E\left(
Q\mid A^*,X\right) g_{a^*}\left( A^*,X\right) +\E\left( Q\mid X
\right) g_{x}\left( X\right)
\end{array}
\right) \right\}.  \label{eq:Q1}
\end{align}}
Here, the left-hand side $\nabla _{t}\psi _{a^*,t}$, is defined as $\lim_{t\to 0}(\psi_{a^*}(\tilde{P}_{t}) - \psi_{a^*}(P))/t$, where $\tilde{P}_t$ represents a perturbed distribution deviating from the true underlying distribution $P$ indexed by $t$.
The first equation is the central identity for influence functions in semiparametric theory \citep{tsiatis2007semiparametric}.
Similarly, in the full data space, we have
{\small
\begin{align}
\nabla _{t}\psi _{a^*,t}&=\E\left[ \text{IF}^F\left\{ g_{y}\left( Y,A^*,X\right) +g_{a}\left(
A,A^*,X\right) +g_{z}\left( Z,A^*,X\right) +g_{a^*}\left(
A^*,X\right) +g_{x}\left( X\right) \right\} \right] \nonumber \\
&=\E\left\{ \left(
\begin{array}{c}
\E\left( \text{IF}^F\mid Y,A^*,X\right) g_{y}\left( Y,A^*,X\right) +\E
\left( \text{IF}^F\mid A,A^*,X\right) g_{a}\left( A,A^*,X\right)  \\
+\E\left( \text{IF}^F\mid Z,A^*,X\right) g_{z}\left( Z,A^*,X\right) +\E
\left( \text{IF}^F\mid A^*,X\right) g_{a^*}\left( A^*,X\right) +\E
\left( \text{IF}^F\mid X\right) g_{x}\left( X\right)
\end{array}
\right) \right\}, \label{eq:Q2}
\end{align}}
where $\text{IF}^F$ denotes any influence function in the full data model, and $g_y, g_a, g_z, g_{a^*}, g_x$ are elements of the spaces $\Lambda^F_y, \Lambda^F_a, \Lambda^F_z, \Lambda^F_{a^*}, \Lambda^F_x$, respectively.
Comparing the $g_x$ terms in \eqref{eq:Q1} and \eqref{eq:Q2}, we obtain
$
\E \{ \E(Q\mid X)g_x(X) \} = \E \{ \E(\text{IF}^F\mid X)g_x(X)\}
$ for all $g_x\in \Lambda^F_x$. Equivalently, $\E [ \{ \E(Q\mid X) - \E(\text{IF}^F\mid X)\} g_x(X) ] = 0$ for all $g_x\in \Lambda^F_x$. Since $\Lambda_x = \{g_x(X): \E(g_x)=0 \}$, choosing $g_x(X) = \E(Q\mid X) - \E(\text{IF}^F\mid X) \in \Lambda^F_x$ yields $\E(Q\mid X) - \E(\text{IF}^F\mid X) = 0$ a.s.
Then comparing the $g_{a^*}$ terms in \eqref{eq:Q1} and \eqref{eq:Q2}, we obtain $ \E \{ \E(Q\mid A^*, X)g_{a^*}(A^*, X) \} = \E \{ \E(\text{IF}^F\mid A^*, X)g_{a^*}(A^*, X)\} $ for all $g_{a^*}\in \Lambda^F_{a^*}$. Equivalently, $ \E [ \{  \E(Q\mid A^*, X) - \E(\text{IF}^F\mid A^*, X)\} g_{a^*}(A^*, X) ] = 0 $ for all $g_{a^*}\in \Lambda^F_{a^*}$. Since $\Lambda^F_{a^*} = \{g_{a^*}(A^*,X): \E(g_{a^*}\mid X)=0 \}$ and we have already established $\E(Q\mid X) - \E(\text{IF}^F\mid X) = 0$, choosing $g_{a^*}(A^*, X) = \E(Q\mid A^*, X) - \E(\text{IF}^F\mid A^*, X)$ yields $\E(Q\mid A^*, X) - \E(\text{IF}^F\mid A^*, X) = 0$ a.s.
Next, by comparing the $g_{y}$ terms in \eqref{eq:Q1} and \eqref{eq:Q2}, we have
$ \E \{ \E(Q\mid Y, A^*, X)g_{y}(Y, A^*, X) \} = \E \{ \E(\text{IF}^F\mid Y, A^*, X)g_{y}(Y, A^*, X)\} $, i.e., $ \E [ \{ \E(Q\mid Y, A^*, X) - \E(\text{IF}^F\mid Y, A^*, X) \} g_{y}(Y, A^*, X) ] = 0 $ for all $g_y\in \Lambda^F_y$. Since $ \Lambda^F_y = \{ g_y(Y,A^*,X): \E \{ g_{y}\mid A^*, X \} = 0 \}$ and we have already established that $\E(Q\mid A^*, X) - \E(\text{IF}^F\mid A^*, X) = 0$, choosing $ g_{y}(Y, A^*, X) = \E(Q\mid Y, A^*, X) - \E(\text{IF}^F\mid Y, A^*, X) $ yields $\E(Q\mid Y, A^*, X) - \E(\text{IF}^F\mid Y, A^*, X) = 0 $ a.s.
By the same argument applied to the $g_y$ term, we further obtain $\E(Q\mid A, A^*, X) - \E(\text{IF}^F\mid A, A^*, X) = 0 $ and $\E(Q\mid Z, A^*, X) - \E(\text{IF}^F\mid Z, A^*, X) = 0 $ a.s.
To sum up, we have that:
\begin{align*}
\begin{split}
\E\left( Q\mid Y,A^*,X\right)  &=\E\left( \text{IF}^{F}\mid Y,A^*,X\right), \\
\E\left( Q\mid A,A^*,X\right)  &=\E\left( \text{IF}^{F}\mid A,A^*,X\right), \\
\E\left( Q\mid Z,A^*,X\right)  &=\E\left( \text{IF}^{F}\mid Z,A^*,X\right),
\end{split}
\end{align*}
which is \eqref{main-eq:if_condition} in the main text.
Conversely, suppose there exists a function $Q = q(Y, A, Z, X)$ that satisfies \eqref{main-eq:if_condition}, then it must also hold that $\E\left( Q\mid A^*,X\right)  =\E\left( \text{IF}^{F}\mid A^*,X\right)$ and $\E\left( Q\mid X\right)  =\E\left( \text{IF}^{F}\mid X\right)$. Then,
{\footnotesize
\begin{align*}
\nabla _{t}\psi _{a^*,t}&=\E\left[ \text{IF}^F\left\{ g_{y}\left( Y,A^*,X\right) +g_{a}\left(
A,A^*,X\right) +g_{z}\left( Z,A^*,X\right) +g_{a^*}\left(
A^*,X\right) +g_{x}\left( X\right) \right\} \right] \nonumber \\
&=\E\left\{ \left(
\begin{array}{c}
\E\left( \text{IF}^F\mid Y,A^*,X\right) g_{y}\left( Y,A^*,X\right) +\E
\left( \text{IF}^F\mid A,A^*,X\right) g_{a}\left( A,A^*,X\right)  \\
+\E\left( \text{IF}^F\mid Z,A^*,X\right) g_{z}\left( Z,A^*,X\right) +\E
\left( \text{IF}^F\mid A^*,X\right) g_{a^*}\left( A^*,X\right) +\E
\left( \text{IF}^F\mid X\right) g_{x}\left( X\right)
\end{array}
\right) \right\} \\
&=\E\left\{ \left(
\begin{array}{c}
\E\left( Q\mid Y,A^*,X\right) g_{y}\left( Y,A^*,X\right) +\E\left(
Q\mid A,A^*,X\right) g_{a}\left( A,A^*,X\right)  \\
+\E\left( Q\mid Z,A^*,X\right) g_{z}\left( Z,A^*,X\right) +\E\left(
Q\mid A^*,X\right) g_{a^*}\left( A^*,X\right) +\E\left( Q\mid X
\right) g_{x}\left( X\right)
\end{array}
\right) \right\} \\
&=\E\left( QS^{F}\right) \\
&=\E\left\{ Q \E\left( S^{F}\mid Y,A,Z,X\right)\right\}
\end{align*}}
which is exactly the defining property of an observed-data influence function. This completes the proof.
\section{Proof of Theorem \ref{main-thm:eif}}
\subsection{Closed-form expression of an influence function}
By \eqref{main-eq:if_condition}, we want to find a function $Q=q(Y,X,A,Z)$ such that
\begin{align*}
\E(Q-\text{IF}^F\mid Y,A^*,X) = \E(Q-\text{IF}^F\mid A,A^*,X) = \E(Q-\text{IF}^F\mid Z,A^*,X) =0,
\end{align*}
that is, $Q - \text{IF}^F \in \Lambda^{F,\perp}$ by \eqref{main-eq:tangentspace_full}. Then by the equivalence of \eqref{main-eq:tangentspace_full} and \eqref{main-eq:tangent_space}, we want to find an $W\in\Lambda^{F,\perp}$ such that $Q = \text{IF}^F + W$ is a function of $Y,X,A,Z$, where $W$ has the form in \eqref{main-eq:tangent_space}.
Let
\begin{align*}
G_A(A,X)&=\frac{A - \E(A\mid A^*=1-a^*,X)}
        {\E(A\mid A^*=a^*,X) - \E(A\mid A^*=1-a^*,X)},\\
G_Z(Z,X)&=\frac{Z - \E(Z\mid A^*=1-a^*,X)}
        {\E(Z\mid A^*=a^*,X) - \E(Z\mid A^*=1-a^*,X)},\\
\pi(X)&=\Pr(A^*=a^*\mid X).
\end{align*}
Let $W = M-\E\left( M\mid Y,A^{\ast },X\right) -\E\left(
M\mid A,A^{\ast },X\right) -\E\left( M\mid Z,A^{\ast },X\right) +2\E\left( M\mid A^{\ast
},X\right)$, where
{\small
\begin{align}
\label{eq:candidate_q}
\begin{split}
M = G_A(A,X)G_Z(Z,X)\frac{1}{\pi(X)}
\{Y-\E(Y\mid a^*,X)\}
+ \E(Y\mid a^*,X) - \psi_{a^*}.
\end{split}
\end{align}}
Note that by rearranging
\begin{align*}
\E(A\mid A^*,X) = \mathbb{I}(A^*=a^*)\E(A\mid A^*=a^*,X) + \{1-\mathbb{I}(A^*=a^*)\}\E(A\mid A^*=1-a^*,X),
\end{align*}
we get
\begin{align*}
\mathbb{I}(A^*=a^*) = \frac{\E(A\mid A^*,X) - \E(A\mid A^*=1-a^*,X)}{\E(A\mid A^*=a^*,X) - \E(A\mid A^*=1-a^*,X)}.
\end{align*}
Similarly,
\begin{align*}
\mathbb{I}(A^*=a^*) = \frac{\E(Z\mid A^*,X) - \E(Z\mid A^*=1-a^*,X)}{\E(Z\mid A^*=a^*,X) - \E(Z\mid A^*=1-a^*,X)}.
\end{align*}
Therefore,
{\footnotesize
\begin{align*}
&\E\{G_A(A,X)G_Z(Z,X)\mid Y,A^*,X\} \\
&= \E\{G_A(A,X)\mid A^*,X\}\E\{G_Z(Z,X)\mid A^*,X\}\\
&=\mathbb{I}_{a^*}(A^*)\mathbb{I}_{a^*}(A^*)\\
&=\mathbb{I}_{a^*}(A^*).
\end{align*}}
We can check that
\begingroup
\allowdisplaybreaks
{\footnotesize
\begin{align*}
\E(M\mid Y,A^*,X)
&= \frac{\mathbb{I}_{a^*}(A^*)}{\pi(X)}\{Y-\E(Y\mid a^*,X)\}
 + \E(Y\mid a^*,X) - \psi_{a^*} = \text{EIF}^F, \\
\E(M\mid A,A^*,X)
&= G_A(A,X)\mathbb{I}_{a^*}(A^*) \frac{1}{\pi(X)}
\{\E(Y\mid A^*,X)-\E(Y\mid a^*,X)\} \\
&\qquad \quad  + \E(Y\mid a^*,X) - \psi_{a^*}\\
&= \E(Y\mid a^*,X) - \psi_{a^*}, \\
\E(M\mid Z,A^*,X)
&= \mathbb{I}_{a^*}(A^*)G_Z(Z,X)\frac{1}{\pi(X)}
\{\E(Y\mid A^*, X)-\E(Y\mid a^*,X)\} \\
&\qquad \quad  + \E(Y\mid a^*,X) - \psi_{a^*}\\
&= \E(Y\mid a^*,X) - \psi_{a^*},\\
\E(M\mid A^*,X)
&=  \mathbb{I}_{a^*}(A^*)\mathbb{I}_{a^*}(A^*)\frac{1}{\pi(X)}
\{\E(Y\mid A^*,X)-\E(Y\mid a^*,X)\} \\
&\qquad \quad + \E(Y\mid a^*,X) - \psi_{a^*} \\
&= \E(Y\mid a^*,X) - \psi_{a^*}.
\end{align*}}
\endgroup
Therefore,
\begin{align*}
Q=&W+ \text{EIF}^F\\
=&M-\E\left( M\mid Y,A^{\ast },X\right) -\E\left(
M\mid A,A^{\ast },X\right) -\E\left( M\mid Z,A^{\ast },X\right) +2\E\left( M\mid A^{\ast
},X\right) + \text{EIF}^F\\
=& M - \E(M\mid Y,A^*,X) + \text{EIF}^F \\
=& M
\end{align*}
is a function of $Y,A,Z,X$. Hence, $M$ constructed in \eqref{eq:candidate_q} is an influence function in the observed data model.
Now we show the triple robustness of \eqref{eq:candidate_q}. Let $P^*$ be a distribution which could be different from the true underlying distribution $P$, and use $\E_*(\cdot)$ and $\Pr_*(\cdot)$ to denote expectation and probability with respect to $P^*$.
We denote
\begin{align*}
g(A,X)
&= \frac{A - \E_*(A\mid A^*=1-a^*,X)}
        {\E_*(A\mid A^*=a^*,X) - \E_*(A\mid A^*=1-a^*,X)},  \\
h(Z,X)
&= \frac{Z - \E_*(Z\mid A^*=1-a^*,X)}
        {\E_*(Z\mid A^*=a^*,X) - \E_*(Z\mid A^*=1-a^*,X)},
\end{align*}
and let
$$
\widehat{M} = g(A,X)h(Z,X)\frac{1}{\Pr_*(A^*=a^*\mid X)}\{Y-\E_*(Y\mid a^*,X)\} + \E_*(Y\mid a^*,X) - \psi_{a^*}.
$$
Denote $\widetilde{h}(A^*,X)=\E\{h(Z,X)\mid A^*,X\}$, $\widetilde{g}(A^*,X)=\E\{g(A,X)\mid A^*,X\}$, $\pi_*(X)=\Pr_*(A^*=a^*\mid X)$, and $\Delta_*(X)=\E(Y\mid a^*,X)-\E_*(Y\mid a^*,X)$.
If $\E_*(A\mid A^*,X)$ is correct, then $\widetilde{g}(A^*,X)=\mathbb{I}_{a^*}(A^*)$, and
\begingroup
\small
\allowdisplaybreaks
\begin{align*}
\E(\widehat{M}\mid A^*,X)
&= \widetilde{h}(A^*,X)\frac{\mathbb{I}_{a^*}\left( A^*\right) }{\pi_*(X)}
\left\{ \E(Y\mid A^*,X)-\E_*\left( Y\mid a^*,X\right) \right\}
+\E_*\left( Y\mid a^*,X\right) -\psi _{a^*}\\
&= \widetilde{h}(A^*,X)\frac{\mathbb{I}_{a^*}(A^*)}{\pi_*(X)}\E(Y\mid A^*,X) \\
& \qquad - \widetilde{h}(A^*,X)\frac{\mathbb{I}_{a^*}( A^*) - \pi_*(X)}{\pi_*(X)}
\E_*(Y\mid a^*,X)\\
& \qquad +\left\{1-\widetilde{h}(A^*,X)\right\}\E_*(Y\mid a^*,X) - \psi_{a^*}\\
&= \widetilde{h}(A^*,X)
\left\{\E(Y\mid a^*,X)
 + \frac{\mathbb{I}_{a^*}( A^*) - \pi_*(X)}{\pi_*(X)}\E(Y\mid a^*,X) \right\} \\
&\qquad - \widetilde{h}(A^*,X)
\frac{\mathbb{I}_{a^*}( A^*) - \pi_*(X)}{\pi_*(X)}
\E_*(Y\mid a^*,X) \\
& \qquad +\left\{1-\widetilde{h}(A^*,X)\right\}\E_*(Y\mid a^*,X) - \psi_{a^*}\\
&= \widetilde{h}(A^*,X) \E(Y\mid a^*,X)
 + \widetilde{h}(A^*,X)\frac{\mathbb{I}_{a^*}( A^*) - \pi_*(X)}{\pi_*(X)}\Delta_*(X)\\
& \qquad +\left\{1-\widetilde{h}(A^*,X)\right\}\E_*(Y\mid a^*,X) - \psi_{a^*} \\
&= \E(Y\mid a^*,X)
 + \widetilde{h}(A^*,X)\frac{\mathbb{I}_{a^*}( A^*) - \pi_*(X)}{\pi_*(X)}\Delta_*(X) \\
& \qquad - \left\{1-\widetilde{h}(A^*,X)\right\}\{\E(Y\mid a^*,X)- \E_*(Y\mid a^*,X)\} - \psi_{a^*} \\
&= \E(Y\mid a^*,X)
 + \frac{\widetilde{h}(A^*,X)\mathbb{I}_{a^*}(A^*) - \pi_*(X)}{\pi_*(X)}\Delta_*(X)
 - \psi_{a^*}.
\end{align*}
\endgroup
Similarly, when $\E(Z\mid A^*,X)$ is correctly specified,
{\small
\begin{align*}
\E(M\mid A^*,X)
&= \E(Y\mid a^*,X)
 + \frac{\widetilde{g}(A^*,X)\mathbb{I}_{a^*}(A^*) - \pi_*(X)}{\pi_*(X)}
\Delta_*(X)
 - \psi_{a^*}.
\end{align*}
}
One can check that $\E(M) = \E\{\E(M\mid A^*,X)\}=0$ if one of the following holds:
\begin{enumerate}
    \item $\E(A\mid A^*, X)$ and $\E(Y\mid A^*, X)$ are correct;
    \item $\E(Z\mid A^*, X)$ and $\E(Y\mid A^*, X)$ are correct;
    \item $\E(A\mid A^*, X)$, $\E(Z\mid A^*, X)$ and $\E(A^*\mid X)$ are correct.
\end{enumerate}
\subsection{Closed-form expression for \texorpdfstring{$\{w^{opt}_j,j=1,2\}$}{w opt, j=1,2}}\label{sec:closed_form_w}
\textbf{Categorical $Y$ case}:\\
We already know that $P_1$ has dimension $k-2$. We now aim to find $k-2$ functions $p_1, \ldots,p_{k-2}\in P_1$ that are linearly independent, so that they can serve as a basis of $P_1$. Here we implicitly assume that $\Pr(Y\mid A^*=1,X)\ne \Pr(Y\mid A^*=0,X)$.
For compact notation in this subsection, write $p_a(j,X)=\Pr(Y=j\mid A^*=a,X)$.
Note that
\begin{align*}
v_1(Y,X) &= \sum_{j=1}^k\mathbb{I}(Y=j)v_1(j,X) \\
&= v_1(k,X) + \sum_{j=1}^{k-1}\mathbb{I}(Y=j)\{v_1(j,X) - v_1(k,X)\}, \\
\E\{v_1(Y,X)\mid A^*=0,X\} &= \sum_{j=1}^k\Pr(Y=j\mid A^*=0,X)v_1(j,X) \\
&= v_1(k,X) + \sum_{j=1}^{k-1}\Pr(Y=j\mid A^*=0,X)\{v_1(j,X) - v_1(k,X)\},
\end{align*}
so we have
\begin{align*}
&\frac{v_1(Y,X)-\E\{v_1(Y,X)\mid A^*=0,X\}}
        {\E\{v_1(Y,X)\mid A^*=1,X\}-\E\{v_1(Y,X)\mid A^*=0,X\}} \\
&= \frac{\sum_{j=1}^{k-1}\{\mathbb{I}(Y=j)-p_0(j,X)\}\{v_1(j,X)- v_1(k,X)\}}
        {\sum_{j=1}^{k-1}\{p_1(j,X)-p_0(j,X)\}\{v_1(j,X)- v_1(k,X)\}}.
\end{align*}
Similarly, we have
\begin{align*}
&\frac{v_0(Y,X)-\E\{v_0(Y,X)\mid A^*=1,X\}}
        {\E\{v_0(Y,X)\mid A^*=0,X\}-\E\{v_0(Y,X)\mid A^*=1,X\}} \\
&= \frac{\sum_{j=1}^{k-1}\{\mathbb{I}(Y=j)-p_1(j,X)\}\{v_0(j,X)- v_0(k,X)\}}
        {\sum_{j=1}^{k-1}\{p_0(j,X)-p_1(j,X)\}\{v_0(j,X)- v_0(k,X)\}}.
\end{align*}
Let $v_1(j,X) = v_1(k,X)$ for $j=2,\ldots,k-1$, then
\begin{align*}
&\frac{v_1(Y,X)-\E\{v_1(Y,X)\mid A^*=0,X\}}
        {\E\{v_1(Y,X)\mid A^*=1,X\}-\E\{v_1(Y,X)\mid A^*=0,X\}} \\
&= \frac{\mathbb{I}(Y=1)-p_0(1,X)}{p_1(1,X)-p_0(1,X)}.
\end{align*}
Let $v_0^{(t)}(j,X) = v_0^{(t)}(k,X)$ for $j\in[k-1]\setminus\{t \}$, where $t=2,\ldots,k-1$, we have
\begin{align*}
&\frac{v_0^{(t)}(Y,X)-\E\{v_0^{(t)}(Y,X)\mid A^*=1,X\}}
        {\E\{v_0^{(t)}(Y,X)\mid A^*=0,X\}-\E\{v_0^{(t)}(Y,X)\mid A^*=1,X\}} \\
&= \frac{\sum_{j=1}^{k-1}\{\mathbb{I}(Y=j)-p_1(j,X)\}\{v_0^{(t)}(j,X)- v_0^{(t)}(k,X)\}}
        {\sum_{j=1}^{k-1}\{p_0(j,X)-p_1(j,X)\}\{v_0^{(t)}(j,X)- v_0^{(t)}(k,X)\}}  \\
&= \frac{\mathbb{I}(Y=t)-p_1(t,X)}{p_0(t,X)-p_1(t,X)}.
\end{align*}
Then for $t=2,\ldots,k-1$,
{\small
\begin{align*}
& p_{t-1}(Y,X)\\
:=&\frac{v_1(Y,X)-\E\{v_1(Y,X)\mid A^*=0,X\}}{\E\{v_1(Y,X)\mid A^*=1,X\}-\E\{v_1(Y,X)\mid A^*=0,X\}} \\
&\qquad + \frac{v_0^{(t)}(Y,X)-\E\{v_0^{(t)}(Y,X)\mid A^*=1,X\}}{\E\{v_0^{(t)}(Y,X)\mid A^*=0,X\}-\E\{v_0^{(t)}(Y,X)\mid A^*=1,X\}} -1 \\
&= \frac{\mathbb{I}(Y=1)-p_0(1,X)}{p_1(1,X)-p_0(1,X)}
 + \frac{\mathbb{I}(Y=t)-p_1(t,X)}{p_0(t,X)-p_1(t,X)} - 1.
\end{align*}}
We now check that $p_1,\ldots,p_{k-2}$ are linearly independent. To see this, let
{\small
\begin{align*}
\sum_{t=2}^{k-1}c_t\left\{
\frac{\mathbb{I}(Y=1)-p_1(1,X)}{p_1(1,X)-p_0(1,X)}
+ \frac{\mathbb{I}(Y=t)-p_1(t,X)}{p_0(t,X)-p_1(t,X)}
\right\}=0.
\end{align*}}
When $Y=1$, we have
{\small
\begin{align}\label{eq:linear_ind1}
\sum_{t=2}^{k-1}c_t\left\{
\frac{1-p_1(1,X)}{p_1(1,X)-p_0(1,X)}
- \frac{p_1(t,X)}{p_0(t,X)-p_1(t,X)}
\right\}=0.
\end{align}}
When $Y=j$, $j=2,\ldots,k-1$,
{\small
\begin{align}\label{eq:linear_ind2}
\begin{split}
& \sum_{t=2}^{k-1}c_t\left\{
- \frac{p_1(1,X)}{p_1(1,X)-p_0(1,X)}
- \frac{p_1(t,X)}{p_0(t,X)-p_1(t,X)}
\right\} \\
&\qquad\qquad\qquad + \frac{c_j}{p_0(j,X)-p_1(j,X)}=0.
\end{split}
\end{align}}
Subtracting \eqref{eq:linear_ind2} from \eqref{eq:linear_ind1} gives
{\small
\begin{align*}
\sum_{t=2}^{k-1}\frac{c_t}{p_1(1,X)-p_0(1,X)}
&= \frac{c_j}{p_0(j,X)-p_1(j,X)}, \qquad j=2,\ldots,k-1.
\end{align*}}
Denote the LHS as $m$, we have
\begin{align*}
c_j = m\{p_0(j,X)-p_1(j,X)\},\quad j=2,\ldots,k-1.
\end{align*}
Therefore,
\begin{align*}
\sum_{t=2}^{k-1} m\frac{p_0(t,X)-p_1(t,X)}{p_1(1,X)-p_0(1,X)} = m.
\end{align*}
The above equation holds only when $m=0$, i.e., $c_t=0$ for $t=2,\ldots,k-1$. Therefore, $p_1,\ldots,p_{k-2}$ form a basis for $P_1$.
Let $B_0=\{A-\E(A\mid A^*=0,X)\}\{Z-\E(Z\mid A^*=1,X)\}$ and $B_1=\{A-\E(A\mid A^*=1,X)\}\{Z-\E(Z\mid A^*=0,X)\}$.
By \eqref{main-eq:ortho} in Theorem \ref{main-thm:ortho_tangent_space}, $\Lambda^\bot$ is therefore spanned by
{\footnotesize
\begin{align*}
V = \Bigl(B_0p_1,\ldots,B_0p_{k-2},B_1p_1,\ldots,B_1p_{k-2}\Bigr)^\T.
\end{align*}}
Therefore,
\begin{align*}
w^{opt}_1 &= [\E(\phi_{a^*} V^\T) \{\E(V V^\T)\}^{-1}]_{1:(k-2)}(p_1,\ldots,p_{k-2})^\T,\\
w^{opt}_2 &= [\E(\phi_{a^*} V^\T) \{\E(V V^\T)\}^{-1}]_{(k-1):(2k-4)}(p_1,\ldots,p_{k-2})^\T.
\end{align*}
The efficient influence function $\phi_{a^*}^\text{eff}$ is given by the orthogonal projection of  $\phi_{a^*}$ onto the observed data tangent space of the hidden treatment model, i.e.,
\begin{align*}
\begin{split}
\phi_\text{eff}&=\Pi(\phi_{a^*}\mid \Lambda) \\
&= \phi_{a^*} - \Pi(\phi_{a^*}\mid \Lambda^\bot) \\
&= \phi_{a^*} - \E(\phi_{a^*} V^\T) \{\E(V V^\T)\}^{-1} V.
\end{split}
\end{align*}
An estimator of the efficient influence function is
\begin{align}\label{eq:est_eif}
\widehat{\phi}_\text{eff} = \widehat{\phi}_{a^*} - \widehat{\E}(\widehat{\phi}_{a^*} \widehat{V}^\T) \{\E(\widehat{V} \widehat{V}^\T)\}^{-1} \widehat{V}.
\end{align}
The asymptotic efficiency bound is $\E\left( [\phi_{a^*} - \E(\phi_{a^*} V^\T) \{\E(V V^\T)\}^{-1} V]^2 \right)$.
\vspace{5mm}
\noindent\textbf{Continuous $Y$ case:}\\
When $Y$ is continuous, $P_1$ is an infinite dimensional space. We assume that $|Y|<L$ where $L$ is a constant. For a positive integer $K$, we let $I^K=\{ I_1^K,\ldots,I_K^K \}$ be a partition of the interval $(-L,L]$, where $I_j^K = (-L+2(j-1)L/K, -L+2jL/K], j=1,\ldots,K$.
Write $p^K_a(j,X)=\Pr(Y\in I_j^K\mid A^*=a,X)$.
Let $Q^K$ be a subspace of $L^2(Y,X)$ consisting of all functions satisfying
\begin{align*}
b^K(Y,X) = \sum_{j=1}^K \mathbb{I}(Y\in I_j^K) d_j(X),
\end{align*}
where $d_j(X)\in L^2(X)$.
Analogous to the categorical $Y$ case,
\begin{align*}
P_1^K := \{ h(Y,X): \E\{ h(Y,X)\mid A^*,X \}=0\}\cap Q^K
\end{align*}
is $K-2$ dimensional, and it is equivalent to
{\small
\begin{align*}
P_2^K := \Biggl\{v(Y,X):\
v(Y,X)
&= \frac{v_1(Y,X)-\E\{v_1(Y,X)\mid A^*=0,X\}}
        {\E\{v_1(Y,X)\mid A^*=1,X\}-\E\{v_1(Y,X)\mid A^*=0,X\}} \\
&\quad +  \frac{v_0(Y,X)-\E\{v_0(Y,X)\mid A^*=1,X\}}
        {\E\{v_0(Y,X)\mid A^*=0,X\}-\E\{v_0(Y,X)\mid A^*=1,X\}} - 1,\\
&\hspace{42mm} v_1(Y,X),v_0(Y,X)\in Q^K \Biggr\}.
\end{align*}}
We let $v_1^K(Y,X) = v^K_{1,1}(X)$ for $Y\in I_1^K$, and $v_1^K(Y,X) = v^K_{1,0}(X)$ for $Y\in I_2^K\cup\cdots\cup I_K^K$. One can see $v_1^K(Y,X)\in Q^K$. Then
\begin{align*}
&\frac{v_1^K(Y,X)-\E\{v_1^K(Y,X)\mid A^*=0,X\}}
        {\E\{v_1^K(Y,X)\mid A^*=1,X\}-\E\{v_1^K(Y,X)\mid A^*=0,X\}} \\
&= \frac{\sum_{j=1}^{K-1}\{\mathbb{I}(Y\in I_j^K)-p^K_0(j,X)\}\{v_1^K(j,X)- v_1^K(K,X)\}}
        {\sum_{j=1}^{K-1}\{p^K_1(j,X)-p^K_0(j,X)\}\{v_1^K(j,X)- v_1^K(K,X)\}} \\
&= \frac{\mathbb{I}(Y\in I_1^K)-p^K_0(1,X)}{p^K_1(1,X)-p^K_0(1,X)}.
\end{align*}
Let $v_0^{K,(t)}(Y,X) = v_{0,1}^{K,(t)}(X)$ for $Y\in I^K_t$, and $v_0^{K,(t)}(Y,X) = v_{0,0}^{K,(t)}(X)$ for $Y\in(-L,L]\setminus I^K_t$, where $t=2,\ldots,K-1$. One can check that $v_0^{K,(t)}(Y,X)\in Q^K$, and we have
\begin{align*}
&\frac{v_0^{K,(t)}(Y,X)-\E\{v_0^{K,(t)}(Y,X)\mid A^*=1,X\}}
        {\E\{v_0^{K,(t)}(Y,X)\mid A^*=0,X\}-\E\{v_0^{K,(t)}(Y,X)\mid A^*=1,X\}} \\
&= \frac{\sum_{j=1}^{K-1}\{\mathbb{I}(Y\in I_j^K)-p^K_1(j,X)\}\{v_0^{K,(t)}(j,X)- v_0^{K,(t)}(K,X)\}}
        {\sum_{j=1}^{K-1}\{p^K_0(j,X)-p^K_1(j,X)\}\{v_0^{K,(t)}(j,X)- v_0^{K,(t)}(K,X)\}}  \\
&= \frac{\mathbb{I}(Y\in I_t^K)-p^K_1(t,X)}{p^K_0(t,X)-p^K_1(t,X)}.
\end{align*}
Therefore, for $t=2,\ldots,K-1$,
{\small
\begin{align*}
&p^K_{t-1}(Y,X)\\
:=&\frac{v_1^K(Y,X)-\E\{v_1^K(Y,X)\mid A^*=0,X\}}{\E\{v_1^K(Y,X)\mid A^*=1,X\}-\E\{v_1^K(Y,X)\mid A^*=0,X\}} \\
&\qquad + \frac{v_0^{K,(t)}(Y,X)-\E\{v_0^{K,(t)}(Y,X)\mid A^*=1,X\}}{\E\{v_0^{K,(t)}(Y,X)\mid A^*=0,X\}-\E\{v_0^{K,(t)}(Y,X)\mid A^*=1,X\}} -1 \\
&= \frac{\mathbb{I}(Y\in I^K_1)-p^K_0(1,X)}{p^K_1(1,X)-p^K_0(1,X)}
 + \frac{\mathbb{I}(Y\in I^K_t)-p^K_1(t,X)}{p^K_0(t,X)-p^K_1(t,X)} - 1.
\end{align*}}
One can show that $p^K_1,\ldots,p^K_{K-2}$ are linearly independent using a similar argument as in the categorical case. Therefore, $p^K_1,\ldots,p^K_{K-2}$ is a basis system for $P^K_1$.
Note that $\lim_{K\to\infty}P_1^K = P_1$, hence $p^K_1,\ldots,p^K_{K-2},\ldots$ is a dense basis system for $P_1$ as $K\to\infty$. Let
{\footnotesize
\begin{align*}
V^K = \Bigl(B_0p_1^K, B_1p_1^K,\ldots,
B_0p_{K-2}^K, B_1p_{K-2}^K\Bigr)^\T,
\end{align*}}
then $V^K$ is a basis system of functions dense in $\Lambda^\bot$ as $K\to \infty$.
The efficient influence function $\phi_{a^*}^\text{eff}$ is given by the orthogonal projection of  $\phi_{a^*}$ onto the observed data tangent space of the hidden treatment model, i.e.,
\begin{align*}
\Pi(\phi_{a^*}\mid \Lambda) &= \phi_{a^*} - \Pi(\phi_{a^*}\mid \Lambda^\bot) \\
&= \phi_{a^*} - \lim_{K\to\infty}\E\{\phi_{a^*} (V^K)^\T\} \{\E(V^K (V^K)^\T)\}^{-1} V^K.
\end{align*}
As this projection may not be available in closed form , we propose to use a finite-$K$ approximation of the efficient score \begin{align*}
\Pi(\phi_{a^*}\mid \Lambda) &= \phi_{a^*} - \Pi(\phi_{a^*}\mid \Lambda^\bot) \\
&= \phi_{a^*} - \E\{\phi_{a^*} (V^K)^\T\} \{\E(V^K (V^K)^\T)\}^{-1} V^K,
\end{align*}
with approximate efficiency bound given by
$$\E\left( \left[\phi_{a^*} - \E\{\phi_{a^*} (V^K)^\T\} \{\E(V^K (V^K)^\T)\}^{-1} V^K \right]^2 \right).$$
In this case, $\{w^{opt}_j:j=1,2\}$ are approximated by
\begin{align*}
w^{opt,K}_1 &= \left[\E(\phi_{a^*} (V^K)^\T) \{\E(V^K (V^K)^\T)\}^{-1}\right]_{1,3,\ldots,2K-5}(p_1,\ldots,p_{K-2})^\T,\\
w^{opt,K}_2 &= \left[\E(\phi_{a^*} (V^K)^\T) \{\E(V^K (V^K)^\T)\}^{-1}\right]_{2,4,\ldots,2K-4}(p_1,\ldots,p_{K-2})^\T.
\end{align*}
The above procedure can be viewed as discretizing $Y$ and then applying the categorical approach. As the discretization becomes sufficiently fine, the approach recovers the efficiency bound asymptotically.
\section{Relationship between an oracle IF and the corresponding observed data IF}\label{sec:relationship_if}
To see that the conditional mean of \eqref{main-eq:substitute} given $A^*$ and $X$ matches the indicator function $\mathbb{I}_{a^*}(A^*)$:
\begin{align*}
&\E\{G_A(A,X)G_Z(Z,X)\mid A^*,X\} \\
&= \E\{G_A(A,X)\mid A^*,X\}\E\{G_Z(Z,X)\mid A^*,X\}\\
&= \mathbb{I}_{a^*}(A^*),
\end{align*}
where we have used the fact that $(A,Y,Z)$ are mutually independent given $A^*,X$ as well as the fact that
$$G_A(A,X)$$ and $$G_Z(Z,X)$$ are conditionally unbiased for $\mathbb{I}_{a^*}(A^*)$.
\section{Moment equation based estimator}\label{sec:me_estimator}
We denote the nuisance parameters by $
\eta_1(X) = \E(A\mid 1-a^*,X),
\eta_2(X) = \E(A\mid a^*,X) - \E(A\mid 1-a^*,X),
\eta_3(X) = \E(Z\mid 1-a^*,X),
\eta_4(X) = \E(Z\mid a^*,X) - \E(Z\mid 1-a^*,X),
\eta_5(X) = 1/\Pr(a^*\mid X),
\eta_6(X) = \E(Y\mid a^*,X)
$. If we perturb any of $\eta_j,j=1,\ldots,6$ in  \eqref{main-eq:IF_observed}, the expectation of the perturbed EIF is still zero by Theorem \ref{main-thm:eif}. As an example, we let
\begin{align*}
\left.\frac{\partial \phi_{a^*}(\eta_1+t\delta_1,\cdot)}{\partial t}\right|_{t=0}= \frac{-\delta_1(X)}{\E(A\mid a^*,X)-\E(A\mid 1-a^*,X)}\frac{Z-\E(Z\mid 1-a^*,X)}{\E(Z\mid a^*,X)-\E(Z\mid 1-a^*,X)}\\
\frac{1}{\Pr(a^*\mid X)}\{Y-\E(Y\mid a^*,X)\}.
\end{align*}
Then $\E\left\{\left.\frac{\partial \phi_{a^*}(\eta_1+t\delta_1,\cdot)}{\partial t}\right|_{t=0}\right\}=0$ implies
\begin{align} \label{eq:moment1}
    \E\left[\{Z-\E(Z\mid 1-a^*,X)\}\{Y-\E(Y\mid a^*, X)\}\mid X\right]=0.
\end{align}
This moment equation could be used to estimate the nuisance parameters. Other moment equations can be similarly derived from perturbing $\eta_2,\ldots, \eta_6$:
\begin{align}
&\E\left[\{A-\E(A\mid 1-a^*,X)\}\{Y-\E(Y\mid a^*, X)\}\mid X\right]=0, \label{eq:moment2}\\
&\E\left[\{Z-\E(Z\mid 1-a^*,X)\}\{A-\E(A\mid 1-a^*,X)\}\{Y-\E(Y\mid a^*, X)\}\mid X\right]=0, \label{eq:moment3}\\
&\E\left[\{Z-\E(Z\mid 1-a^*,X)\}\{A-\E(A\mid 1-a^*, X)\}\mid X\right] \label{eq:moment4}\\
&\quad=\{\E(A\mid a^*,X)-\E(A\mid 1-a^*,X)\}\{\E(Z\mid a^*,X)-\E(Z\mid 1-a^*,X)\}\Pr(a^*\mid X). \nonumber
\end{align}
Note that we only get 4 moment equations from 6 perturbations. This is because some perturbations reduce to the same equation. However, we could get extra moment equations by perturbing the following two functions:
{\small
\begin{align*}
S_1(Y,A,X)&= \frac{A-\E(A\mid 1-a^*,X)}{\E(A\mid a^*,X)-\E(A\mid 1-a^*,X)}\frac{1}{\Pr(a^*\mid X)}\{Y-\E(Y\mid a^*,X)\} + \E(Y\mid a^*,X) - \psi_{a^*},\\
S_2(Y,Z,X)&= \frac{Z-\E(Z\mid 1-a^*,X)}{\E(Z\mid a^*,X)-\E(Z\mid 1-a^*,X)}\frac{1}{\Pr(a^*\mid X)}\{Y-\E(Y\mid a^*,X)\} + \E(Y\mid a^*,X) - \psi_{a^*}.
\end{align*}}
The functions $S_1(Y,A,X)$ and $S_2(Y,A,X)$ belong to the tangent space, but they are not influence functions. It can be verified that $\E(S_1)=0$ when $\E(A\mid A^*,X)$ and $\E(a^*\mid X)$ are correct, or when $\E(Y\mid a^*,X)$ is correct. By perturbing $\E(Y\mid a^*,X)$, we get
\begin{align}\label{eq:moment5}
\E\{A - \E(A\mid 1-a^*,X)\mid X\} = \{\E(A\mid a^*,X)-\E(A\mid 1-a^*,X)\}\Pr(a^*\mid X).
\end{align}
Similarly by perturbing $\E(Y\mid a^*,X)$ in $S_2(Y,A,X)$, we get
\begin{align}\label{eq:moment6}
\E\{Z - \E(Z\mid 1-a^*,X)\mid X\} = \{\E(Z\mid a^*,X)-\E(Z\mid 1-a^*,X)\}\Pr(a^*\mid X).
\end{align}
\begin{lemma}
    The two sets of solutions to \eqref{eq:moment1}-\eqref{eq:moment6} are given by
\begin{align}\label{eq:solution}
\text{Solution I:}\quad
\E(Z\mid a^*,X)
&=\frac{-F-G\E(Y\mid X)+H\E(Z\mid X)-\sqrt{\Delta}}{2H}, \nonumber\\
\E(Z\mid 1-a^*,X)
&=\frac{-F-G\E(Y\mid X)+H\E(Z\mid X)+\sqrt{\Delta}}{2H}, \nonumber\\
\E(Y\mid a^*,X)
&=\frac{-F-H\E(Z\mid 1-a^*,X)}{G}, \nonumber\\
\E(Y\mid 1-a^*,X)
&=\frac{-F-H\E(Z\mid a^*,X)}{G}, \nonumber\\
\E(A\mid a^*,X)
&=\frac{\E(A\mid X)\E(Y\mid a^*,X)-\E(YA\mid X)}
        {\E(Y\mid a^*,X)-\E(Y\mid X)}, \nonumber\\
\E(A\mid 1-a^*,X)
&=\frac{\E(A\mid X)\E(Y\mid 1-a^*,X)-\E(YA\mid X)}
        {\E(Y\mid 1-a^*,X)-\E(Y\mid X)}, \nonumber\\
\Pr(a^*\mid X)
&=\frac{\E\{Z-\E(Z\mid 1-a^*,X)\mid X\}
        \E\{A-\E(A\mid 1-a^*,X)\mid X\}}
        {\E\left[\{Z-\E(Z\mid 1-a^*,X)\}
        \{A-\E(A\mid 1-a^*, X)\}\mid X\right]}.
\end{align}
or
\begin{align}
\text{Solution II:}\quad
\E(Z\mid a^*,X)
&=\frac{-F-G\E(Y\mid X)+H\E(Z\mid X)+\sqrt{\Delta}}{2H}, \nonumber\\
\E(Z\mid 1-a^*,X)
&=\frac{-F-G\E(Y\mid X)+H\E(Z\mid X)-\sqrt{\Delta}}{2H}, \nonumber\\
\E(Y\mid a^*,X)
&=\frac{-F-H\E(Z\mid 1-a^*,X)}{G}, \nonumber\\
\E(Y\mid 1-a^*,X)
&=\frac{-F-H\E(Z\mid a^*,X)}{G}, \nonumber\\
\E(A\mid a^*,X)
&=\frac{\E(A\mid X)\E(Y\mid a^*,X)-\E(YA\mid X)}
        {\E(Y\mid a^*,X)-\E(Y\mid X)}, \nonumber\\
\E(A\mid 1-a^*,X)
&=\frac{\E(A\mid X)\E(Y\mid 1-a^*,X)-\E(YA\mid X)}
        {\E(Y\mid 1-a^*,X)-\E(Y\mid X)}, \nonumber\\
\Pr(a^*\mid X)
&=\frac{\E\{Z-\E(Z\mid 1-a^*,X)\mid X\}
        \E\{A-\E(A\mid 1-a^*,X)\mid X\}}
        {\E\left[\{Z-\E(Z\mid 1-a^*,X)\}
        \{A-\E(A\mid 1-a^*, X)\}\mid X\right]}.
\end{align}
where
\begin{align*}
F &= \E(AZY\mid X) - \E(ZY\mid X)\E(A\mid X),\\
G &= \E(Z\mid X)\E(A\mid X) - \E(AZ\mid X),\\
H &= \E(Y\mid X)\E(A\mid X) - \E(AY\mid X), \\
\Delta &= \{F+G\E(Y\mid X) - H\E(Z\mid X)\}^2 + 4H\{F\E(Z\mid X)+G\E(ZY\mid X)\}.
\end{align*}
Under Assumption \ref{main-asp:label}, only one of them is the valid solution.
\end{lemma}
We summarize the estimation procedure as follows:
\begin{enumerate}
    \item Estimate  $\E(Y\mid X)$, $\E(A\mid X)$, $\E(Z\mid X)$, $\E(YA\mid X)$, $\E(YZ\mid X)$, $\E(AZ\mid X)$, $\E(YAZ\mid X)$ using nonparametric methods.
    \item Plug the estimates from Step 1 into \eqref{eq:solution} to get $\widehat{\E}(Z\mid a^*,X)$, $\widehat{\E}(Z\mid 1-a^*,X)$, $\widehat{\E}(Y\mid a^*,X)$, $\widehat{\E}(Y\mid 1-a^*,X)$, $\widehat{\E}(A\mid a^*,X)$, $\widehat{\E}(A\mid 1-a^*,X)$, $\widehat{\Pr}(a^*\mid X)$. Use Assumption \ref{main-asp:label} to decide which set of solution to keep.
    \item Plug all the estimated nuisance parameters into \eqref{main-eq:IF_observed} to get $\widehat{\psi}_{a^*}$.
\end{enumerate}
Since we plug $\widehat{\Pr}(Y,A,Z\mid X)$ into \eqref{eq:solution} to get the nuisance parameters, the solutions can be unstable if $\widehat{\Pr}(Y,A,Z\mid X)$ deviates moderately from the true values.
\section{Logit-scale update formulas for the iterative algorithm}
\label{sec:iterative_details}
We provide the componentwise expressions for the logit-scale update in \eqref{main-eq:logit_update}. At iteration \(t\), define
\[
Z_{a^*}^{(t)}=\frac{Z-p_{z,1-a^*}^{(t)}(X)}{p_{z,a^*}^{(t)}(X)-p_{z,1-a^*}^{(t)}(X)},\qquad
A_{a^*}^{(t)}=\frac{A-p_{a,1-a^*}^{(t)}(X)}{p_{a,a^*}^{(t)}(X)-p_{a,1-a^*}^{(t)}(X)},
\]
and
\[
Y_{a^*}^{(t)}=\frac{Y-p_{y,1-a^*}^{(t)}(X)}{p_{y,a^*}^{(t)}(X)-p_{y,1-a^*}^{(t)}(X)},
\qquad a^*=0,1.
\]
For the \(j\)th nuisance component \(\theta_j\) updated on the logit scale, write
\[
\ell_j^{(t)}(X)=\logit\{\theta_j^{(t)}(X)\}.
\]
The logit-scale update takes the form
\[
\ell_j^{(t+1)}=\mathcal P_{k(n),n}\left[\ell_j^{(t)}+\eta L_j^{(t)}\right],
\]
where
\[
L_j^{(t)}=\frac{\rho_j(O,\theta^{(t)})}{\theta_j^{(t)}(X)\{1-\theta_j^{(t)}(X)\}}.
\]
For the six conditional mean components, the update directions are
\begin{align*}
L_{z,1}^{(t)}&=\frac{\{Z-p_{z,1}^{(t)}(X)\}A_1^{(t)}Y_1^{(t)}}{p_{a^*}^{(t)}(X)p_{z,1}^{(t)}(X)\{1-p_{z,1}^{(t)}(X)\}},\\
L_{z,0}^{(t)}&=\frac{\{Z-p_{z,0}^{(t)}(X)\}A_0^{(t)}Y_0^{(t)}}{\{1-p_{a^*}^{(t)}(X)\}p_{z,0}^{(t)}(X)\{1-p_{z,0}^{(t)}(X)\}},\\
L_{a,1}^{(t)}&=\frac{\{A-p_{a,1}^{(t)}(X)\}Z_1^{(t)}Y_1^{(t)}}{p_{a^*}^{(t)}(X)p_{a,1}^{(t)}(X)\{1-p_{a,1}^{(t)}(X)\}},\\
L_{a,0}^{(t)}&=\frac{\{A-p_{a,0}^{(t)}(X)\}Z_0^{(t)}Y_0^{(t)}}{\{1-p_{a^*}^{(t)}(X)\}p_{a,0}^{(t)}(X)\{1-p_{a,0}^{(t)}(X)\}},\\
L_{y,1}^{(t)}&=\frac{\{Y-p_{y,1}^{(t)}(X)\}Z_1^{(t)}A_1^{(t)}}{p_{a^*}^{(t)}(X)p_{y,1}^{(t)}(X)\{1-p_{y,1}^{(t)}(X)\}},\\
L_{y,0}^{(t)}&=\frac{\{Y-p_{y,0}^{(t)}(X)\}Z_0^{(t)}A_0^{(t)}}{\{1-p_{a^*}^{(t)}(X)\}p_{y,0}^{(t)}(X)\{1-p_{y,0}^{(t)}(X)\}}.
\end{align*}
For \(p_{a^*}(X)\), define
\[
R_{a^*}^{(t)}=Z_1^{(t)}A_1^{(t)}Y_1^{(t)}-p_{a^*}^{(t)}(X)-
\frac{Z_1^{(t)}A_1^{(t)}}{p_{y,1}^{(t)}(X)-p_{y,0}^{(t)}(X)}\{Y-p_{y,1}^{(t)}(X)\}
\]
\[
\hspace{2.5cm}-\frac{A_1^{(t)}Y_1^{(t)}}{p_{z,1}^{(t)}(X)-p_{z,0}^{(t)}(X)}\{Z-p_{z,1}^{(t)}(X)\}-\frac{Z_1^{(t)}Y_1^{(t)}}{p_{a,1}^{(t)}(X)-p_{a,0}^{(t)}(X)}\{A-p_{a,1}^{(t)}(X)\}.
\]
Then the logit-scale update direction for \(p_{a^*}(X)\) is
\[
L_{a^*}^{(t)}=\frac{R_{a^*}^{(t)}}{p_{a^*}^{(t)}(X)\{1-p_{a^*}^{(t)}(X)\}}.
\]
\section{Proof of Theorem \ref{main-thm:finite_dim_equi}}
\label{sec:proof_finite_dim_equi}
\begin{proof}
Fix \(K=k(n)\) and the sample \(\{O_i\}_{i=1}^n\). For simplicity, write \(\mathcal P_{K,n}\) for the empirical projection operator, applied componentwise. By the previous discussion, the sieve MD criterion satisfies \(\widehat Q_n^{\mathrm{MD}}(\theta)\ge 0\), with equality if and only if \(\mathcal P_{K,n}[\rho(O,\theta)]=0\) on the sample support of \(X\). Therefore, if the empirical projected moment equation has a unique solution \(\theta^\star\in\mathcal H_K^7\), then this solution is the unique minimizer of \(\widehat Q_n^{\mathrm{MD}}(\theta)\). Hence \(\widehat\theta^{\mathrm{MD}}=\theta^\star\).
It remains to identify the limit of the iterative sequence. Suppose
\(\theta^{(t)}\to\widehat\theta^{\mathrm{it}}\in\mathcal H_K^7\). Taking limits in the iteration
\(
\theta^{(t+1)}=\mathcal P_{K,n}\left[\theta^{(t)}+\eta\rho(O,\theta^{(t)})\right]
\)
is valid under Assumption~\ref{main-asp:separation}, which ensures that the components of \(\rho(O,\theta)\) are continuous in \(\theta\). Thus,
\[
\widehat\theta^{\mathrm{it}}=\mathcal P_{K,n}\left[\widehat\theta^{\mathrm{it}}
+\eta\rho(O,\widehat\theta^{\mathrm{it}})\right].
\]
Since \(\widehat\theta^{\mathrm{it}}\in\mathcal H_K^7\), \(\mathcal P_{K,n}[\widehat\theta^{\mathrm{it}}]=\widehat\theta^{\mathrm{it}}\). By linearity of \(\mathcal P_{K,n}\) and \(\eta>0\), it follows that
\[
\mathcal P_{K,n}[\rho(O,\widehat\theta^{\mathrm{it}})]=0.
\]
Thus \(\widehat\theta^{\mathrm{it}}\) solves the empirical projected moment equation. By uniqueness, \(\widehat\theta^{\mathrm{it}}=\theta^\star\). Combining this with \(\widehat\theta^{\mathrm{MD}}=\theta^\star\) yields \[\widehat\theta^{\mathrm{it}}=\widehat\theta^{\mathrm{MD}}=\theta^\star\].
\end{proof}
\section{Consistency of the sieve MD estimator}
\label{sec:sieve_md_consistency}
\begin{theorem}[Consistency of the sieve MD estimator]
Let \(\widehat\theta^{\rm MD}\) be the sieve minimum distance estimator defined in \eqref{main-eq:sieve_md_estimator}. Suppose Assumption \ref{main-asp:separation} holds. In addition, assume that:
\begin{enumerate}
\item[(i)] The true nuisance vector \(\theta_0\) belongs to the closure of the sieve spaces: there exists \(\Pi_n\theta_0\in \Theta_n\) such that
\(
\|\Pi_n\theta_0-\theta_0\|=o(1).
\)
\item[(ii)] The sieve dimension satisfies \(k(n)\to\infty\) and \(k(n)/n\to0\). Moreover,
\(\Theta_n=\Theta\cap\mathcal H_{k(n)}^7\) is closed and bounded for each \(n\).
\item[(iii)] The series least-squares estimator \(\widehat m(X,\theta)\) satisfies the sample criterion conditions in Assumption 3.3 of \cite{chen2012estimation}, with \(\eta_{0,n}=o(1)\) and \(\bar\delta_{m,n}=o(1)\).
\end{enumerate}
Then
\[
\|\widehat\theta^{\rm MD}-\theta_0\|=o_p(1).
\]
\end{theorem}
\begin{proof}
We verify the conditions of Theorem 3.1 of \cite{chen2012estimation} for the unpenalized sieve MD estimator. Their theorem applies to penalized sieve MD estimators using slowly growing finite-dimensional sieves and allows the unpenalized case \(\lambda_n=0\). In our notation,
\[
\widehat Q_n^{\rm MD}(\theta)=\frac1n\sum_{i=1}^n\|\widehat m(X_i,\theta)\|_2^2,
\]
so the estimator is exactly the penalized sieve MD estimator with identity weighting matrix and no penalty.
First, by construction of the moment equations,
\(
m(X,\theta_0)=0\text{ a.s.}
\)
Moreover, the conditional moment system \(m(X,\theta)=0\) identifies \(\theta_0\) up to labeling.
Hence, for every fixed \(\epsilon>0\), the population criterion separates \(\theta_0\) from sieve elements at distance at least \(\epsilon\). That is, the separation quantity
\[
g(k(n),\epsilon)=\inf_{\substack{\theta\in\Theta_n:
\|\theta-\theta_0\|\ge \epsilon}}\mathbb E\left\{\|m(X,\theta)\|_2^2\right\}
\]
is bounded away from zero for every fixed \(\epsilon>0\). This verifies the
identification/separation condition in Theorem 3.1 of
\citet{chen2012estimation}.
Second, by the sieve approximation assumption,
\(
\|\Pi_n\theta_0-\theta_0\|=o(1).
\)
Since the moment map is continuous at \(\theta_0\) under Assumption \ref{main-asp:separation}, and \(m(\cdot,\theta_0)=0\), it follows that
\[
\E\{\|m(X,\Pi_n\theta_0)\|_2^2\}=\|m(\cdot,\Pi_n\theta_0)-m(\cdot,\theta_0)\|^2=o(1).
\]
This verifies the sieve approximation condition in Assumption 3.1(iv) of
\cite{chen2012estimation}.
Third, since \(\lambda_n=0\), their penalty condition Assumption 3.2(a) holds automatically. No penalty condition is needed.
Fourth, we use the series least-squares estimator of \(m(X,\theta)\), obtained by projecting \(\rho(O,\theta)\) onto the sieve basis. \cite{chen2012estimation} verify that this estimator satisfies their Assumption 3.3 under standard regularity conditions. Therefore, under the same regularity conditions, we have
\[
\frac{1}{n}\sum_{i=1}^n \|\widehat m(X_i,\Pi_n\theta_0)\|_2^2 \le c_0\mathbb \E\{\|m(X,\Pi_n\theta_0)\|_2^2\} +O_p(\eta_{0,n}),
\]
with \(\eta_{0,n}=o(1)\), and
\[
\frac{1}{n}\sum_{i=1}^n \|\widehat m(X_i,\theta)\|_2^2 \ge c\,\mathbb E\{\|m(X,\theta)\|_2^2\} -O_p(\bar\delta_{m,n}^2)
\]
uniformly over the relevant sieve class, with \(\bar\delta_{m,n}=o(1)\).
Combining the preceding displays, for every fixed \(\epsilon>0\),
\[
\max\left\{\eta_{0,n},
\E\{\|m(X,\Pi_n\theta_0)\|_2^2\},\bar\delta_{m,n}^2,\lambda_n\right\}=o\{g(k(n),\epsilon)\}.
\]
Indeed, \(\lambda_n=0\), the first and third terms are \(o(1)\) by the sample criterion condition, the second term is \(o(1)\) by sieve approximation and local Lipschitz continuity of \(m\), and \(g(k(n),\epsilon)\) is bounded away from zero by the global identification argument above.
Theorem 3.1 of \cite{chen2012estimation} therefore implies
\[
\|\widehat\theta^{\rm MD}-\theta_0\|=o_p(1).
\]
This proves the claim.
\end{proof}
\section{Local well-posedness of the conditional moment map}\label{sec:well_posed}
\begin{lemma}\label{lem:well_posed}
Let \(\theta_0\) denote the true nuisance vector. Under Assumption \ref{main-asp:separation},
the conditional moment map \(\theta \mapsto m(\cdot,\theta)\) is Fréchet differentiable at \(\theta_0\) as a map from \(L^2(P_X)^7\) to \(L^2(P_X)^7\).
Moreover, for any admissible perturbation \(\Delta\),
\[
D m(\cdot,\theta_0)[\Delta]=\left.\frac{d}{dt}m(\cdot,\theta_0+t\Delta)\right|_{t=0}=-\Delta .
\]
Consequently, there exist constants \(c>0\) and \(\delta>0\) such that, whenever
\(\|\theta-\theta_0\|_\infty\le \delta\),
\[
\|m(\cdot,\theta)\|\ge c\|\theta-\theta_0\|.
\]
\end{lemma}
\begin{proof}
We first show that the conditional moment map is locally differentiable.
By Assumption \ref{main-asp:separation}, all denominators appearing in \(\rho(O,\theta)\) are
bounded away from zero at \(\theta_0\). Hence, if
\(\|\theta-\theta_0\|_\infty\) is sufficiently small, these denominators
remain uniformly bounded away from zero. On this neighborhood,
\(\rho(O,\theta)\) is a smooth function of the nuisance components.
Therefore, for any admissible perturbation \(\Delta\),
\[
\rho(O,\theta_0+\Delta)=\rho(O,\theta_0)+D\rho(O,\theta_0)[\Delta]+r(O,\Delta),
\]
where, under the boundedness assumptions on the variables entering
\(\rho(O,\theta)\),
\[
\left\|\mathbb E\{r(O,\Delta)\mid X=\cdot\}
\right\|\le C\|\Delta\|_\infty\|\Delta\|=o(\|\Delta\|)
\]
as \(\|\Delta\|_\infty\to0\). Taking conditional expectations gives
\[
m(\cdot,\theta_0+\Delta)=m(\cdot,\theta_0)+\mathbb E\{D\rho(O,\theta_0)[\Delta]\mid X=\cdot\}+o(\|\Delta\|).
\]
Thus \(\theta\mapsto m(\cdot,\theta)\) is Fréchet differentiable at
\(\theta_0\), with derivative
\[
Dm(\cdot,\theta_0)[\Delta]=\mathbb E\{D\rho(O,\theta_0)[\Delta]\mid X=\cdot\}.
\]
We next compute the derivative at \(\theta_0\). For the moment corresponding to
\(p_{z,1}\),
\[
m_{z,1}(X,\theta)=\mathbb E\left[\{Z-p_{z,1}(X)\}A_1Y_1\{p_{a^*}(X)\}^{-1}\mid X\right].
\]
For a perturbation \(\Delta\),
\[
D m_{z,1}(\cdot,\theta_0)[\Delta]=-\Delta_{z,1}(X)\mathbb E\left[A_1Y_1\{p_{a^*}(X)\}^{-1}\mid X\right],
\]
because all remaining first-order terms are multiplied by \(Z-p_{z,1}(X)\), and their corresponding conditional moments vanish at \(\theta_0\). Since
\[
\mathbb E(A_1Y_1\mid X)=p_{a^*}(X),
\]
we obtain
\[
D m_{z,1}(\cdot,\theta_0)[\Delta]=-\Delta_{z,1}(X).
\]
The same calculation yields
\[
D m_{z,0}(\cdot,\theta_0)[\Delta]=-\Delta_{z,0}(X),\qquad
D m_{a,1}(\cdot,\theta_0)[\Delta]=-\Delta_{a,1}(X),
\]
\[
D m_{a,0}(\cdot,\theta_0)[\Delta]=-\Delta_{a,0}(X),\qquad
D m_{y,1}(\cdot,\theta_0)[\Delta]=-\Delta_{y,1}(X),
\]
and
\[
D m_{y,0}(\cdot,\theta_0)[\Delta]=-\Delta_{y,0}(X).
\]
For the adjusted moment corresponding to \(p_{a^*}\), the augmentation terms remove the first-order sensitivity to the other six nuisance components, so
\[
D m_{a^*}(\cdot,\theta_0)[\Delta]=-\Delta_{a^*}(X).
\]
Combining the seven components gives
\[
D m(\cdot,\theta_0)[\Delta]=-\Delta .
\]
Now let \(\Delta=\theta-\theta_0\). By the preceding differentiability result,
\[
m(\cdot,\theta)-m(\cdot,\theta_0)=-\Delta+R(\theta),
\]
where
\[
\|R(\theta)\|=o\bigl(\|\Delta\|\bigr)
\]
as \(\|\theta-\theta_0\|_\infty\to0\). Since \(m(\cdot,\theta_0)=0\) by construction, for sufficiently small \(\delta>0\),
\[
\|R(\theta)\|\le\frac12\|\Delta\|
\]
whenever \(\|\theta-\theta_0\|_\infty\le\delta\). Therefore,
\[
\begin{aligned}
\|m(\cdot,\theta)\|
&=\|-\Delta+R(\theta)\|  \\
&\ge\|\Delta\|-\|R(\theta)\|  \\
&\ge\frac12\|\theta-\theta_0\|.
\end{aligned}
\]
Taking \(c=1/2\) proves the claim.
\end{proof}
\section{Local convergence of the iterative algorithm}
\label{sec:local_convergence}
We study the local convergence of the sample-level iteration
\[
\theta^{(t+1)}=T_n(\theta^{(t)}):=\mathcal P_{k(n),n}\{\theta^{(t)}+\eta\rho(O,\theta^{(t)})\}.
\]
Let
\[
m(X,\theta)=\mathbb E\{\rho(O,\theta)\mid X\}
\]
be the population conditional moment map, and let
\[
T(\theta)=\mathcal P_{k(n)}\{\theta+\eta m(X,\theta)\}
\]
be the corresponding population projected update. We first verify local contraction for \(T\), and then transfer it to the sample update \(T_n\).
Let \(\theta_0\) denote the target nuisance vector, so that
\(m(X,\theta_0)=0\). Under Assumption~\ref{main-asp:separation}, all denominators appearing in
\(\rho(O,\theta)\) are bounded away from zero in a neighborhood of
\(\theta_0\). Hence \(m(X,\theta)\) is continuously differentiable in
\(\theta\) in this neighborhood.
We next compute the derivative of \(m\) at \(\theta_0\). For example, the
moment for \(p_{z,1}\) is
\[
m_{Z,1}(X,\theta)
=
\mathbb E\left[
\{Z-p_{z,1}(X)\}
A_1Y_1\{p_{a^*}(X)\}^{-1}
\mid X
\right].
\]
Let \(\Delta\) be a perturbation of \(\theta\). Differentiating at
\(\theta_0\) gives
\[
Dm_{Z,1}(\theta_0)[\Delta]
=
-\Delta_{z,1}(X)
\mathbb E\left[
A_1Y_1\{p_{a^*}(X)\}^{-1}
\mid X
\right],
\]
because the remaining derivative terms are multiplied by the residual
\(Z-p_{z,1}(X)\), whose relevant conditional moments vanish at \(\theta_0\).
Since, at the target,
\[
\mathbb E(A_1Y_1\mid X)=p_{a^*}(X),
\]
we obtain
\[
Dm_{Z,1}(\theta_0)[\Delta]=-\Delta_{z,1}(X).
\]
The same calculation gives
\[
Dm_{Z,0}(\theta_0)[\Delta]=-\Delta_{z,0}(X),\qquad
Dm_{A,1}(\theta_0)[\Delta]=-\Delta_{a,1}(X),
\]
\[
Dm_{A,0}(\theta_0)[\Delta]=-\Delta_{a,0}(X),\qquad
Dm_{Y,1}(\theta_0)[\Delta]=-\Delta_{y,1}(X),
\]
and
\[
Dm_{Y,0}(\theta_0)[\Delta]=-\Delta_{y,0}(X).
\]
For the adjusted moment for \(p_{a^*}\), the augmentation terms remove the
first-order sensitivity to the six conditional mean components. Therefore,
\[
Dm_{a^*}(\theta_0)[\Delta]=-\Delta_{a^*}(X).
\]
Consequently,
\[
Dm(\cdot,\theta_0)[\Delta]=-\Delta .
\]
It follows that the derivative of the population update at \(\theta_0\) is
\[
DT(\theta_0)
=
\mathcal P_{k(n)}\{I+\eta Dm(\cdot,\theta_0)\}
=
(1-\eta)\mathcal P_{k(n)}.
\]
Since \(\mathcal P_{k(n)}\) is an orthogonal projection,
\(\|\mathcal P_{k(n)}\|_{\mathrm{op}}=1\). Thus
\[
\|DT(\theta_0)\|_{\mathrm{op}}=|1-\eta|.
\]
Hence, for any step size \(0<\eta<2\), the population update is locally
contractive at \(\theta_0\).
By continuity of \(Dm(\cdot,\theta)\), this contraction extends to a
neighborhood of \(\theta_0\): there exist \(\delta>0\) and \(q_0<1\) such that
\[
\sup_{\theta\in B_\delta(\theta_0)}
\|DT(\theta)\|_{\mathrm{op}}\le q_0.
\]
It remains to transfer this contraction to the sample update. We require the
standard uniform convergence of the sample derivative to the population
derivative:
\[
\sup_{\theta\in B_\delta(\theta_0)}
\|DT_n(\theta)-DT(\theta)\|_{\mathrm{op}}=o_p(1).
\]
This condition follows under the usual boundedness and sieve-growth conditions
for the empirical series projection; in particular, Assumption~\ref{main-asp:separation} ensures that
the derivatives of \(\rho(O,\theta)\) are uniformly bounded in the local
neighborhood.
Therefore, with probability tending to one,
\[
\sup_{\theta\in B_\delta(\theta_0)}
\|DT_n(\theta)\|_{\mathrm{op}}\le q
\]
for some \(q<1\). Hence \(T_n\) is a contraction on \(B_\delta(\theta_0)\).
By the contraction mapping theorem, if the sample update has a fixed point
\(\widehat\theta^{\mathrm{it}}\in B_\delta(\theta_0)\), then this fixed point is
unique in the neighborhood, and for any initial value
\(\theta^{(0)}\in B_\delta(\theta_0)\),
\[
\|\theta^{(t)}-\widehat\theta^{\mathrm{it}}\|
\le
q^t\|\theta^{(0)}-\widehat\theta^{\mathrm{it}}\|.
\]
Thus the iteration converges geometrically to
\(\widehat\theta^{\mathrm{it}}\).
Finally, any fixed point of \(T_n\) satisfies
\[
\theta=\mathcal P_{k(n),n}\{\theta+\eta\rho(O,\theta)\}.
\]
Since \(\theta\in\mathcal H_{k(n)}^7\), this implies
\[
\mathcal P_{k(n),n}\{\rho(O,\theta)\}=0.
\]
Therefore the fixed point solves the empirical projected moment equation. If
this equation has a unique solution in the sieve space, Theorem~\ref{main-thm:finite_dim_equi} implies that
\[
\widehat\theta^{\mathrm{it}}=\widehat\theta^{\mathrm{MD}}.
\]
\section{Starting values for the iterative algorithm}
\label{sec:starting_values}
We initialize the iterative algorithm using constant marginal latent-class estimators. The idea is to ignore the dependence on \(X\) in the first step and estimate the marginal quantities
\(
\pi=\Pr(A^*=1),\
\mu_{A,a^*}=\mathbb E(A\mid A^*=a^*),\
\mu_{Z,a^*}=\mathbb E(Z\mid A^*=a^*),\
\)
and
\(
\mu_{Y,a^*}=\mathbb E(Y\mid A^*=a^*),\  a^*=0,1.
\)
These seven quantities are estimated from unconditional moment restrictions implied by the latent-class representation.
Let
\[
d_A=\mu_{A,1}-\mu_{A,0},\qquad
d_Z=\mu_{Z,1}-\mu_{Z,0},\qquad
d_Y=\mu_{Y,1}-\mu_{Y,0}.
\]
For a candidate value
\[
\vartheta=(\mu_{A,1},\mu_{A,0},\mu_{Z,1},\mu_{Z,0},
\mu_{Y,1},\mu_{Y,0},\pi),
\]
we use the following unconditional moments:
\[
\mathbb E\left[
(A-\mu_{A,1})\frac{Z-\mu_{Z,0}}{d_Z}
\frac{Y-\mu_{Y,0}}{d_Y}
\right]=0,
\]
\[
\mathbb E\left[
(A-\mu_{A,0})\frac{Z-\mu_{Z,1}}{-d_Z}
\frac{Y-\mu_{Y,1}}{-d_Y}
\right]=0,
\]
\[
\mathbb E\left[
(Z-\mu_{Z,1})\frac{A-\mu_{A,0}}{d_A}
\frac{Y-\mu_{Y,0}}{d_Y}
\right]=0,
\]
\[
\mathbb E\left[
(Z-\mu_{Z,0})\frac{A-\mu_{A,1}}{-d_A}
\frac{Y-\mu_{Y,1}}{-d_Y}
\right]=0,
\]
\[
\mathbb E\left[
(Y-\mu_{Y,1})\frac{A-\mu_{A,0}}{d_A}
\frac{Z-\mu_{Z,0}}{d_Z}
\right]=0,
\]
\[
\mathbb E\left[
(Y-\mu_{Y,0})\frac{A-\mu_{A,1}}{-d_A}
\frac{Z-\mu_{Z,1}}{-d_Z}
\right]=0,
\]
together with the marginal mean equations
\[
\mathbb E\{A-\mu_{A,0}-d_A\pi\}=0,\qquad
\mathbb E\{Z-\mu_{Z,0}-d_Z\pi\}=0,
\]
and
\[
\mathbb E\{Y-\mu_{Y,0}-d_Y\pi\}=0.
\]
In implementation, the empirical analogues of these nine moments are solved by
GMM for the seven unknowns in \(\vartheta\). This gives an overidentified
marginal system. To respect the parameter constraints, we optimize over the
logit transforms of \(\mu_{A,1},\mu_{A,0},\mu_{Z,1},\mu_{Z,0}\), and \(\pi\);
for binary outcomes, \(\mu_{Y,1}\) and \(\mu_{Y,0}\) are also truncated to lie in
\((0,1)\).
The resulting estimates are then used as constant starting values for all sample
points:
\[
p_{a^*}^{(0)}(X_i)=\widehat\pi,\qquad
p_{a,a^*}^{(0)}(X_i)=\widehat\mu_{A,a^*},\qquad
p_{z,a^*}^{(0)}(X_i)=\widehat\mu_{Z,a^*},
\]
and
\[
p_{y,a^*}^{(0)}(X_i)=\widehat\mu_{Y,a^*},
\qquad a^*=0,1,\quad i=1,\ldots,n.
\]
These values are subsequently transformed to the logit scale when the
logit-scale version of the iterative update is used.
\section{Proof of Theorem \ref{main-thm:consistency}}\label{sec:proof_asymptotic}
Let $\varphi = \sum^2_{j=1}  \mathbb{I}\{d=j\}S_j$ and $\widehat{\varphi} = \sum^2_{j=1}  \mathbb{I}\{\hat{d}=j\}\widehat{S}_j$,
\begin{align*}
\widehat{\psi}_1 - \psi_1 &= \mathbb{P}_n\widehat{\varphi} - \mathbb{P}\varphi \\
&=(\mathbb{P}_n - \mathbb{P})(\widehat{\varphi}-\varphi) + \mathbb{P}(\widehat{\varphi}-\varphi) + (\mathbb{P}_n-\mathbb{P})\varphi\\
&=R_1 + R_2 + (\mathbb{P}_n-\mathbb{P})\varphi,
\end{align*}
where $R_1 = (\mathbb{P}_n - \mathbb{P})(\widehat{\varphi}-\varphi)$, $R_2 = \mathbb{P}(\widehat{\varphi}-\phi)$.
\vspace{5mm}
\noindent \textbf{Term $R_1$}
For \(R_1\), by the assumed \(P\)-Donsker condition, the class containing
\(\widehat\varphi\) and \(\varphi\) is \(P\)-Donsker with probability
approaching one. Since the nuisance estimators are consistent,
\[
\|\widehat\varphi-\varphi\|=o_p(1).
\]
By stochastic equicontinuity of the empirical process,
\[
\sqrt n(\mathbb P_n-\mathbb P)(\widehat\varphi-\varphi)=o_p(1).
\]
Thus \(R_1=o_p(n^{-1/2})\).
\noindent\textbf{Term $R_2$}
Let
{\small
\begin{align*}
\widehat{S}_j =& \frac{A-\widehat{\E}(A\mid 1-a^*_j,X)}{\widehat{\E}(A\mid a^*_j,X)-\widehat{\E}(A\mid 1-a^*_j,X)}\frac{Z-\widehat{\E}(Z\mid 1-a^*_j,X)}{\widehat{\E}(Z\mid a^*_j,X)-\widehat{\E}(Z\mid 1-a^*_j,X)}\frac{1}{\widehat{\Pr}(a^*_j\mid X)}\{Y-\widehat{\E}(Y\mid a^*_j,X)\}\\
& \qquad + \widehat{\E}(Y\mid a^*_j,X), \\
S_j =& \frac{A-\E(A\mid 1-a^*_j,X)}{\E(A\mid a^*_j,X)-\E(A\mid 1-a^*_j,X)}\frac{Z-\E(Z\mid 1-a^*_j,X)}{\E(Z\mid a^*_j,X)-\E(Z\mid 1-a^*_j,X)}\frac{1}{\Pr(a^*_j\mid X)}\{Y-\E(Y\mid a^*_j,X)\} \\
&\qquad + \E(Y\mid a^*_j,X).
\end{align*}}
\begin{align*}
R_2 =& \mathbb{P}(\hat{\varphi}-\varphi) \\
=&\sum^2_{j=1}\E\Bigl[ \mathbb{I}\{\widehat{d}=j\} \widehat{S}_j -  \mathbb{I}\{d=j\}S_j  \Bigr] \\
=& \sum^2_{j=1}\E\Bigl[ \mathbb{I}\{\widehat{d}=j\} (\widehat{S}_j-S_j) + ( \mathbb{I}\{\widehat{d}=j\} - \mathbb{I}\{d=j\})S_j  \Bigr] \\
\lesssim & \max_{j\in\{1,2\}}\E\left(\widehat{S}_j-S_j\right)  + \E\left[ \left|\mathbb{I}\{\widehat{d}=j\} - \mathbb{I}\{d=j\}\right|\right] \\
\lesssim & \max_{j=1,2}\E|\widehat S_j-S_j|+\Pr(\widehat d\ne d),
\end{align*}
where $\epsilon_1$ is defined in Assumption \ref{main-asp:separation}.
\[
R_2=\mathbb P(\widehat\varphi-\varphi)
=
\sum_{j=1}^2
\mathbb P\left[
\mathbb I\{\widehat d=j\}\widehat S_j
-
\mathbb I\{d=j\}S_j
\right].
\]
Adding and subtracting \(\mathbb I\{\widehat d=j\}S_j\), we obtain
\[
\begin{aligned}
|R_2|
&\le
\sum_{j=1}^2
\mathbb P\left[
\mathbb I\{\widehat d=j\}|\widehat S_j-S_j|
\right]
+
\sum_{j=1}^2
\mathbb P\left[
\left|
\mathbb I\{\widehat d=j\}-\mathbb I\{d=j\}
\right|\,|S_j|
\right]  \\
&\lesssim
\max_{j=1,2}\mathbb P|\widehat S_j-S_j|
+
\Pr(\widehat d\neq d),
\end{aligned}
\]
where the last inequality uses the boundedness of \(S_j\).
It remains to bound \(\Pr(\widehat d\neq d)\). Under Condition 1 or Condition 2
of Assumption~\ref{main-asp:label}, together with Assumption~\ref{main-asp:separation}, the population label ordering is
separated: for some \(\epsilon_1>0\),
\[
\mathbb P_n\{\mathbb \E(A\mid A^*=a_d^*,X)-\mathbb \E(A\mid A^*=a_{d^c}^*,X)\}
\ge \epsilon_1 .
\]
On the event \(\{\widehat d\neq d\}\), the empirical ordering is reversed, so
\[
\mathbb P_n\{\widehat{\mathbb E}(A\mid A^*=a_{d^c}^*,X)\}
\ge
\mathbb P_n\{\widehat{\mathbb E}(A\mid A^*=a_d^*,X)\}.
\]
Therefore, on \(\{\widehat d\neq d\}\),
\[
\begin{aligned}
\epsilon_1
&\le\mathbb P_n\{\mathbb \E(A\mid A^*=a_d^*,X)\}-\mathbb P_n\{\mathbb \E(A\mid A^*=a_{d^c}^*,X)\} \\
&\le\mathbb P_n\{\mathbb \E(A\mid A^*=a_d^*,X)-\widehat{\mathbb E}(A\mid A^*=a_d^*,X)\} \\
&\quad+\mathbb P_n\{\widehat{\mathbb E}(A\mid A^*=a_{d^c}^*,X) - \mathbb \E(A\mid A^*=a_{d^c}^*,X)\} \\
&\le2\max_{j=1,2}\left|\mathbb P_n\{\widehat{\mathbb E}(A\mid A^*=a_j^*,X) - \mathbb \E(A\mid A^*=a_j^*,X)\}\right|.
\end{aligned}
\]
Since
\[
\left|\mathbb P_n\{\widehat{\mathbb E}(A\mid A^*=a_j^*,X)-\mathbb \E(A\mid A^*=a_j^*,X)\}
\right|\le\left\|\widehat{\mathbb E}(A\mid A^*=a_j^*,X)-\mathbb \E(A\mid A^*=a_j^*,X)\right\|_\infty ,
\]
we have, for any \(r>0\),
\[
\mathbb I\{\widehat d\neq d\}\le\mathbb I\left[\max_{j=1,2}\left\|\widehat{\mathbb E}(A\mid A^*=a_j^*,X)-\mathbb \E(A\mid A^*=a_j^*,X)\right\|_\infty\ge \epsilon_1/2\right]
\]
\[
\le\left(\frac{2}{\epsilon_1}\right)^r\max_{j=1,2}\left\|\widehat{\mathbb E}(A\mid A^*=a_j^*,X)-\mathbb \E(A\mid A^*=a_j^*,X)\right\|_\infty^r .
\]
Combining this with the preceding bound for \(R_2\), we obtain
\[
|R_2|\lesssim\max_{j=1,2}\E|\widehat S_j-S_j|+\max_{j=1,2}\left\|\widehat{\mathbb E}(A\mid A^*=a_j^*,X)-\mathbb \E(A\mid A^*=a_j^*,X)\right\|_\infty^r .
\]
Note that there is a trade-off between the value of $r$ and $\epsilon_1$. When $\epsilon_1$ goes to zero, $r$ must also go to zero in order for $(2/\epsilon_1)^{r}$ to be finite.
For the term $\E\left(\left|\widehat{S}_j-S_j\right|\right)$, we denote
\begin{align*}
\frac{\E(A\mid A^*,X)-\widehat{\E}(A\mid 1-a^*,X)}
     {\widehat{\E}(A\mid a^*,X)-\widehat{\E}(A\mid 1-a^*,X)}
&= \frac{\widehat{\E}(A\mid A^*,X)-\widehat{\E}(A\mid 1-a^*,X)}
        {\widehat{\E}(A\mid a^*,X)-\widehat{\E}(A\mid 1-a^*,X)} \\
&\quad + \frac{\E(A\mid A^*,X)-\widehat{\E}(A\mid A^*,X)}
        {\widehat{\E}(A\mid a^*,X)-\widehat{\E}(A\mid 1-a^*,X)} \\
&= \mathbb{I}_{a^*}(A^*) + R_A(A^*), \\
\frac{\E(Z\mid A^*,X)-\widehat{\E}(Z\mid 1-a^*,X)}
     {\widehat{\E}(Z\mid a^*,X)-\widehat{\E}(Z\mid 1-a^*,X)}
&= \frac{\widehat{\E}(Z\mid A^*,X)-\widehat{\E}(Z\mid 1-a^*,X)}
        {\widehat{\E}(Z\mid a^*,X)-\widehat{\E}(Z\mid 1-a^*,X)} \\
&\quad + \frac{\E(Z\mid A^*,X)-\widehat{\E}(Z\mid A^*,X)}
        {\widehat{\E}(Z\mid a^*,X)-\widehat{\E}(Z\mid 1-a^*,X)} \\
&= \mathbb{I}_{a^*}(A^*) + R_Z(A^*).
\end{align*}
Write
\begin{align*}
D_A(a)&=\widehat{\E}(A\mid a^*,X)-\widehat{\E}(A\mid 1-a^*,X),\\
D_Z(a)&=\widehat{\E}(Z\mid a^*,X)-\widehat{\E}(Z\mid 1-a^*,X),\\
\Delta_A(a)&=\E(A\mid a,X)-\widehat{\E}(A\mid a,X),\\
\Delta_Z(a)&=\E(Z\mid a,X)-\widehat{\E}(Z\mid a,X),\\
\Delta_Y(a)&=\E(Y\mid a,X)-\widehat{\E}(Y\mid a,X),\\
\Delta_Y^{\dagger}(a)&=\E(Y\mid 1-a,X)-\widehat{\E}(Y\mid a,X),\\
\widehat G_A&=\frac{A-\widehat{\E}(A\mid 1-a^*,X)}{D_A(a)},\\
\widehat G_Z&=\frac{Z-\widehat{\E}(Z\mid 1-a^*,X)}{D_Z(a)},\\
\pi(X)&=\Pr(a^*\mid X),\qquad \widehat\pi(X)=\widehat{\Pr}(a^*\mid X).
\end{align*}
\begingroup
\allowdisplaybreaks
\footnotesize
Note that
\begin{align*}
&\E(\widehat{S}_j-S_j) \\
&= \E\Biggl[
    \frac{\widehat G_A\widehat G_Z}{\widehat\pi(X)}
    \{Y-\widehat{\E}(Y\mid a^*,X)\}
    + \widehat{\E}(Y\mid a^*,X) \Biggr]\\
& \qquad  - \psi_{a^*} \\
&= \E\left\{\E\left[
    \frac{\widehat G_A\widehat G_Z}{\widehat\pi(X)}
    \{Y-\widehat{\E}(Y\mid a^*,X)\}
    \Bigm| A^*,X\right]\right\} \\
& \qquad + \E\{\widehat{\E}(Y\mid a^*,X)\} - \E\{\E(Y\mid a^*,X)\} \\
&= \E\Biggl\{
    \frac{[\mathbb{I}_{a^*}(A^*) + R_A][\mathbb{I}_{a^*}(A^*)+R_Z]}
         {\widehat\pi(X)}
    [\E(Y\mid A^*,X)-\widehat{\E}(Y\mid a^*,X)] \Bigg\} \\
&\qquad + \E\{\widehat{\E}(Y\mid a^*,X) - \E(Y\mid a^*,X)\} \\
&= \E\Biggl\{
    \frac{\mathbb{I}_{a^*}(A^*) + \mathbb{I}_{a^*}(A^*)R_A
          + \mathbb{I}_{a^*}(A^*)R_Z + R_AR_Z}
         {\widehat\pi(X)}
    [\E(Y\mid A^*,X)-\widehat{\E}(Y\mid a^*,X)] \Bigg\} \\
&  \qquad+ \E\{\widehat{\E}(Y\mid a^*,X) - \E(Y\mid a^*,X)\} \\
=& \E \left[
    \frac{\pi(X)-\widehat\pi(X)}{\widehat\pi(X)}
    \Delta_Y(a^*)\right] \\
& \qquad + \E \left[
    \frac{\pi(X)}{\widehat\pi(X)}
    \frac{\Delta_A(a^*)}{D_A(a)}
    \Delta_Y(a^*) \right] \\
& \qquad + \E \left[
    \frac{\pi(X)}{\widehat\pi(X)}
    \frac{\Delta_Z(a^*)}{D_Z(a)}
    \Delta_Y(a^*) \right] \\
&  \qquad+ \E \left[
    \frac{\pi(X)}{\widehat\pi(X)}
    \frac{\Delta_A(a^*)}{D_A(a)}
    \frac{\Delta_Z(a^*)}{D_Z(a)}
    \Delta_Y(a^*) \right] \\
&  \quad+ \E\Biggl[
    \frac{1-\pi(X)}{\widehat\pi(X)}
    \frac{\Delta_A(1-a^*)}{D_A(a)}
    \frac{\Delta_Z(1-a^*)}{D_Z(a)} \\
&  \hspace{38mm}\times
    \Delta_Y^\dagger(a^*) \Biggr].
\end{align*}
\endgroup
Therefore,
\begin{align*}
\E\left(\widehat{S}_j-S_j\right)
& \lesssim
\|\widehat{\Pr}(a^*\mid X)-\Pr(a^*\mid X)\|
\|\widehat{\E}(Y\mid a^*,X)-E(Y\mid a^*,X)\| \\
& \quad + \|\widehat{\E}(A\mid a^*,X)-\E(A\mid a^*,X)\|
\|\widehat{\E}(Y\mid a^*,X)-\E(Y\mid a^*,X)\| \\
& \quad + \|\widehat{\E}(Z\mid a^*,X)-\E(Z\mid a^*,X)\|
\|\widehat{\E}(Y\mid a^*,X)-\E(Y\mid a^*,X)\| \\
& \quad + \| \E(A\mid 1-a^*,X) - \widehat{\E}(A\mid 1-a^*,X) \| \\
& \qquad \times
\| \E(Z\mid 1-a^*,X) - \widehat{\E}(Z\mid 1-a^*,X) \|.
\end{align*}
\section{Proof of Theorem \ref{main-thm:extension}}\label{sec:proof_extension}
It suffices to show that
\begin{align*}
\E\{\phi(Y,A,Z,X) \mid Y,A^*,X \} &= \E\{\phi^F(Y,A^*,X) \mid Y,A^*,X  \}, \\
\E\{\phi(Y,A,Z,X) \mid A,A^*,X \} &= \E\{\phi^F(Y,A^*,X) \mid A,A^*,X  \}, \\
\E\{\phi(Y,A,Z,X) \mid Z,A^*,X \} &= \E\{\phi^F(Y,A^*,X) \mid Z,A^*,X  \}.
\end{align*}
Note that $Y\bigCI A\bigCI Z \mid (A^*,X)$ and we have
\begin{align*}
\E\{ M_{a^*,A}(A,X)\mid Y,A^*,Z,X\} = \E\{ M_{a^*,Z}(Z,X)\mid Y,A^*,A,X\} = \mathbb{I}(A^*=a^*),
\end{align*}
which implies
\begin{align*}
&\E\{ M_{a^*,A}(A,X)M_{a^*,Z}(Z,X)\mid Y,A^*,X\} \\
&\qquad =
\E\{ M_{a^*,A}(A,X)\mid A^*,X\}
\E\{ M_{a^*,Z}(Z,X)\mid A^*,X\}\\
&\qquad = \mathbb{I}(A^*=a^*).
\end{align*}
Therefore, the first equality holds.
The second and third equalities are isomorphic so it suffices to prove one of them. Note that
\begin{align*}
& \E\{\phi(Y,A,Z,X) \mid Y, A, A^*,X\}\\
= &\sum_{a^*=0,1}\Bigl\{
q_{0,a^*}(X) h_{0,a^*}(X) g_{1,a^*}(Y,X)
M_{a^*,A} \mathbb{I}(A^*=a^*) \\
&\qquad
+q_{0,a^*}(X) g_{2,a^*}( Y,X)
M_{a^*,A} \mathbb{I}(A^*=a^*) \\
&\qquad
+ h_{0,a^*}(X) g_{3,a^*}( Y,X) +g_{4,a^*}( Y,X) \Bigr\}.
\end{align*}
Furthermore, for $a^*=0,1$,
\begin{align*}
&\E\Bigl[
q_{0,a^*}(X) h_{0,a^*}(X) g_{1,a^*}(Y,X)
M_{a^*,A} \mathbb{I}(A^*=a^*)\\
&\qquad
+q_{0,a^*}(X) g_{2,a^*}( Y,X)
M_{a^*,A} \mathbb{I}(A^*=a^*)
\mid A, A^*,X \Bigr] \\
=& M_{a^*,A}q_{0,a^*}(X)\mathbb{I}(A^*=a^*)
\E\left[h_{0,a^*}(X) g_{1,a^*}(Y,X) + g_{2,a^*}( Y,X)
\mid A, A^*,X\right] \\
=& M_{a^*,A}q_{0,a^*}(X)\mathbb{I}(A^*=a^*)
\E\left[h_{0,a^*}(X) g_{1,a^*}(Y,X) + g_{2,a^*}( Y,X)
\mid A^*=a^*,X\right]\\
=& 0,
\end{align*}
where the last line makes use of Proposition 2 of \cite{ghassami2022minimax},
\begin{align*}
 \E\left[h_{0,a^*}(X) g_{1,a^*}(Y,X) + g_{2,a^*}( Y,X) \mid A^*=a^*,X\right] = 0.
\end{align*}
This implies that
\begin{align*}
& \E\{\phi(Y,A,Z,X) \mid A, A^*,X\}\\
=& \E\left[\E\{\phi(Y,A,Z,X) \mid Y, A, A^*, X\} \mid A,A^*,X \right] \\
=& \sum_{a^*=0,1} \E \left[ h_{0,a^*}(X) g_{3,a^*}( Y,X) +g_{4,a^*}( Y,X) \mid A,A^*,X \right] \\
=& \E\{\phi^F(Y,A^*,X) \mid A, A^*, X\},
\end{align*}
which concludes the proof.
We also provide a more general theorem, of which Theorem \ref{main-thm:extension} is a special case.
\begin{theorem}\label{thm:dr_if2}
Suppose the influence functions for the parameter of interest $\psi$ in the full data model $\mathcal{M}^*$ are of the form
\begin{align*}
\begin{split}
\phi^F(Y,A^*,X,M) &=\sum_{a^*=0,1}\Bigl\{q_{0,a^*}(V_q) h_{0,a^*}(V_h) g_{1,a^*}(V) \mathbb{I}( A^*=a^*)  \\
&\quad +q_{0,a^*}(V_q) g_{2,a^*}( V) \mathbb{I}(A^*=a^*) +h_{0,a^*}(V_h) g_{3,a^*}(V) +g_{4,a^*}(Y,X) \Bigr\},
\end{split}
\end{align*}
where $V=(Y,X,W)$, $X\subset V_q\subset V$, $X\subset V_h\subset V$; $q_{0,a^*}(V_q)$ and $h_{0,a^*}(V_h) $ are unknown nuisance functions and $g_{j,a^*}$ are known functions of $V$.
Then
\begin{align*}
\begin{split}
\phi(Y,A,Z,X,M) &=\sum_{a^*=0,1}\Bigl\{q_{0,a^*}(V_q) h_{0,a^*}(V_h) g_{1,a^*}(V) M_{a^*,A}M_{a^*,Z}  \\
&\quad +q_{0,a^*}(V_q) g_{2,a^*}( V) M_{a^*,A}M_{a^*,Z} +h_{0,a^*}(V_h) g_{3,a^*}(V) +g_{4,a^*}(Y,X) \Bigr\}
\end{split}
\end{align*}
is an observed data influence function for $\psi$ in the hidden treatment model, where $M_{a^*,A}$ and $M_{a^*,Z}$ are defined in \eqref{main-eq:m_indicator}.
\end{theorem}
This class includes all examples we consider in the paper (with no $W$ and $V_q=V_h=X$) but also includes some other examples such as proximal causal inference \citep{cui2024semiparametric}, where treatment and outcome proxies are included in $W$.
To prove Theorem \ref{thm:dr_if2}, we note that $(A,Z)\bigCI (Y,W) \mid (A^*,X)$ and $A\bigCI Z\mid (A^*,X)$, therefore
\begin{align*}
& \E\{ q_{0,a^*}(V_q) h_{0,a^*}(V_h) g_{1,a^*}(V) M_{a^*,A}M_{a^*,Z}\mid Y,A^*,X \}\\
=& \E\left[ \E\{ q_{0,a^*}(V_q) h_{0,a^*}(V_h) g_{1,a^*}(V) M_{a^*,A}M_{a^*,Z}\mid Y,A^*,X,W \}\mid Y,A^*,X\right] \\
=& \E\left[  q_{0,a^*}(V_q) h_{0,a^*}(V_h) g_{1,a^*}(V) \E\{M_{a^*,A}M_{a^*,Z}\mid A^*,X \}\mid Y,A^*,X\right] \\
=& \E\left[  q_{0,a^*}(V_q) h_{0,a^*}(V_h) g_{1,a^*}(V) \mathbb{I}( A^*=a^*) \mid Y,A^*,X\right].
\end{align*}
Similarly,
\begin{align*}
    \E\{ q_{0,a^*}(V_q) g_{2,a^*}( V) M_{a^*,A}M_{a^*,Z} \mid Y,A^*X \} = \E\{ q_{0,a^*}(V_q) g_{2,a^*}( V) \mathbb{I}(A^*=a^*) \mid Y,A^*X \}.
\end{align*}
So it holds that $\E\{\phi(Y,A,Z,X,M) \mid Y,A^*,X \} = \E\{\phi^F(Y,A^*,X,M) \mid Y,A^*,X  \}$.
Now we prove the second equality.
\begin{align*}
& \E\{\phi(Y,A,Z,X,W) \mid Y, A, A^*,X,W\} \\
= &\sum_{a^*=0,1}\Bigl\{q_{0,a^*}(V_q) h_{0,a^*}(V_h) g_{1,a^*}(V) M_{a^*,A} \E\{M_{a^*,Z}(Z,X)\mid Y, A, A^*,X,W\} \\
& +q_{0,a^*}(V_q) g_{2,a^*}( Y,X) M_{a^*,A}\E\{M_{a^*,Z}(Z,X)\mid Y, A, A^*,X,W\}  + h_{0,a^*}(V_h) g_{3,a^*}( V) +g_{4,a^*}( V) \Bigr\}\\
=& \sum_{a^*=0,1}\Bigl\{q_{0,a^*}(V_q) h_{0,a^*}(V_h) g_{1,a^*}(V) M_{a^*,A} \mathbb{I}(A^*=a^*) \\
& +q_{0,a^*}(V_q) g_{2,a^*}( Y,X) M_{a^*,A}\mathbb{I}(A^*=a^*)  + h_{0,a^*}(V_h) g_{3,a^*}( V) +g_{4,a^*}( V)\Bigr\}.
\end{align*}
Furthermore,
\begin{align*}
&\E\left[q_{0,a^*}(V_q) h_{0,a^*}(V_h) g_{1,a^*}(V) M_{a^*,A} \mathbb{I}(A^*=a^*) +q_{0,a^*}(V_q) g_{2,a^*}(V) M_{a^*,A} \mathbb{I}(A^*=a^*) \mid A, A^*,V_q \right] \\
=& M_{a^*,A}q_{0,a^*}(V_q)\mathbb{I}(A^*=a^*) \E\left[h_{0,a^*}(V_h) g_{1,a^*}(V) + g_{2,a^*}(V) \mid A, A^*,V_q\right] \\
=& M_{a^*,A}q_{0,a^*}(V_q)\mathbb{I}(A^*=a^*) \E\left[h_{0,a^*}(V_h) g_{1,a^*}(V) + g_{2,a^*}(V) \mid A^*=a^*,V_q\right]\\
=& 0,
\end{align*}
where the last line is by Proposition 2 of \cite{ghassami2022minimax},
\begin{align*}
 \E\left[h_{0,a^*}(V_h) g_{1,a^*}(V) + g_{2,a^*}(V) \mid A^*=a^*,V_q\right] = 0.
\end{align*}
This implies that
\begin{align*}
& \E\{\phi(Y,A,Z,X,W) \mid A, A^*, X\}\\
=& \E\left(\E\left[\E\{\phi(Y,A,Z,X,W) \mid Y, A, A^*,X,W\} \mid A,A^*,V_q \right]\mid A,A^*,X\right) \\
=& \sum_{a^*=0,1} \E \left[ h_{0,a^*}(V_h) g_{3,a^*}(V) +g_{4,a^*}(V) \mid A,A^*,X \right] \\
=& \E\{\phi^F(Y,A^*,X,W) \mid A, A^*, X\}.
\end{align*}
The proof of the third equation follows the same method as that of the second equation.
\section{Details in simulation study}\label{sec:simu_details}
The data generating process is described in the main text.
The true target values used for simulation summaries were computed by Monte Carlo integration with 200,000 draws and seed 999.
\subsection{Estimator implementation}
The nuisance functions were updated by the iterative procedure described in the main text. Probability-valued nuisance functions were updated on the log-odds scale; for continuous outcomes, the conditional outcome means were updated on their original scale. At each iteration, the updated target functions were projected onto a tensor-product cubic regression spline sieve using \texttt{mgcv::gam} with
\[
\texttt{te}(X_1,X_2,\texttt{bs}=c(\texttt{"cr"},\texttt{"cr"}),\texttt{k}=c(K_1,K_2),\texttt{fx}=TRUE).
\]
Thus the basis was fixed and unpenalized.
Starting values were obtained using the randomized GMM initialization routine. For replication \(k\), the data seed was \(1100+k\), and the restart seed base was \(100000+1000k\). The initialization jitter was 0.01. Each failed attempt was restarted with a new randomized initialization.
\subsection{Tuning parameters}
Tables~\ref{tab:simulation-tuning-core} and~\ref{tab:simulation-tuning-diagnostics} list the tuning parameters used for the final simulation results. The notation used in these tables is defined in Table~\ref{tab:tuning-notation}. Briefly, each Monte Carlo replication was estimated by repeated attempts. An attempt starts from randomized GMM initial values and iterates the nuisance-function updates. The attempt is declared converged only after the burn-in period and only if all convergence diagnostics are below their thresholds for \(c\) consecutive iterations. If an attempt fails, a new attempt is started with a new randomized initialization, up to \(R\) attempts.
\begin{table}[!ht]
\centering
\caption{Notation for tuning parameters.}
\label{tab:tuning-notation}
\begin{tabular}{lp{0.72\linewidth}}
\hline
Symbol & Meaning \\
\hline
\(K_1,K_2\) & Tensor-product spline basis dimensions for \(X_1\) and \(X_2\). \\
\(\eta\) & Step size in the iterative nuisance-function update. \\
\(M\) & Maximum number of iterations within one attempt. \\
\(\tau_{\Delta}\) & Threshold for the maximum state change between consecutive iterations. \\
\(B\) & Burn-in iterations before convergence can be declared. \\
\(c\) & Required number of consecutive iterations satisfying convergence criteria. \\
\(R\) & Maximum number of randomized restart attempts per Monte Carlo replication. \\
\(\tau_g\) & Guard threshold for abandoning numerically unstable attempts. \\
\(\tau_m\) & Threshold for the mean estimating-equation diagnostic. \\
\(\tau_c\) & Threshold for the conditional estimating-equation diagnostic. \\
\(\tau_f\) & Failure threshold for the conditional diagnostic after \(B/2\) iterations. \\
\hline
\end{tabular}
\end{table}
At iteration \(t\), let \(\Delta_t\) denote the maximum absolute change across the updated nuisance-function states. Let \(D_{\mathrm{mean},t}\) and \(D_{\mathrm{cond},t}\) denote the mean and conditional estimating-equation diagnostics. An attempt is counted as converged if, for \(c\) consecutive iterations after iteration \(B\),
\[
\Delta_t < \tau_{\Delta},\qquad
D_{\mathrm{mean},t} < \tau_m,\qquad
D_{\mathrm{cond},t} < \tau_c.
\]
An attempt is abandoned and restarted if it becomes numerically unstable, if \(\Delta_t>\tau_g\), if it reaches \(M\) iterations without satisfying the convergence rule, or if \(D_{\mathrm{cond},t}>\tau_f\) after \(B/2\) iterations.
\begin{table}[!ht]
\centering
\caption{Core iteration and sieve tuning parameters for the final simulation results.}
\label{tab:simulation-tuning-core}
\begin{tabular}{llrrrrrr}
\hline
Outcome & \(n\) & \(K_1\) & \(K_2\) & \(\eta\) & \(M\) & \(\tau_{\Delta}\) & \(B\) \\
\hline
Binary & 500  & 3 & 3 & 0.001 & 1000 & 0.02 & 200 \\
Binary & 1000 & 3 & 3 & 0.001 & 1000 & 0.02 & 200 \\
Binary & 2000 & 4 & 4 & 0.001 & 1000 & 0.02 & 200 \\
Continuous & 500  & 3 & 3 & 0.001 & 1000 & 0.02 & 200 \\
Continuous & 1000 & 3 & 3 & 0.001 & 1000 & 0.02 & 200 \\
Continuous & 2000 & 4 & 4 & 0.001 & 1000 & 0.02 & 200 \\
\hline
\end{tabular}
\end{table}
\begin{table}[!ht]
\centering
\caption{Restart and convergence-diagnostic tuning parameters for the final simulation results.}
\label{tab:simulation-tuning-diagnostics}
\begin{tabular}{llrrrrrr}
\hline
Outcome & \(n\) & \(c\) & \(R\) & \(\tau_g\) & \(\tau_m\) & \(\tau_c\) & \(\tau_f\) \\
\hline
Binary & 500  & 30 & 500 & 0.05 & 0.10 & 1.5 & 2.0 \\
Binary & 1000 & 30 & 500 & 0.05 & 0.10 & 1.5 & 2.0 \\
Binary & 2000 & 30 & 500 & 0.05 & 0.10 & 1.5 & 2.0 \\
Continuous & 500  & 30 & 500 & 0.05 & 0.05 & 1.0 & 1.5 \\
Continuous & 1000 & 30 & 500 & 0.05 & 0.05 & 1.0 & 1.5 \\
Continuous & 2000 & 30 & 500 & 0.05 & 0.05 & 1.0 & 1.5 \\
\hline
\end{tabular}
\end{table}
The following implementation choices were common to all scenarios. Initial nuisance values were generated by the randomized GMM initialization described above, with Gaussian perturbation scale 0.01. All probability-valued nuisance estimates were truncated to lie in \([10^{-6},1-10^{-6}]\). Denominators involving contrasts between latent classes, such as differences in conditional means or probabilities, were thresholded away from zero at 0.05; variance-like denominators in the update equations were also thresholded at 0.05. For binary outcomes, attempts producing \(\hat\psi_1\) or \(\hat\psi_0\) outside \([0,1]\) were rerun with a new randomized initialization. For continuous outcomes, the corresponding bounds were the observed range of \(Y\) in that replication expanded by 5\% on each side. No separate bound was imposed on the estimated average treatment effect. The output matrix in each combined result file has eight columns: success indicator, \(\hat\psi_1\), standard error for \(\hat\psi_1\), \(\hat\psi_0\), standard error for \(\hat\psi_0\), estimated average treatment effect \(\hat\psi_1-\hat\psi_0\), standard error for the average treatment effect, and the number of restart attempts.
\section{ADNI data usage acknowledgment}\label{sec:adni_acknowledgment}
Data used in the preparation of this article were obtained from the Alzheimer’s Disease Neuroimaging Initiative (ADNI) database (adni.loni.usc.edu). The ADNI was launched in 2003 as a public-private partnership, led by Principal Investigator Michael W. Weiner, MD. The primary goal of ADNI has been to test whether serial magnetic resonance imaging (MRI), positron emission tomography (PET), other biological markers, and clinical and neuropsychological assessment can be combined to measure the progression of mild cognitive impairment and early Alzheimer’s disease (AD). For up-to-date information, see www.adni-info.org.
Data collection and sharing for this project was funded by the Alzheimer’s Disease Neuroimaging Initiative (ADNI) (National Institutes of Health Grant U01 AG024904) and DOD ADNI (Department of Defense award number W81XWH-12-2-0012). ADNI is funded by the National Institute on Aging, the National Institute of Biomedical Imaging and Bioengineering, and through generous contributions from the following: AbbVie, Alzheimer’s Association; Alzheimer’s Drug Discovery Foundation; Araclon Biotech; BioClinica, Inc.; Biogen; Bristol-Myers Squibb Company; CereSpir, Inc.; Cogstate; Eisai Inc.; Elan Pharmaceuticals, Inc.; Eli Lilly and Company; EuroImmun; F. Hoffmann-La Roche Ltd and its affiliated company Genentech, Inc.; Fujirebio; GE Healthcare; IXICO Ltd.; Janssen Alzheimer Immunotherapy Research \& Development, LLC.; Johnson \& Johnson Pharmaceutical Research \& Development LLC.; Lumosity; Lundbeck; Merck \& Co., Inc.; Meso Scale Diagnostics, LLC.; NeuroRx Research; Neurotrack Technologies; Novartis Pharmaceuticals Corporation; Pfizer Inc.; Piramal Imaging; Servier; Takeda Pharmaceutical Company; and Transition Therapeutics. The Canadian Institutes of Health Research is providing funds to support ADNI clinical sites in Canada. Private sector contributions are facilitated by the Foundation for the National Institutes of Health (www.fnih.org). The grantee organization is the Northern California Institute for Research and Education, and the study is coordinated by the Alzheimer’s Therapeutic Research Institute at the University of Southern California. ADNI data are disseminated by the Laboratory for Neuro Imaging at the University of Southern California.